\documentclass[reprint, aps, prd, superscriptaddress, nofootinbib]{revtex4-2}

\usepackage{graphicx}
\usepackage{amsmath}
\usepackage{amssymb}
\usepackage{orcidlink}
\usepackage{ulem}
\usepackage[export]{adjustbox}
\usepackage{xcolor, soul}

\newcommand{\du}{\mathrm{d}}   
\newcommand{\iu}{\mathrm{i}}   

\newcommand*{\defeq}{\mathrel{\vcenter{\baselineskip0.5ex \lineskiplimit0pt
                     \hbox{\scriptsize.}\hbox{\scriptsize.}}}%
                     =}

\newcommand{\jena}{Institute for Theoretical Physics, University of Jena, Fröbelstieg 1, 07743, Jena, Germany}
\newcommand{\damtp}{Department of Applied Mathematics and Theoretical Physics, Centre for Mathematical Sciences, University of Cambridge, Wilberforce Road, Cambridge CB3 0WA, United Kingdom}
\newcommand{\jhu}{Department of Physics and Astronomy, Johns Hopkins University, 3400 N Charles Street, Baltimore, Maryland 21218, USA}
\newcommand{\caltech}{Theoretical Astrophysics 350-17, California Institute of Technology, 1200 E California Boulevard, Pasadena, California 91125, USA}

\begin{document}

\normalem

\title{Boson stars in massless and massive scalar-tensor gravity}

\author{Tamara Evstafyeva
\orcidlink{0000-0002-2818-701X}}
\email{te307@maths.cam.ac.uk}
\affiliation{\damtp}

\author{Roxana Ro\textcommabelow{s}ca-Mead \orcidlink{0000-0001-5666-1033}}
\email{roxana.rosca-mead@uni-jena.de}
\affiliation{\jena}

\author{Ulrich Sperhake \orcidlink{0000-0002-3134-7088}}
\email{U.Sperhake@damtp.cam.ac.uk}
\affiliation{\damtp}
\affiliation{\jhu}
\affiliation{\caltech}

\author{Bernd Br{\"u}gmann
\orcidlink{0000-0003-4623-0525}}
\email{bernd.bruegmann@uni-jena.de}
\affiliation{\jena}

\date{\today}

\begin{abstract}
We study phenomenological features and stability of boson stars in
massless and massive scalar-tensor theory of gravity with
Damour-Esposito-Far\`ese coupling. This coupling between the tensor and
scalar sectors of the theory leads to a phenomenon called spontaneous
scalarization, the onset of which we investigate by numerically
computing families of boson-star models using shooting and relaxation
algorithms. We systematically explore the effects of the theory's
coupling, the mass of the gravitational scalar and the choice of the
bosonic potential on the structure of weakly and strongly scalarized
solutions. Scalarized boson-star models share many common features
with neutron stars in the same scalar-tensor theory of gravity. In
particular, scalarization can result in boson stars with significantly
larger radii and masses, which tend to be energetically favored
over their weakly or non-scalarized counterparts. Overall, we find
that boson stars are not quite as susceptible to scalarization as
neutron stars.
\end{abstract}

\maketitle

\section{Introduction}\label{sec:introduction}
Boson stars (BSs) are equilibrium configurations of hypothetical
bosonic particles, first proposed by Kaup \cite{Kaup:1968zz}. The
stars are balanced by the Heisenberg uncertainty principle inhibiting
gravitational collapse and by the gravitational attraction counteracting
the dispersion. These theoretical objects have recently gained
significant interest as dark matter candidates \cite{Guver:2012ba,
Alcubierre:2001ea, Hu:2000ke, Macedo:2013qea} and as astrophysical
models mimicking in certain regimes the observational properties
of black holes (BHs) and neutrons stars (NSs) \cite{Sennett:2017etc,
Herdeiro:2021lwl, Rosa:2022tfv, Khlopov:1985jw, Brihaye:2020klz, Rosa:2023qcv, Rosa:2022toh}.
BSs\footnote{Throughout this work, with the term ``boson star'' we
refer to self-gravitating equilibrium configurations composed of a
complex {\it scalar} field.} have been widely studied in the context
of general relativity (GR), both as single BS spacetimes
\cite{Colpi:1986ye,Seidel:1990jh,Kobayashi:1994qi,Ryan:1996nk,Schunck:1996he,Balakrishna:1997ej,Yoshida:1997qf,Schunck:1999zu,Schunck:2003kk,Balakrishna:2006ru,Balakrishna:2007mr,Hartmann:2012da,Siemonsen:2020hcg},
in terms of their formation
\cite{Schunck:1999pm,Sanchis-Gual:2019ljs,Siemonsen:2023hko} and
as binary systems
\cite{Palenzuela:2006wp,Palenzuela:2007dm,Palenzuela:2017kcg,Helfer:2021brt,Sanchis-Gual:2020mzb,Bezares:2022obu,Cardoso:2022vpj,Croft:2022bxq,Sanchis-Gual:2022zsr,Evstafyeva:2022bpr,
Siemonsen:2023age}; see also
Refs.~\cite{Jetzer:1991jr,Mielke:1997re,Liebling:2012fv,Visinelli:2021uve}
for reviews.  Their understanding in modified theories of gravity,
however, remains scarce. BSs in Brans-Dicke type theories with
various couplings inspired from cosmology have been considered in
Refs.~\cite{Torres:1997np, Balakrishna:1997ek, Comer:1997ns}, in
Palatini $f(\mathcal{R})$ gravity in Refs.~\cite{Maso-Ferrando:2021ngp,
Maso-Ferrando:2023nju} and in $f(T)$ extended theory of gravity in Ref.~\cite{Ilijic:2020vzu}.

ST theories of gravity are one of the most natural and simplest
extensions of GR that preserve the universality of a free fall
\cite{Will:1993}. The key ingredient of ST gravity is the presence
of a scalar field non-minimally coupled to the metric such that
gravity is not only mediated by a spin-2 graviton but also by the
spin-0 scalar field. The inclusion of the scalar field is a common
by-product of the compactification of higher dimensional Kaluza-Klein
type theories, like string theory, \cite{Taylor:1988nw,
1988MPLA....3..243M, Damour:1994zq}, making ST theories well-motivated
gravity theories to study.

The theoretical implications of ST theories of gravity in various
areas of fundamental physics are vast. For instance, on cosmological
scales, ST theories have been proposed as models for dark energy,
replacing the role of the cosmological constant \cite{Riazuelo:2001mg,
Schimd:2004nq}. Furthermore, the additional scalar degree of freedom
introduces a new channel of radiation emission in the form of scalar
gravitational waves (GWs) also known as a ``breathing mode''. The
perfect laboratories for tests and searches of these scalar waves
are compact objects and binary systems composed thereof. In particular,
it has been conjectured that in the process of \textit{spontaneous
scalarization}, the scalar radiation can be detected with future
gravitational wave detectors \cite{6d97dcf5be1a4f88aed4ff8b4d1635b7}.
The phenomenon of spontaneous scalarization is caused by a linear
tachyonic instability inside the star that triggers the growth of the
scalar field around it. Once the instability is quenched, the star
becomes endowed with "scalar hair"; see Ref.~\cite{Doneva:2022ewd}
for a comprehensive review. This remarkable feature was first
identified in the pioneering work of Damour and Esposito-Far\`ese
\cite{Damour:1992we, Damour:1993hw, Damour:1996ke}, who studied
spontaneous scalarization of neutron stars in massless ST theory
with a pair of parameters $(\alpha_0, \beta_0)$, inspired by
Post-Newtonian (PN) theory and controlling the onset of scalarization.

However, stringent constraints have been placed on massless ST
theories by numerous observations, including the Cassini bound
$\alpha_0 < 3.4\times 10^{-3}$ \cite{2003Natur.425..374B} and
pulsar–white dwarf binaries which rule out models with $\beta_0
\lesssim -5$ \cite{Freire:2012mg,Antoniadis:2013pzd}; for a summary
of constraints on a broad range of modified theories of gravity
obtained from gravitational-wave (GW) observations see also
\cite{Yunes:2016jcc}.  As a consequence, massive ST theories have
acquired a great deal of attention, mainly for two reasons.  First,
the screening of the massive field maintains their compatibility
with current observations and leaves a significant portion of the
parameter space unconstrained. Second, because the mass term endows
the scalar GW signal with a highly conspicuous long-lived, inverse
chirp character
\cite{Ramazanoglu:2016kul,2017PhRvL.119t1103S,2020PhRvD.102d4010R,Aurrekoetxea:2022ika,Blazquez-Salcedo:2021exm,Kuan:2022oxs,Rosca-Mead:2023tdc,Kuroda:2023zbz,
Yazadjiev:2016pcb, Morisaki:2017nit, PhysRevD.93.064005}; see also
Refs.~\cite{Rosca-Mead:2019seq,Huang:2021tpu,Asakawa:2023obq} for
studies of massive ST theory with self-interacting fields.

So far NSs have been extensively explored in this class of ST
theories as ideal compact matter objects, in whose environment
scalarization can occur in this class of ST theories \cite{Ma:2023sok,
Yagi:2021loe, Barausse:2012da} (see also Ref.~\cite{Odintsov:2021nqa} on NSs in ST theory with quadratic conformal coupling). An intriguing extension is to investigate how the presence of a non-minimally coupled scalar
field affects the phenomenology of other compact matter objects,
such as BSs. The phenomena of spontaneous and induced scalarization of BSs have been previously considered in Refs.~\cite{Whinnett:1999ma, Whinnett:1999sc, Ruiz:2012jt, Alcubierre:2010ea, Brihaye:2019puo}. In this work we extend the study of BSs
in scalar-tensor (ST) theories of gravity with Damour-Esposito-Far\`ese
coupling \cite{PhysRevLett.70.2220} by focusing on
the scalarization of ground-state equilibrium BSs for both
massless and massive gravitational scalar fields as well as the
stars' stability for various BS potentials.

This paper is structured as follows. In Sec.~\ref{sec:theory}, we
present an overview of ST gravity with the specific
Damour-Esposito-Far\`ese coupling function employed in our work,
and formulate the system of first-order differential equations with
the bosonic field in the matter sector. In Sec.~\ref{sec:results},
we present the results of our exploration of the parameter space
and describe the main features of BS models in ST theory of gravity.
We span the solution space for different parameters of the coupling
function, mass values of the gravitational scalar field and BS
potentials. In Sec.~\ref{sec:stability}, we assess the stability
of these models.  Finally, we summarize our results together with
an outlook on future research directions in Sec.~\ref{sec:conclusions}.
In Appendix \ref{app:asymptotic_behaviour}, we study in detail the
asymptotic behavior of BS solutions in ST theory of gravity and GR,
highlighting the main inconsistencies that can arise in a {\it
flat-field} treatment.  Our numerical methods are described in
Appendix \ref{app:numerics} and in Appendix \ref{app:BSfamilies},
we illustrate the structure of BS families of solutions for a broad
range of BS potentials.  Unless specified otherwise, we employ units
where the speed of light and Planck's constant satisfy $c=1=\hbar$,
so that the Planck mass is given by $M_{\rm Pl}=1/\sqrt{G}$.

\section{Theory and formulation}\label{sec:theory}
\subsection{Scalar-tensor action} 
\label{subsec:action}
We start by considering a general class of ST theories of gravity
with a single, real-valued gravitational scalar field $\varphi$
with potential $W(\varphi)$, whose action in the Einstein frame is
given by \cite{Salgado:2005hx},
\begin{eqnarray}
  S &=& \int \du x^4 \frac{\sqrt{-\bar{g}}}{16\pi}
  \left[
  \frac{\bar{R}}{G}-2\bar{g}^{\mu\nu} \partial_{\mu} \varphi\,
  \partial_{\nu}\varphi-4W(\varphi)
  \right]
  + S_{\rm M}\,,
  \nonumber \\[10pt]
  S_{\rm M} &=& \int \du x^4 \sqrt{-g}
  \left\{-\frac{1}{2}g^{\mu\nu}\nabla_{\mu}\psi^*
  \nabla_{\nu}\psi + V(\psi)]\right\}\,.
  \label{eq:S}
\end{eqnarray}
Here, $\bar{g}_{\mu\nu}$ is a conformal metric, with corresponding
Ricci scalar $\bar{R}$, related to the physical or Jordan-frame
metric $g_{\mu\nu}$ by\footnote{In our notation for the conformal
factor we follow Salgado \cite{Salgado:2005hx}; a common and
equivalent alternative is to write Eq.~(\ref{eq:ggbar}) as
$g_{\mu\nu}=a^2(\varphi)\bar{g}_{\mu\nu}$ where $a^2=1/F$.}
\begin{equation}
  \bar{g}_{\mu\nu} = F(\varphi) g_{\mu\nu},
  \label{eq:ggbar}
\end{equation}
where $F(\varphi)$ is the \textit{coupling function}, which needs
to be positive to ensure that the graviton carries positive energy
\cite{Damour:1992we}. For the matter sources, represented in
Eq.~(\ref{eq:S}) by the term $S_{\rm M}$, we consider a relativistic
boson condensate, modelled as a massive complex scalar field
$\psi=\psi_R+\iu \psi_I$ with complex conjugate $\psi^*$ and potential
$V(\psi)$. Note that by Eq.~(\ref{eq:S}) the matter moves according
to the geometry of the {\it physical} metric $g_{\mu\nu}$; this is
the price tag for achieving the minimal coupling between the
gravitational scalar and the conformal metric $\bar{g}_{\mu\nu}$
of the Einstein frame.  Although the scalar field does not directly
interact with the ordinary matter (which guarantees the weak
equivalence principle), it affects the motion of test particles
through its presence in Eq.~\eqref{eq:ggbar}.  The appearance of
two distinct scalar fields in our physical setup inevitably holds
potential for confusion; whenever the meaning is not self-evident,
we will therefore explicitly refer to the two fields as the {\it
gravitational scalar} $\varphi$ and the {\it BS scalar} $\psi$ with
amplitude $A\defeq |\psi|$.

The covariant field equations are obtained from varying the action
(\ref{eq:S}) with respect to $\bar{g}_{\mu\nu}$, $\varphi$ and
$\psi$. This leads to
\begin{align}
  &\frac{\bar{G}_{\alpha\beta}}{G} = 2\partial_{\alpha}\varphi
  \partial_{\beta}\varphi
  -\bar{g}_{\alpha\beta} \bar{g}^{\mu\nu}
  \partial_{\mu}\varphi\,\partial_{\nu}\varphi
  - 2W \bar{g}_{\alpha\beta}
  + 8\pi \bar{T}_{\alpha\beta}
  \,,
  \nonumber \\[10pt]
  &\bar{\nabla}^{\mu}\bar{\nabla}_{\mu} \varphi
  = 2\pi \frac{F_{,\varphi}}{F}\bar{T} + W_{,\varphi}\,,
  \nonumber \\[10pt]
  &\bar{\nabla}^{\mu}\bar{\nabla}_{\mu}\psi
  =
  \frac{1}{2F} \left(
  \frac{\partial V}{\partial \psi_R}+\iu\frac{\partial V}{\partial \psi_I}
  \right)
  +\bar{g}^{\mu\nu} \frac{\partial_{\mu}F}{F}\partial_{\nu}\psi\,,
  \label{eq:FEQE}
\end{align}
where an overbar denotes tensors and operators in the Einstein
frame; in particular, the energy-momentum tensor is given by
\begin{eqnarray} \label{eq:em}
  \bar{T}_{\alpha\beta}
  &\!\!=\!\!&
  \frac{2}{\sqrt{-\bar{g}}} \frac{\delta S_m}{\delta\bar{g}^{\alpha\beta}}
  ~=~ \frac{1}{F} T_{\alpha\beta}
  \\[10pt]
  &\!\!=\!\!&
  \frac{1}{F}
  \left\{
  \partial_{(\alpha}\psi^* \partial_{\beta)}\psi
  -\frac{1}{2 F}\bar{g}_{\alpha\beta}
  [F\bar{g}^{\mu\nu}\partial_{\mu}\psi^*\partial_{\nu}\psi
  +V(\psi)]
  \right\}.
  \nonumber
\end{eqnarray}
%

\subsection{Spherically symmetric solutions}
In this work, we consider stationary, spherically symmetric BS
models and employ a metric ansatz using radial gauge and polar
slicing. Specifically, we write
\begin{eqnarray} \label{eq:line_element}
    \du \bar{s}^2 &=&
  \bar{g}_{\mu\nu}dx^{\mu}dx^{\nu}
  \nonumber \\[10pt]
  &=&
  -F(\varphi)\alpha^2\du t^2
  +F(\varphi)X^2\du r^2
  +r^2\du \Omega^2\,,
\end{eqnarray}
where $r$ is the areal radius in the Einstein frame and $\du
\Omega^2\defeq \du \vartheta^2 + \sin^2\vartheta \du \phi^2$ the
standard angular line element.  In spherical symmetry, it is
convenient to express the complex BS scalar field in terms of
amplitude $A$ and frequency $\omega$,
\begin{equation}
    \psi = Ae^{\iu \omega t}\,.
\end{equation}

At this point, we have three mass scales in our problem: the two
scalar masses, $\mu_{\varphi}$ and $\mu$ for $\varphi$ and $\psi$
respectively, and the Planck mass $M_{\rm{pl}} = \sqrt{1/G}$ that
appears in the Einstein equation \eqref{eq:FEQE}.  All dimensional
quantities have units of some power of $\mu$ or $M_{\rm Pl}$ or a
combination thereof.  To see this explicitly, we still need to
specify the potential functions $V(\psi)$ and $W(\varphi)$.  The
particular choices for these functions will be detailed in
Sec.~\ref{sec:models} below, but we note already at this stage that
they will take the form $V(A) \sim \mu^2 A^2$ and $W(\varphi)
\sim \mu_{\varphi}^2 \varphi^2$, hence acquiring units of $\mu^2
M^2_{\rm{Pl}}$.  One can now eliminate all dimensional factors in
the covariant equations (\ref{eq:FEQE}) by introducing dimensionless
variables according to
\begin{align}
    \hat{t}&=\mu t, ~~~ \hat{r}=\mu r, ~~~ \hat{m}= \mu m, ~~~ \hat{\omega}=\frac{\omega}{\mu}, \\
    \hat{\mu}_{\varphi} &= \frac{\mu_{\varphi}}{\mu},~~~\hat{\varphi}= \frac{\varphi}{M_{\rm{Pl}}},  ~~~  \hat{A}= \frac{A}{M_{\rm{Pl}}}\,.
    \label{eq:dimless}
\end{align}

These dimensionless variables can be converted back into SI units
by using
\begin{equation}
  \hbar c = 1.97327\times 10^{-10}~{\rm eV}\,{\rm km}\,,
\end{equation}
so that
\begin{equation}
  \frac{\mu}{\hbar c} = \frac{\mu}{1.97327\times 10^{-10}\,
  {\rm eV\,km}}\,.
\end{equation}
Hence, if we set $\hbar=1=c$, a rescaled radius $\hat{r}= \mu r$
and frequency $\hat{\omega}= \omega/\mu$ translate into SI units
according to
\begin{eqnarray}
  r &=& \frac{\hat{r}}{\mu}
  = \left(\frac{\mu}{\hbar c}\right)^{-1}\hat{r}
  = \hat{r} \left(
  \frac{\mu}{1.97327\times 10^{-10}\,{\rm eV}}
  \right)^{-1}{\rm km}\,,
  \nonumber \\[10pt]
  \omega
  &=&
  \frac{\mu}{\hbar}\hat{\omega}
  = \hat{\omega}\,
  \frac{\mu}{6.58212 \times 10^{-16}\,{\rm eV}}\, s^{-1}\,.
  \label{eq:rrhat}
\end{eqnarray}
and likewise for the mass $m$ and other dimensional quantities.
The scalar field amplitudes, in turn, are measured in units of the
Planck mass, $A=\hat{A}M_{\rm Pl}$ and $\varphi=\hat{\varphi}M_{\rm
Pl}$.  In words, for a BS scalar mass $\mu=1.97327\times 10^{-10}\,{\rm
eV}$, the numerical value of our radial coordinate gives the BS
radius in km. For other scalar masses, the radius scales accordingly,
becoming larger for smaller scalar masses and vice versa. From now
on, we will work with dimensionless variables only and we will drop
the caret in their notation for simplicity.
\subsection{The field equations in spherical symmetry}
Our physical system is completely described by $\varphi$, $A$,
$\alpha$ and $X$, which are all functions of radius $r$. It is
sometimes convenient to express the metric functions $\alpha$ and $X$ in terms
of the alternative variables
\begin{equation}
  \Phi \defeq \ln(F\sqrt{\alpha})\,,
  ~~~~~
  m\defeq \frac{r}{2}\left(
  1-\frac{1}{FX^2}\right)\,,
  \label{eq:Phim}
\end{equation}
which, when combined with Eqs.~(\ref{eq:FEQE})-(\ref{eq:line_element}),
results in a set of ordinary differential equations (ODEs) that can
be written as
\begin{widetext}
\begin{eqnarray}
  \partial_r \Phi
  &=&
  \frac{FX^2-1}{2r}
  -rFX^2W
  +\frac{r}{2}\frac{X^2}{\alpha^2}\kappa^2
  +\frac{2\pi r X^2}{F\alpha^2}
  \left(
  \omega^2 A^2-\alpha^2V+F^2\theta^2
  \right)\,,
  \nonumber
  \\[10pt]
  \frac{\partial_r X}{X}
  &=&
  -\frac{FX^2-1}{2r}
  +rFX^2W
  -\frac{X}{2\alpha}\frac{F'}{F} \kappa
  +\frac{r}{2}\frac{X^2}{\alpha^2}\kappa^2
  +\frac{2\pi r X^2}{F\alpha^2}
  \left(
  \omega^2 A^2+ \alpha^2V+F^2\theta^2
  \right)\,,
  \nonumber \\[10pt]
  \partial_r \varphi
  &=&
  \frac{X}{\alpha} \kappa\,,
  \nonumber \\[10pt]
  \partial_r \kappa
  &=&
  -2\frac{\kappa}{r}
  +2\alpha XFW_{,\varphi^2}\varphi
  +2\pi \frac{XF'}{\alpha F^2}
  \left(
  \omega^2 A^2-2\alpha^2 V-F^2\theta^2
  \right)\,,
  \nonumber \\[10pt]
  \partial_r A
  &=&
  \frac{XF}{\alpha}\theta\,,
  \nonumber \\[10pt]
  \partial_r \theta
  &=&
  -2\frac{\theta}{r}
  +\frac{X}{\alpha F}A
  \left(
  \alpha^2 V_{,A^2} - \omega^2
  \right)\,.
  \label{eq:allr}
\end{eqnarray}
\end{widetext}
Here, we have introduced the auxiliary variables $\kappa$ and
$\theta$ in order to write the two scalar-field equations in
first-order form and used the notation
\begin{equation}
  W_{,\varphi^2} \defeq \frac{\du W}{\du (\varphi^2)}\,,
  ~~~~~
  V_{,A^2} \defeq \frac{\du V}{\du (A^2)}\,.
  \nonumber
\end{equation}
The system (\ref{eq:allr}) is complemented by the boundary conditions
\begin{eqnarray}
  &&X(0) = \frac{1}{\sqrt{F}}\,, 
  ~
  A(0) = A_{\rm ctr}\,,
  ~
  \varphi(0) = \varphi_{\rm ctr}\,,
  ~
  \kappa(0) = 0\,,
  \nonumber \\[10pt]
  &&\theta(0) = 0\,,
  ~
  \Phi(\infty) =  \Phi_0\,,
  ~
  A(\infty) = 0\,,
  ~
  \varphi(\infty) = 0\,.
  \label{eq:BCs}
\end{eqnarray}
The vanishing of the scalar fields at infinity can only be achieved
for specific values of $\omega$ and $\varphi_{\rm{ctr}}$ which
characterizes Eq.~(\ref{eq:allr}) as an eigenvalue problem. More
specifically, the scalar fields have an asymptotic behaviour
containing exponential modes
\begin{equation} \label{eq:asymp_wrong}
  \varphi \sim e^{\pm h r}\,,
  ~~~~~
  A \sim e^{\pm k r}\,,
\end{equation}
where $h$ and $k$ depend on the scalar masses and frequency $\omega$,
and the unphysical, growing modes $e^{+hr}$, $e^{+kr}$ only vanish
for discrete values of $\varphi_{\rm ctr}$ and $\omega$. In the
literature, the asymptotic behaviour of the BS scalar is sometimes
given as
\begin{equation}
  \varphi \sim \frac{e^{\pm h r}}{r}\,,
  ~~~~~A \sim \frac{e^{\pm k r}}{r}\,,
  \label{eq:wrongasymptotics}
\end{equation}
but this only holds in the complete absence of gravity and is
incorrect even for $r\rightarrow \infty$ if the scalar field couples
to gravity, either in GR or ST theory.  As we show in Appendix
\ref{app:asymptotic_behaviour}, instead of Eq.~(\ref{eq:wrongasymptotics}),
the asymptotic behaviour with gravity is
\begin{equation}
  \varphi \sim \frac{e^{\pm h r}}{r^{1+\delta}}\,,
  ~~~~~A \sim \frac{e^{\pm k r}}{r^{1+\epsilon}}\,,
\end{equation}
where $h$, $k$, $\delta$ and $\epsilon$ are given in
Eqs.~(\ref{eq:asymptotics_massive}) and (\ref{eq:asymptotics_massless})
for a massive and massless $\varphi$, respectively.

The ground-state BS models, that we focus on in this study, are
furthermore characterized by monotonic profiles $A(r)$ with no zero
crossings except for their vanishing at infinity.
%

\subsection{Choice of model} \label{sec:models}
The field equations derived in the previous section hold for arbitrary
potentials $W(\varphi)$, $V(\psi)$ and conformal function $F(\varphi)$.
We now discuss the specific choices made in our work, starting with
the conformal factor $F(\varphi)$.  We consider the
class of ST theories given by the Damour-Esposito-Far{\`e}se function
\cite{Damour:1993hw},
\begin{eqnarray}
    F(\varphi) 
    &=&
    e^{-2\alpha_0 \varphi-\beta_0\varphi^2},
\end{eqnarray}
with constants $\alpha_0 \geq 0$ and $\beta_0$.  The choice of
Damour and Esposito-Far{\`ese} is in part motivated by the fact
that the modifications of gravity at first post-Newtonian order are
completely determined by the asymptotic values of the first and
second derivatives of $\ln F$ in the far-field limit
\cite{Damour:1992we,Damour:1996ke,Chiba:1997ms}. In our notation,
these derivatives are
\begin{equation}
  \alpha_0 = -\frac{1}{2}\left.
  \frac{\du \ln F}{\du \varphi}\right|_{\varphi=0}\,,
  ~~~~~
  \beta_0 = -\frac{1}{2}\left.
  \frac{\du^2 \ln F}{\du \varphi^2} \right|_{\varphi=0}\,.
\end{equation}
In particular, this implies an effective Newtonian gravitational
constant $\tilde{G}=G(1+\alpha_0^2)$. The parameters $\alpha_0$,
$\beta_0$ furthermore completely determine the Eddington parametrized
post-Newtonian parameters $\beta^{\rm PPN}$ and $\gamma^{\rm PPN}$
\cite{Eddington:1923,Will:1993}; see also Eq.~(3.7) and the surrounding
discussion in Ref.~\cite{Gerosa:2016fri}.  For $\beta_0=0$, we
recover Brans-Dicke theory \cite{Brans:1961sx}, whereas Damour and
Esposito-Far{\`e}se's inclusion of the quadratic term leads to the
spontaneous-scalarization phenomenon for sufficiently \textit{negative}
$\beta_0$.

The pioneering work on spontaneous scalarization focused on massless
ST gravity, which, however, has been constrained significantly by
the Cassini mission and binary pulsar observations, not least because
of the dramatic effect of the spontaneous scalarization mechanism.
This has resulted in an increasing amount of attention devoted to
{\it massive} ST theories which, due to the screening effected by
the mass term $\mu_{\varphi}$, largely remain compatible with the
observations.  Roughly speaking, the Yukawa fall-off of a massive
scalar suppresses the impact of the scalar sector on the binary
pulsar dynamics for sufficiently large separations of the binary
constituents; this leaves ST theories with $\mu_{\varphi} \gtrsim
10^{-15}\,{\rm eV}$ in agreement with present observations
\cite{Ramazanoglu:2016kul}. We note, however, that $\mu_{\varphi}$
may also be constrained through other types of observations.
Specifically, a recent study on equilibrium neutron stars conjectures
that the GW event 170817 is suggestive of a larger mass $\mu_{\varphi}
\gtrsim 10^{-11}\,{\rm eV}$ \cite{Kuan:2023hrh} while the observation
of highly spinning black holes, combined with their superradiant
interaction with surrounding massive fields, may exclude the mass
range $10^{-13}\,{\rm eV} \lesssim \mu_{\varphi} \lesssim 10^{-11}\,{\rm
eV}$  \cite{Narayan:2013gca,Brito:2015oca}.  Our analysis is largely
motivated by theoretical considerations and the goal to obtain a
comprehensive understanding of BS scalarization. While bearing in
mind the possible constraints above, we therefore consider a wide
range of mass values $\mu_{\varphi} \ge 0$ as well as coupling
parameters $\alpha_0,~\beta_0$ with no further prejudice.  In our
theoretical formalism we include the gravitational scalar mass
through a quadratic potential
\begin{eqnarray}
  W(\varphi)
  &=&
  \frac{1}{2}\mu_{\varphi}^2 \varphi^2.
  \label{eq:WVds2}
\end{eqnarray}
where $\mu_{\varphi}\ge 0$ measures the gravitational scalar mass
in units of the BS scalar mass parameter $\mu$.  As we will see
later both the massless and the massive cases will share common
phenomenological features.

\begin{figure}[t]
  \includegraphics[width=0.45\textwidth]{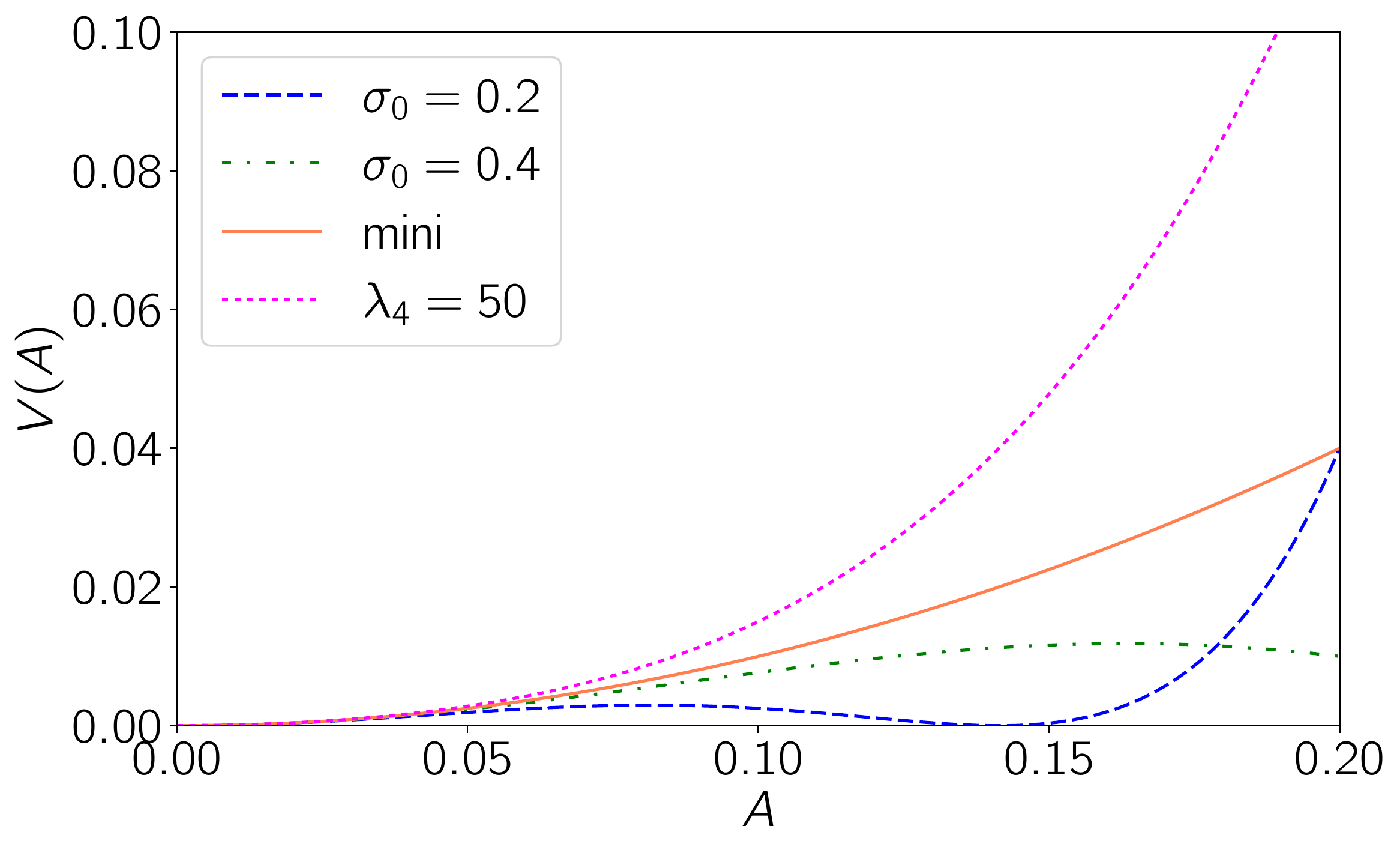}
  \caption{Four example cases of the BS scalar potential function
  $V(A)$ defined in Eq.~(\ref{eq:BSpotentials}).  The non-interacting
  potential for mini BSs is recovered for $\sigma_0 \rightarrow
  \infty$ or $\lambda_4=0$. Note the false vacuum state for solitonic
  potentials and the systematic increase of $V(A)$ from small
  $\sigma_0$ to the mini-BS limit and larger $\lambda_4$.
  \label{fig:potentials}
  }
\end{figure}

Finally, to describe the bosonic sector, we focus on repulsive and
solitonic potentials,
\begin{eqnarray}
V_{\rm rep} &=& A^2 +                 \lambda_4 A^4\, \nonumber , \\[5pt]
V_{\rm sol} &=&                   A^2\left(1-2\frac{A^2}                 {\sigma_0^2}\right)^2, 
  \label{eq:BSpotentials}
\end{eqnarray}
where $\lambda_4$ and $\sigma_0$ are self-interaction terms of the
two individual potentials.  Note that both potential functions
include the case of {\it mini BSs} in the respective limits
$\lambda_4 \to 0$ and $\sigma_0 \to \infty$.  We graphically
illustrate the features of these potentials in Fig.~\ref{fig:potentials}
for two solitonic potentials with $\sigma_0=0.2$ and $\sigma_0=0.4$,
the case of mini BSs and a repulsive potential with $\lambda_4=50$
over the range $A \in [0,\,0.2]$. We clearly see the false vacuum
states for $\sigma_0=0.2$ and also note that in this range of $A$,
the potential values systematically increase in the order $\sigma_0=0.2$,
$\sigma_0=0.4$, the mini BS limit and $\lambda_4=50$.
%

\subsection{Numerics and Diagnostics} 
Accurately capturing the exponential behaviour of the scalars and
determining the eigenvalues $\varphi_{\rm{ctr}}$ and $\omega$ is a
non-trivial numerical challenge and we detail in Appendices
\ref{app:asymptotic_behaviour} and \ref{app:numerics} how we compute
the solutions using shooting and relaxation techniques.  The
conceptual summary of our calculations is as follows.
\begin{enumerate}
\item
Specify the ST theory through the parameters
$\alpha_0$, $\beta_0$ and $\mu_{\varphi}$.
\item
Specify the BS scalar potential by fixing $\sigma_0$ or $\lambda_4$.
\item
Parametrize the sequence of BS solutions inside this framework by
varying a {\it control parameter}, typically $A_{\rm ctr}$, but
sometimes also $\omega$ or $\Phi_{\rm ctr}\defeq\Phi(0)$.
\item
For each value of this control parameter, we iteratively determine
the eigenvalues $\varphi_{\rm ctr}$ and $\omega$ leading to a
physical (i.e.~asymptotically flat) BS solution.
For some values of the control parameter, there exist multiple
solutions with different degrees of gravitational scalarization.
In order to find all possible solutions, we perform the iterative
procedure using different initial guesses for $\omega$ and $\varphi_{\rm
ctr}$, usually obtained from neighbouring models with similar degrees
of scalarization.
\end{enumerate}
\begin{figure*}
  \includegraphics[width=1\textwidth,clip=true]{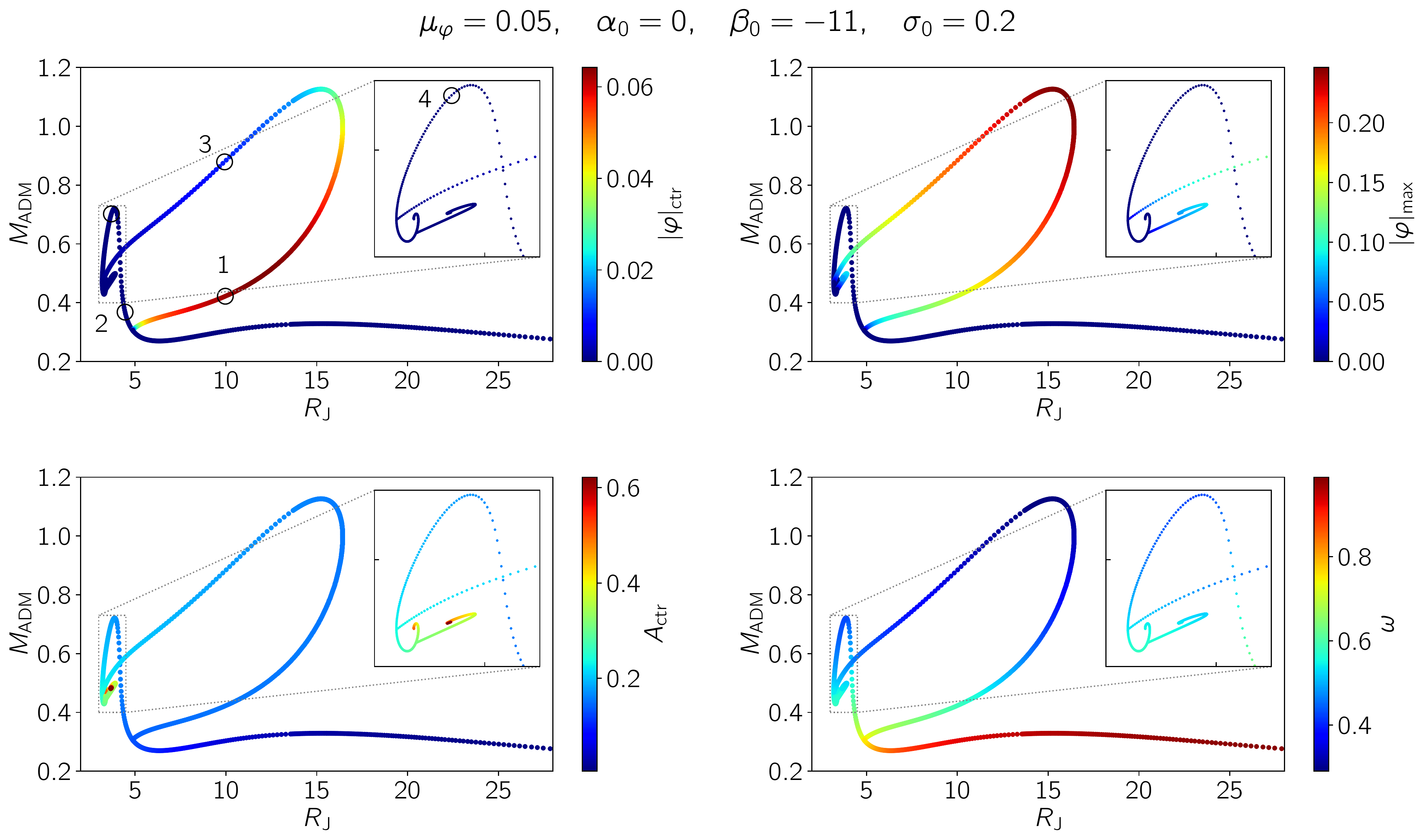}
  \caption{
  BS models in ST gravity with $\mu_{\varphi}=0.05$, $\alpha_0=0$
  and $\beta_0=-11$ for a solitonic potential with $\sigma_0=0.2$.
  Each point in these mass-radius diagrams represents a BS model.
  The location of the point represents the star's mass $M_{\rm ADM}$
  and radius $R_{\rm J}$, and the color coding in the respective
  panels displays the central gravitational scalar $|\varphi|_{\rm
  ctr}$, the maximum (over radius) gravitational scalar magnitude $|\varphi|_{\rm
  max}$, the central BS scalar amplitude $A_{\rm ctr}$ and the
  frequency $\omega$.  The four circles labelled 1 to 4 mark the
  models displayed in more detail in Fig.~\ref{fig:prototype_profiles}.
 }
  \label{fig:mvph005_a0m00_bm11_Soli020}
\end{figure*}
The typical result of such a calculation for a given ST theory and
potential function is a set of one-parameter branches of strongly
and weakly scalarized BS solutions which can be displayed, for
example, as curves in the mass-radius plane.  Whereas the mass of
a BS is well defined as the Arnowitt-Deser-Misner \cite{Arnowitt:1962hi}
or ADM mass, given in our formalism by $m(r\rightarrow \infty)$,
the radius is not; BSs formally extend to infinity, albeit with
exponentially decaying scalar amplitude $A$. In this work, we define
the radius $R_{\rm J}$ of a BS as the Jordan frame radial coordinate
$r/\sqrt{F}$ where the mass function $m$ reaches $99\,\%$ of the
ADM mass, i.e.~where $m(r)=0.99\,m(\infty)$.  Alternative measures
for the radius, using a different percentage or other physical
diagnostics, do not change our main results beyond minor differences
in the radius values.

\section{Scalarized boson stars}\label{sec:results}
\subsection{A prototypical example} \label{sec:prototypical_example}
We begin our discussion of the results with a representative family
of BS models that illustrates the main features of their spontaneous
scalarization. This set of models is obtained for ST parameters
$\mu_{\varphi}=0.05$, $\alpha_0=0$ and $\beta_0=-11$ as well as a
solitonic potential $V_{\rm sol}$ (\ref{eq:BSpotentials}) with
$\sigma_0=0.2$ for the BS matter. The main diagnostic quantities
we display for this family of models are the ADM mass $M_{\rm{ADM}}$,
the radius $R_{\rm J}$, the central and maximal values $\varphi_{\rm
ctr}$, $\varphi_{\rm max}$ of the gravitational scalar, the central
magnitude of the complex BS scalar $A_{\rm ctr}$ and the frequency
$\omega$.

The resulting BS models are graphically illustrated in the form of
mass-radius diagrams in Fig.~\ref{fig:mvph005_a0m00_bm11_Soli020}
with the other diagnostic quantities encoded in color.  We note in
this context that for $\alpha_0=0$ the BS models are invariant under
the change $\varphi(r) \rightarrow -\varphi(r)$ and the profile
$\varphi(r)$ has no zero crossings. Unless stated otherwise, we
display in this section the solutions with positive $\varphi=|\varphi|$.
The breaking of this degeneracy for non-zero $\alpha_0$ will be
explored in Sec.~\ref{sec:alpha0ne0} below.  The main features
visible in this figure are as follows.
\begin{enumerate}
\item
The BS models of GR are recovered as a one-parameter family with
vanishing gravitational scalar $\varphi$; this set of models is
conspicuous as the dark blue branch in the top right panel of
Fig.~\ref{fig:mvph005_a0m00_bm11_Soli020}.  For the vanishing
$\alpha_0$ of our example, this branch exactly equals the BS models
computed in GR as shown, for example, in Fig.~1 of
Ref.~\cite{Helfer:2021brt} for the same $\sigma_0$.
\item
Scalarized models appear most prominently in the form of a second
branch forming a loop starting from about a radius $R_{\rm J}\approx
5$ towards larger radii and masses and eventually reconnecting with
the GR branch at $R_{\rm J}\approx 3$.  As can be seen in the inset
of the top right panel of Fig.~\ref{fig:mvph005_a0m00_bm11_Soli020},
there exists a second segment of the scalarized branch extending
to the right, a bit further down.  As we will see in
Sec.~\ref{sec:mass_onset}, the fine structure of the high-compactness
(upper) end of the scalarized branch can change with $\beta_0$.
\item
Bearing in mind our sign convention for $\varphi$, the upper two
panels in Fig.~\ref{fig:mvph005_a0m00_bm11_Soli020} demonstrate
that for many of the scalarized models, the gravitational scalar
$\varphi$ reaches its maximum away from the origin: $|\varphi|_{\rm
ctr}$ is typically smaller than $|\varphi|_{\rm max}$.
\item
As also observed for neutron stars \cite{Damour:1993hw,Rosca-Mead:2020bzt},
the scalarized models can reach significantly larger masses and
radii\footnote{Alternative definitions of BS radii exhibit the same
qualitative behaviour in spite of quantitative differences.}\,.
\item
As the central BS scalar amplitude increases, the BS models become
increasingly compact, as illustrated in the bottom left panel of
Fig.~\ref{fig:mvph005_a0m00_bm11_Soli020}.
\item
As the BS compactness increases, the frequency $\omega$ generally
decreases from $\lim_{A_{\rm ctr} \rightarrow 0} \omega=1$. This
trend is only mildly reversed at the extreme end for highly compact
models; cf.~the inset in the bottom right panel.  The smallest
frequencies are obtained in the high-mass regime of the scalarized
branch.
\end{enumerate}

\begin{figure}
  \includegraphics[width=0.45\textwidth]{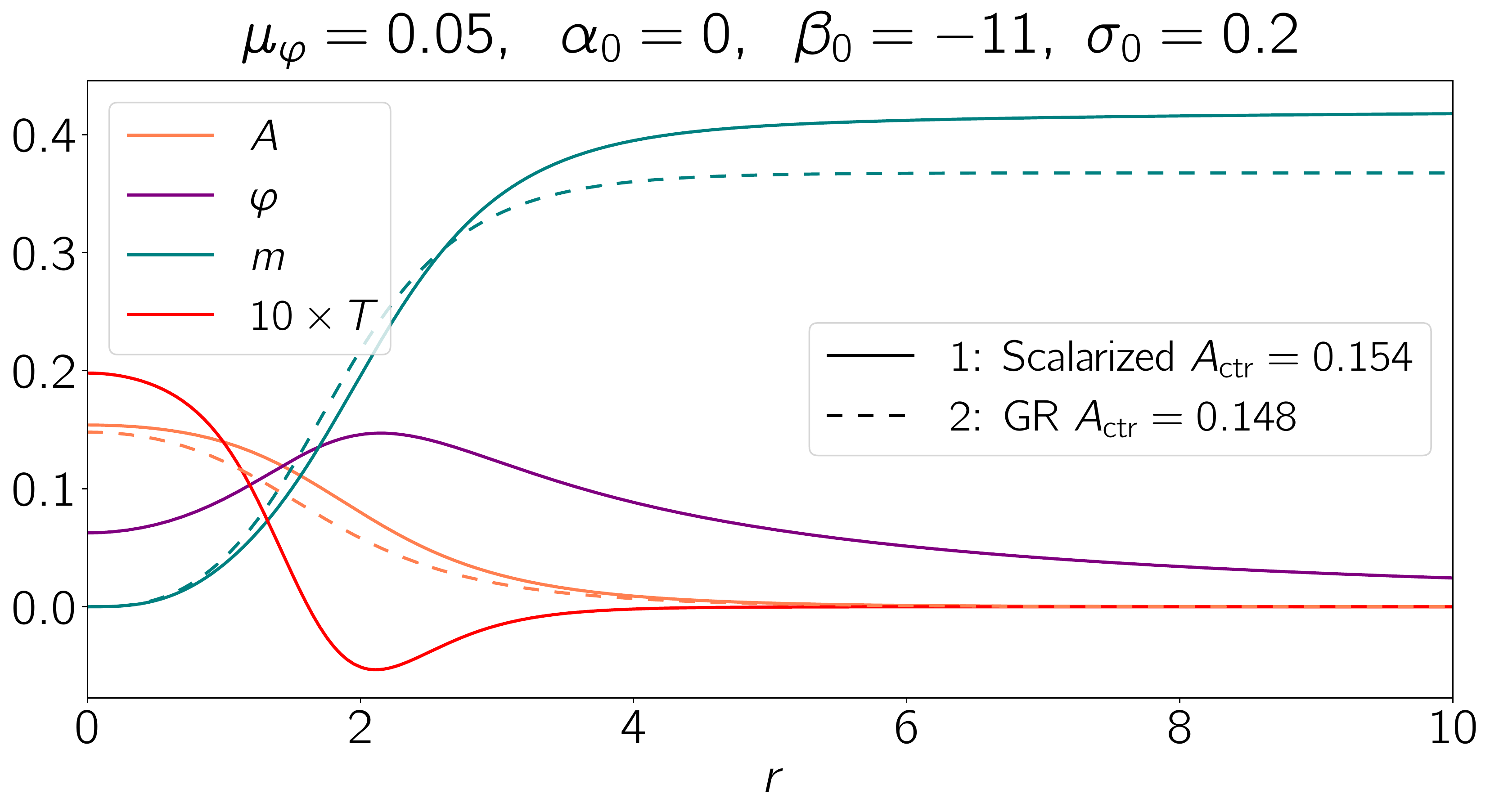}
  \includegraphics[width=0.45\textwidth]{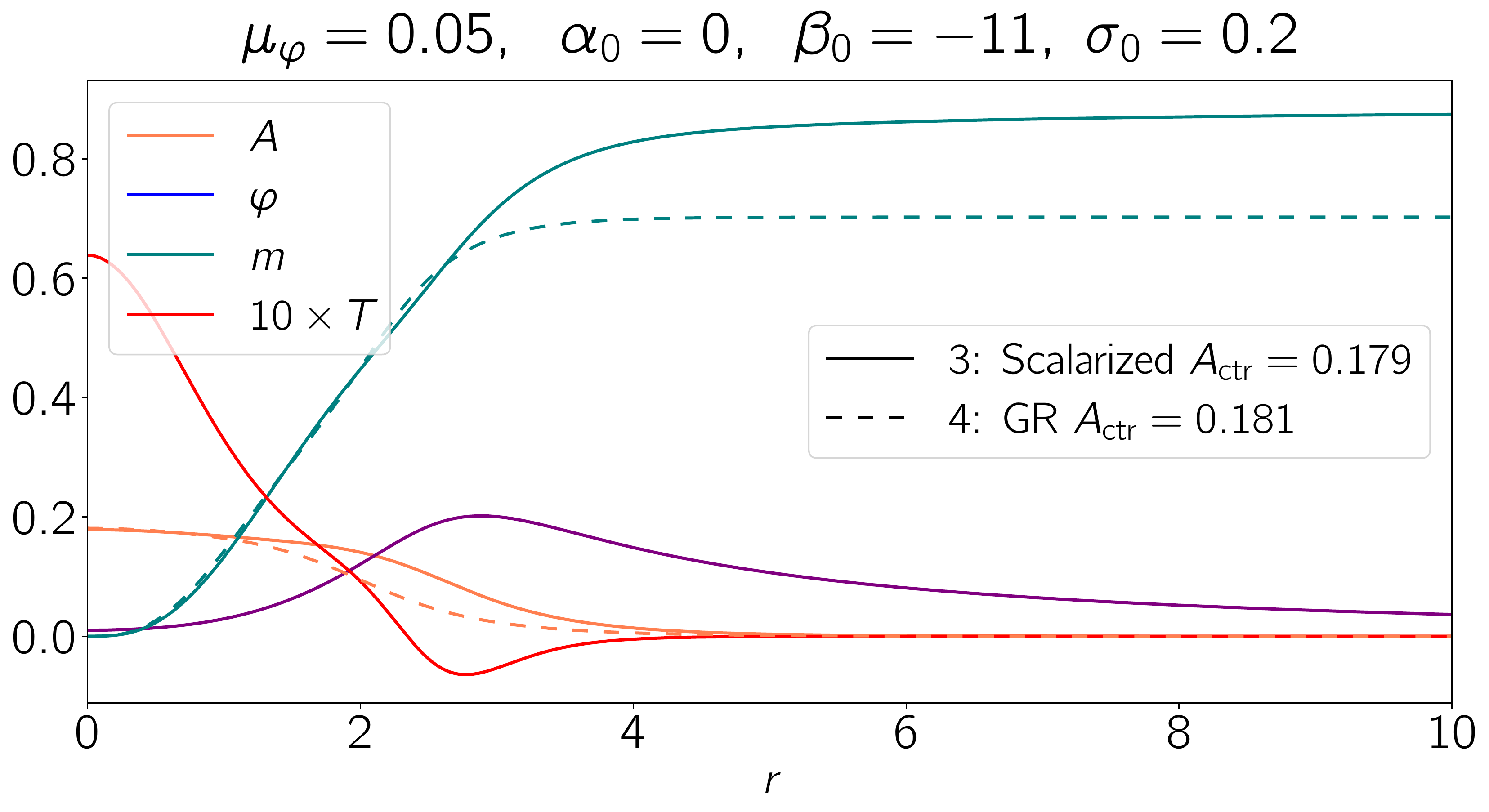}
  \caption{Radial profiles of the BS scalar amplitude $A$, the
  gravitational scalar $\varphi$, the mass function $m$ and the
  trace $T$ of the energy momentum tensor for four selected models
  marked $1$ to $4$ in Fig.~\ref{fig:mvph005_a0m00_bm11_Soli020}.
  The pair of models $1,2$ represents one scalarized and one GR
  star with equal compactness $C=0.117$ and amplitudes $A_{\rm ctr}$
  as indicated in the legend. Likewise, $3,4$ represent a scalarized
  and GR model with $C=0.240$ each. As indicated by the dashed
  curves for GR and the solid lines for scalarized models, the
  scalarization tends to push the BS scalar $A(r)$ towards larger
  radii and results in larger masses. In the region of strongest
  scalarization in $\varphi$, the energy momentum tensor has negative
  trace $T$.
}
  \label{fig:prototype_profiles}
\end{figure}

In Fig.~\ref{fig:prototype_profiles}, we plot the BS scalar amplitude
$A$, the gravitational scalar $\varphi$, the mass aspect $m$ as
well as the trace $T$ of the physical energy momentum tensor as
functions of radius $r$ for four selected models marked by circles
in Fig.~\ref{fig:mvph005_a0m00_bm11_Soli020}. Here, models 1 and 3
are located on the scalarized branch with compactness $C\defeq
\max_{r\in \mathbb{R}} \tfrac{m(r)}{r}=0.117$ and $C=0.240$,
respectively.  The BSs marked 2 and 4 represent GR stars with
respectively equal compactness for comparison.  The models exhibit
quite similar main features despite their different locations in
the mass-radius plane: (i) Whereas the BS scalar amplitude $A$ is
a monotonically decreasing function of $r$ for all cases, the
gravitational scalar of models 1 and 3 reaches its extremal value
away from the origin.  (ii) The mass function increases monotonically
with radius, but levels off at relatively small radius due to the
exponential decrease of $A$.  (iii) The scalarization tends to push
the BS scalar profile $A(r)$ towards larger radii and results in
larger total mass $M_{\rm ADM}$.  (iv) The region of strongest
scalarization coincides with the region of a negative trace of the
energy momentum tensor as expected for the onset of a tachyonic
instability \cite{Ramazanoglu:2016kul}.

In the following sections, we will explore in more detail the
phenomenology of scalarized BS models in ST theory and will highlight
how and when the main features illustrated by this prototype can
alter for different potentials $V(A)$ and ST parameters.

%

\subsection{Scalar mass and the onset of scalarization} \label{sec:mass_onset}
Having seen the main features of scalarized BSs in the previous
section, we next study how the mass $\mu_{\varphi}$ of the gravitational
scalar affects the scalarization phenomenon and, in particular, at
which values of $\beta_0$ scalarization sets in. For this purpose,
we keep the BS potential function $V(A)$ fixed at its solitonic
shape with $\sigma_0=0.2$; the effect of alternative potentials
will be explored in Sec.~\ref{sec:otherpotentials} below.

The majority of spontaneously scalarized neutron-star models have
been found below a threshold for $\beta_0 \lesssim \beta_{0,\rm
thr}=-4.35$ where $\beta_{0,\rm thr}$ turns out to be remarkably
robust against variations of other parameters such as the equation
of state
\cite{Novak:1997hw,Novak:1998rk,Pani:2014jra,Ramazanoglu:2016kul,Rosca-Mead:2020bzt}.
One striking exception to this rule are the strongly scalarized
solutions found by Mendes and collaborators
\cite{Mendes:2014vna,Mendes:2014ufa,Mendes:2016fby} for positive
$\beta_0$ provided the trace of the energy momentum tensor is
positive in the GR limit; see also Ref.~\cite{Ramazanoglu:2016kul}
where this effect is explained in terms of a tachyonic instability.
For the BS models considered in this study, we focus on negative
$\beta_0$, but we will see that the threshold value $\beta_{0,{\rm
thr}}$ can differ from that for NSs and also depends more sensitively
on the BS parameters.

\begin{figure}[t]
  \includegraphics[width=0.45\textwidth]{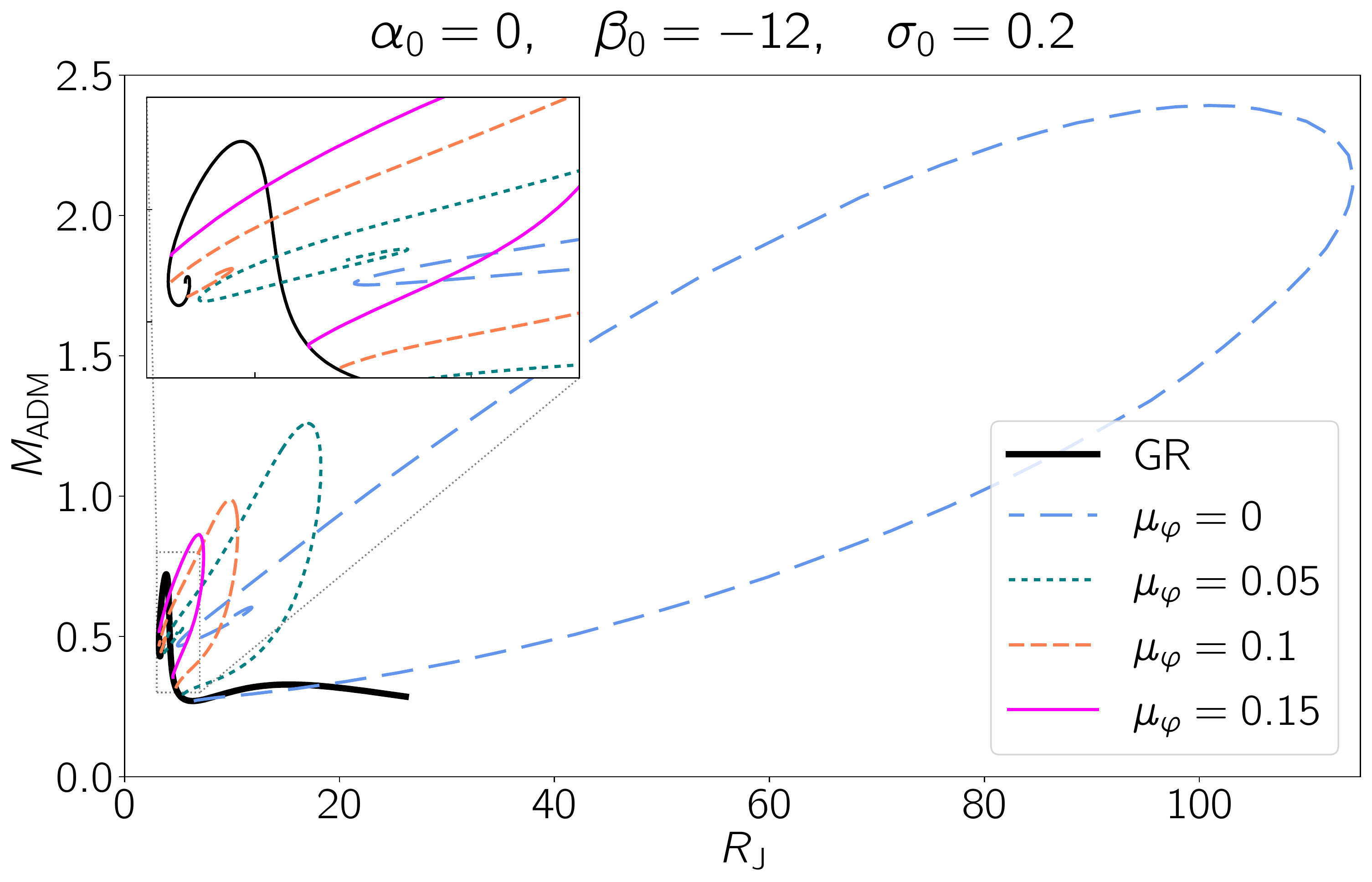}
  \includegraphics[width=0.45\textwidth]{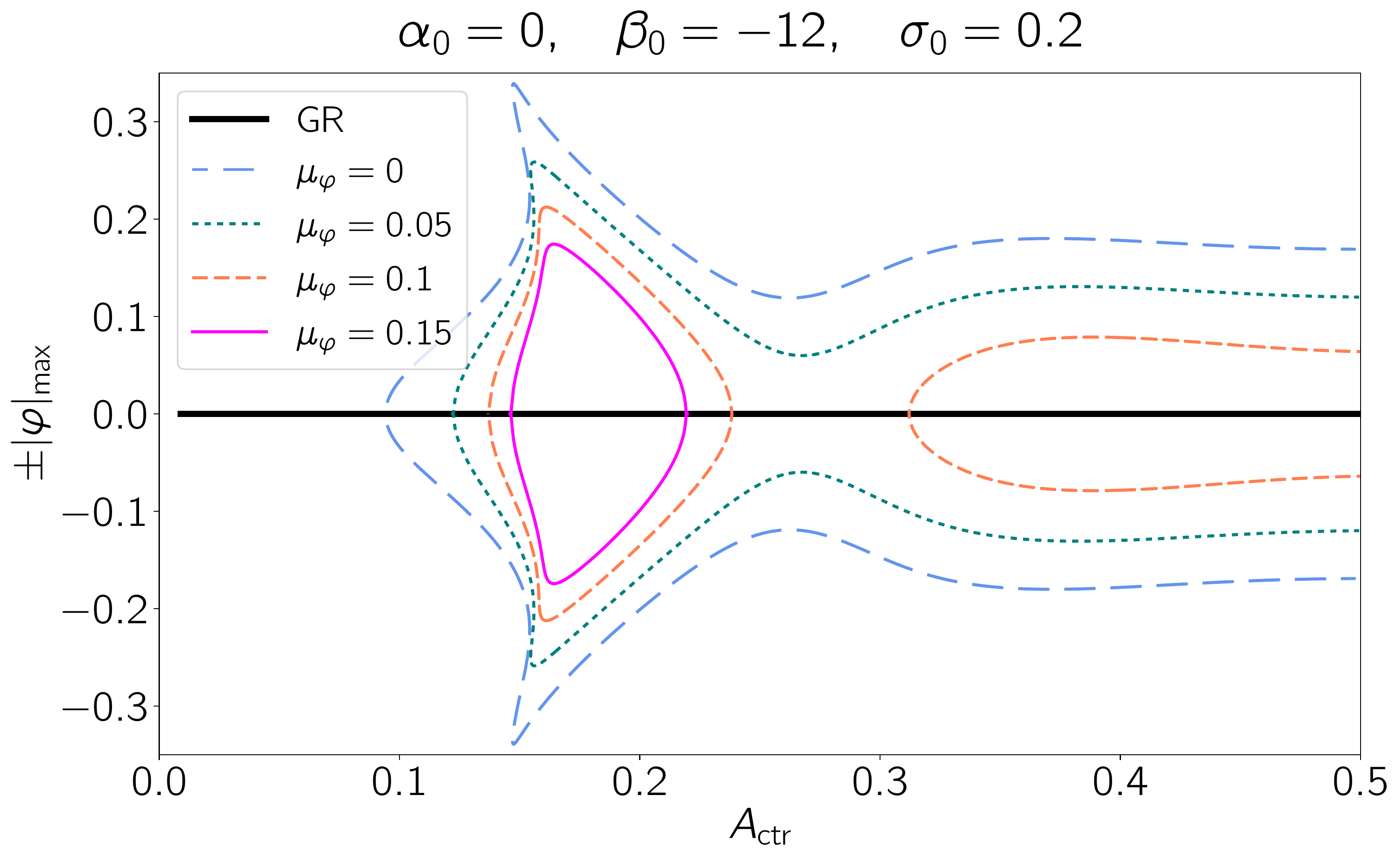}
  \caption{{\it Upper panel}: Mass radius diagram for four values
  of the scalar mass. The scalarized branches decrease in size for
  larger $\mu_{\varphi}$. {\it Lower panel}: The maximal value of
  the gravitational scalar, $|\varphi|_{\rm max}$, is shown as a
  function of the central BS scalar amplitude $A_{\rm ctr}$ for the
  same BS families.  Note that for $\alpha_0=0$, the models for
  positive and negative $\varphi$ are degenerate and we plot both
  extrema, $\pm |\varphi|_{\rm max}$, to illustrate the loop
  structure. Again the loops shrink for larger $\mu_{\varphi}$, but
  we also notice a split into two separate loops at $\mu_{\varphi}=0.1$.
  For $\mu_{\varphi}=0.15$, the second loop containing highly compact
  models with large $A_{\rm ctr}$ is no longer present.
  }
  \label{fig:mvph_MRAvphmax}
\end{figure}
As a general observation we find that increasing the mass $\mu_{\varphi}$
weakens the effect of scalarization.  This is illustrated in
Fig.~\ref{fig:mvph_MRAvphmax} where we display solitonic BS families
for $\alpha_0=0$, $\beta_0=-12$ and $\sigma_0=0.2$, but varying
$\mu_{\varphi}$ from 0 to $0.15$. The upper panel in this figure
demonstrates that the scalarized branches significantly reduce in
terms of the BS radius and mass, getting closer to the GR branch
for larger $\mu_{\varphi}$. We note in this context that the rather
large BS radii found for $\mu_{\varphi}=0$ are a consequence of a
slightly slower fall-off of $A(r)$, likely caused by its coupling
to the now long-ranging $\varphi$. We investigate this effect in
more detail in Sec.~\ref{sec:thinshell} below, but note that the
outer regions even of BSs of radius $R_{\rm J}\approx 100$ are
tenuous with the small energy density merely amplified by the $r^2$
effect of the volume element.

The decrease in scalarization becomes even more evident in the lower
panel of Fig.~\ref{fig:mvph_MRAvphmax} where the same BS families
are represented in the plane spanned by the central BS scalar
amplitude $A_{\rm ctr}$ and the maximal value of $|\varphi(r)|$.
Recalling that the BS models are identical under $\varphi \rightarrow
-\varphi$, we display each BS in this figure in terms of its two
extrema $\pm |\varphi|_{\rm max}$ in order to illustrate more clearly
the loop like structure of the branches. Clearly, the maximal
scalarization decreases for larger $\mu_{\varphi}$, but we also
notice two qualitative changes occuring between $0.05 < \mu_{\varphi}
< 0.1$ and $0.1 < \mu_{\varphi} <0.15$, respectively. First, the
single loop (long-dashed blue and dotted green) splits into two
seprate loops (dashed orange) leaving a finite range $0.25 \lesssim
A_{\rm ctr}\lesssim 0.31$ where only GR models exist. For
$\mu_{\varphi}=0.15$, the second loop of highly compact (large
$A_{\rm ctr}$) models has disappeared; we cannot entirely rule out
that it is shifted to very large $A_{\rm ctr}$, but all our searches
for scalarized models in this regime for $\mu_{\varphi}\ge 0.15$
have yielded no solutions. This behaviour is also reflected in the
mass-radius diagram of the upper panel, where for $\mu_{\varphi}=0$
or $0.05$ the scalarized branch connects to the GR branch only at
the low mass end, but remains separate at the high-mass end as shown
in the inset.  For $\mu_{\varphi}=0.1$, instead, we have two separate
scalarized branch segments, both connected to GR models: note the small
dashed orange branch in the inset near the end of the black GR
branch.  For $\mu_{\varphi}=0.15$ (solid magenta), this small extra
branch is no longer present.

The lower panel of Fig.~\ref{fig:mvph_MRAvphmax} resembles Fig.~3
of Ramanazo\u{g}lu and Pretorius \cite{Ramazanoglu:2017xbl} which
shows $|\varphi|_{\rm max}$ obtained for scalarized neutron-stars.
Their analysis of spontaneous scalarization in terms of a tachyonic
instability leads to an approximate criterion $\lambda_{\varphi} >
\lambda_{\rm eff,\,star}\sim R_{\rm J}/\sqrt{C|\beta_0|}$ -- their
Eq.~(6) -- where $C$ is an estimate of the star's compactness and
$\lambda_{\varphi}$ is the Compton wavelength associated with
$\mu_{\varphi}$. Translated into our variables, this relation becomes
\begin{equation}
  |\beta_0| \gtrsim \frac{R_{\rm J}^2}{4\pi^2 C}\mu_{\varphi}^2\,,
  \label{eq:beta0thr}
\end{equation}
and suggests an approximately quadratic relation between the threshold
for scalarization $\beta_{0,{\rm thr}}$ and the scalar mass
$\mu_{\varphi}$.

\begin{figure}[t]
  \includegraphics[width=0.45\textwidth]{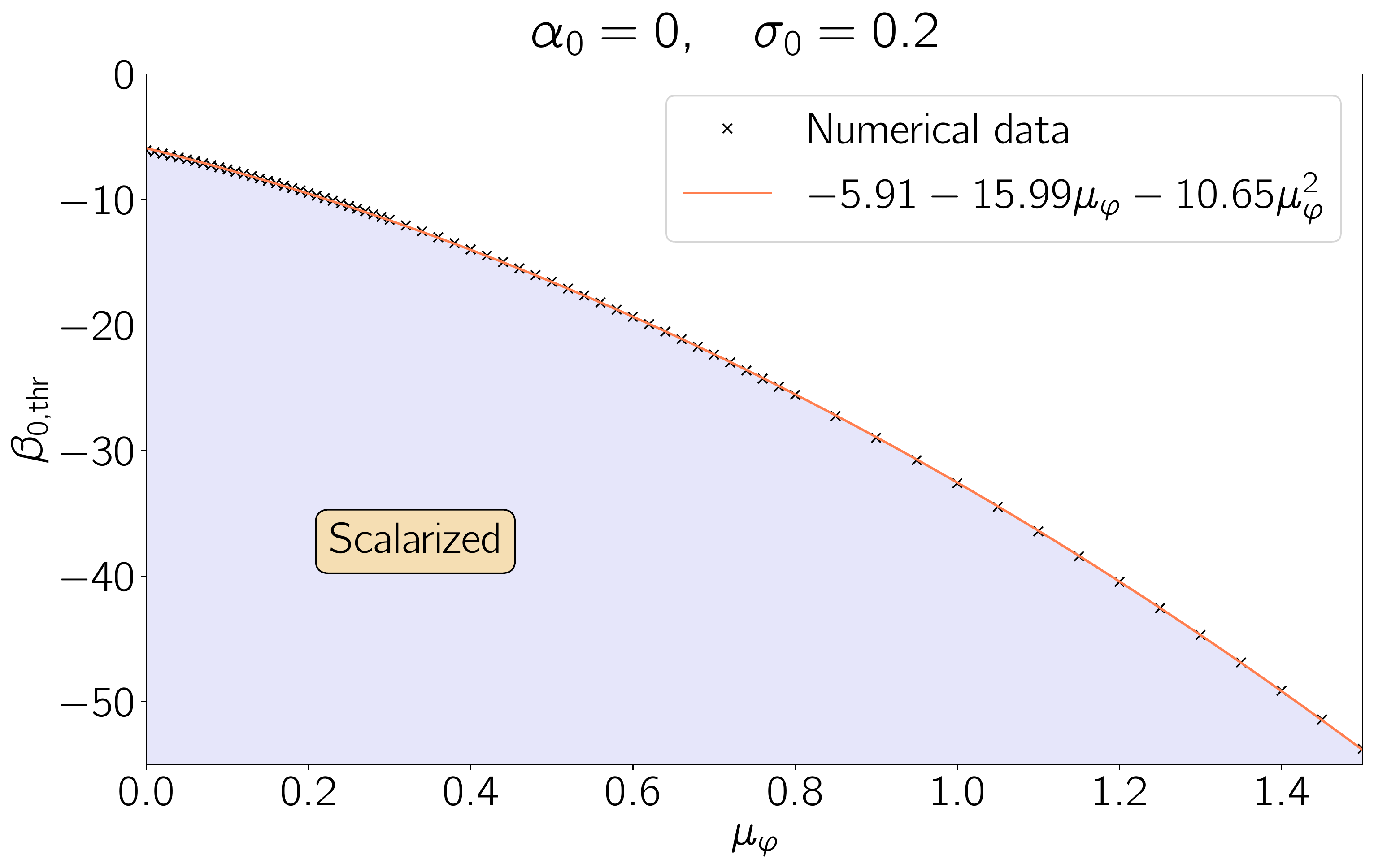}
  \caption{The threshold value of $\beta_0$ for the
  onset of scalarization; for $\beta_0 \leq \beta_{0,{\rm thr}}$, scalarized BS models exist.
  }
  \label{fig:beta0thr}
\end{figure}

We have numerically computed the onset of scalarization for the
families of BS models with $\alpha_0=0$, $\sigma_0=0.2$ by starting
for each fixed $\mu_{\varphi}$ with a sufficiently negative $\beta_0$
that admits scalarized models and then iteratively increasing
$\beta_0$ towards 0 until scalarization disappears. In practice,
we have stopped this search at 4 significant digits in $\beta_0$
and verified that the result remains unchanged under a doubling or
halving of the number of grid points.  The resulting threshold
$\beta_{0,{\rm thr}}$ is plotted as a function of $\mu_{\varphi}$
in Fig.~\ref{fig:beta0thr} together with a quadratic fit.  Typical
values for radius and compactness of the BS models on the verge of
scalarization are $R_{\rm J}\sim 5$ and $C\sim 0.2$, resulting in
a coefficient $R_{\rm J}^2/(4\pi^2 C)\sim 3$ in Eq.~(\ref{eq:beta0thr}).
Bearing in mind its approximate character, the tachyonic analysis
captures the quadratic contribution of our fit rather well.  The
constant and linear contributions to our fit, in turn, are likely
a direct consequence of a finite threshold for scalarization in
massless ST theory: using a series of self-gravity contributions,
Damour and Esposito-Far{\`e}se conclude that the non-perturbative
amplification effect of spontaneous scalarization can set in when
$\beta_0 \lesssim -4$.  For the massless case considered in their
analysis, this remains a remarkably good estimate for BSs.
\begin{figure*}[t]
    \includegraphics[width=8cm, valign=t]{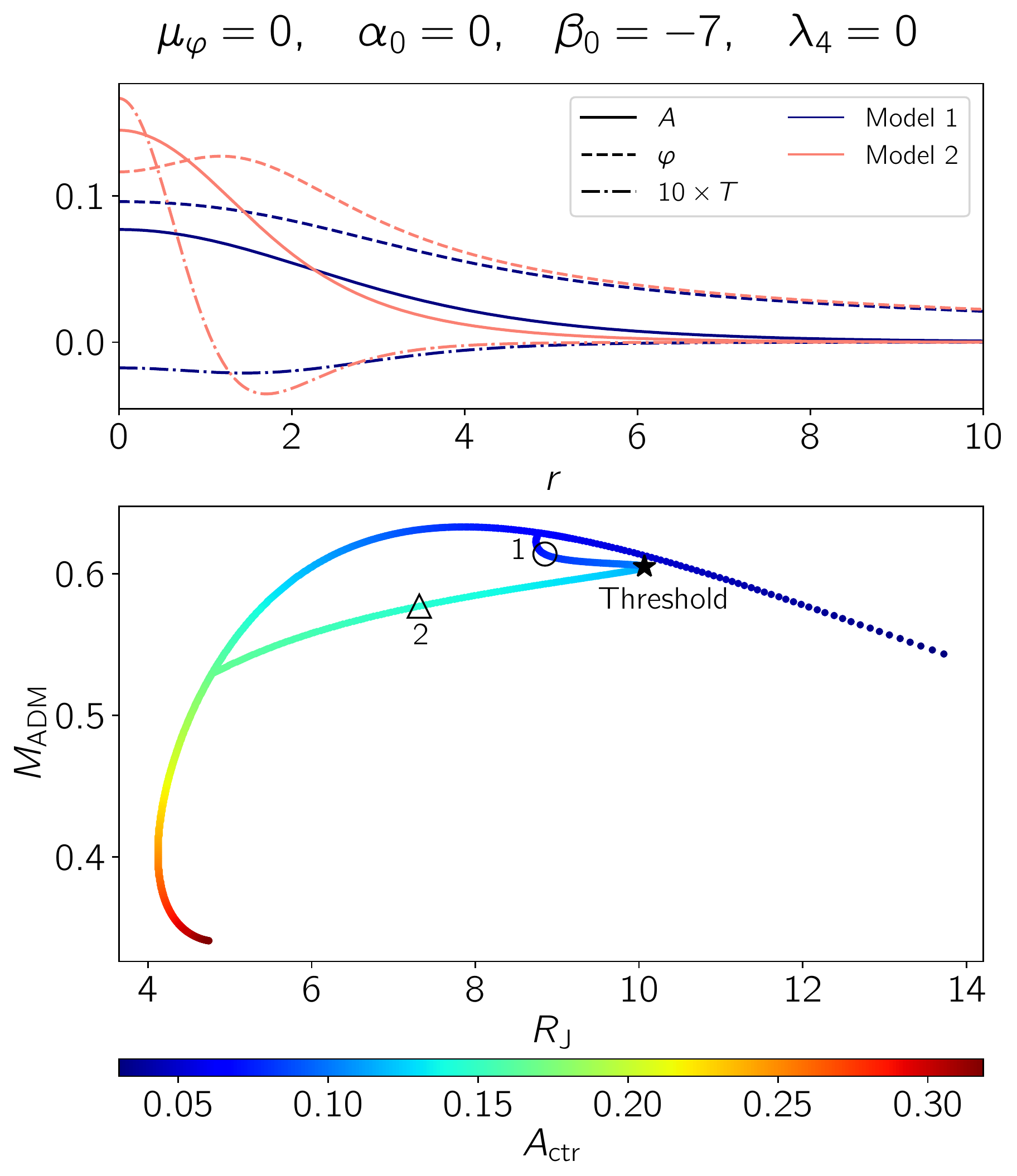}
    \qquad
    \includegraphics[width=8cm, valign=t]{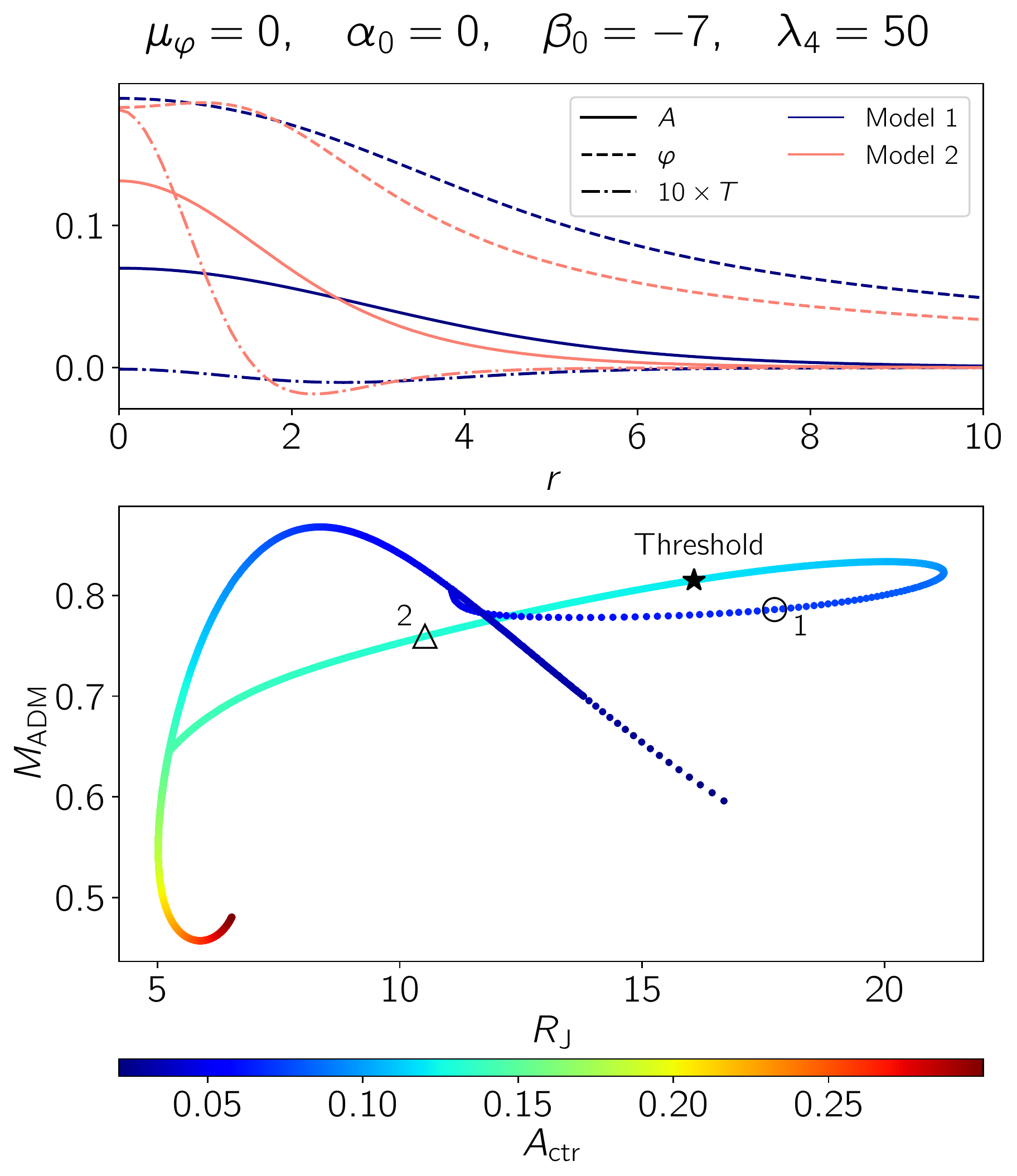}
    \caption{\textit{Upper panels}: Radial profiles of the BS scalar
    amplitude $A$, the gravitational scalar $\varphi$ and the trace
    of the energy-momentum tensor $T$ for the BS  Models 1 and 2
    indicated by a circle and a triangle, respectively, in the
    panels below. The models with $\lambda_4 = 0$ (left) have
    frequencies $\omega_1 = 0.8518$ and $\omega_2 = 0.7729$, whilst
    for models 1 and 2 with $\lambda_4 = 50$ (right) we have $\omega_1
    = 0.8303$ and $\omega_2 = 0.7572$. Note that the trace of the
    energy-momentum tensor has been scaled for presentation purposes.
    As supported by analysis in the main text, $T$ becomes negative
    at small radii for models with a gravitational scalar peaking
    at the origin.  \textit{Lower panels}: Mass-radius plots with
    the colour bar indicating the central amplitude of the BS scalar
    field.  The star marks the threshold at which the gravitational
    scalar starts to peak off centre; we ubiquitously observe that
    $\varphi$ peaks at $r=0$ for less compact stars ($A_{\rm ctr}$
    is lower than for the threshold case, i.e.~dark blue color) and
    $\varphi$ peaks off centre for more compact stars (large $A_{\rm
    ctr}$, i.e.~cyan color). We exemplify this with models 1 (less
    compact with $\varphi$ peaking at $r=0$) and 2 (more compact
    with $\varphi$ peaking off-center) for each of the potentials
    (left with $\lambda_4 = 0$ and right with $\lambda_4 = 50$).
    }
    \label{fig:MofR_diff_pots}
\end{figure*}
%

\subsection{Dependence on the bosonic potential}
\label{sec:otherpotentials}

In this section, we consider the structure of BS solutions in ST
theory of gravity for other potentials of the BS scalar field. We
focus on models governed by the potentials given in
Eq.~\eqref{eq:BSpotentials}, namely a solitonic one with $\sigma_0
\geq 0.2$ and a repulsive potential with $\lambda_4 \geq 0$. We
find that the specific choices we consider here affect the structure
of scalarized BS models, resulting in several differences to the
prototypical example studied in Sec.~\ref{sec:prototypical_example}.
In particular, we find that the gravitational scalar changes its
shape with stellar compactness and that strongly scalarized solutions
may no longer have larger masses and radii than their GR counterparts. Similar to Alcubierre et. al. \cite{Alcubierre:2010ea} we do not find self-interaction terms in the bosonic potential to be necessary for obtaining strong scalarization.
We analyse these features in more detail in the following subsections.

\subsubsection{Massless gravitational scalar field} \label{sec:potentials_massless}
{\it Gravitational scalar profiles:} Previously we noted that a
solitonic potential with $\sigma_0 = 0.2$ leads to the distinct
feature of a gravitational scalar field $\varphi(r)$ peaking
off-centre; cf.~Fig.~\ref{fig:prototype_profiles}.  For repulsive
($\lambda_4>0$) and mini ($\lambda_4=0$) BS models, in contrast,
the gravitational scalar more commonly reaches its global extremum
exactly at the centre. We illustrate this in Fig.~\ref{fig:MofR_diff_pots},
where $\varphi(r)$ gradually changes its behaviour from peaking at
$r=0$ to peaking off-centre, as one moves to models with higher
central amplitudes $A_{\rm ctr}$ of the BS scalar. The same behaviour
has been observed for neutron stars of increasing compactness;
cf.~Fig.~7 in Ref.~\cite{Rosca-Mead:2020bzt}.  In
Fig.~\ref{fig:MofR_diff_pots} we mark by a star the threshold, where
this transition occurs.  We additionally plot in each of the upper
panels the profiles $A(r)$ and $\varphi(r)$ for two models marked
1 and 2 in the mass-radius diagram. These models are located on
either side of the threshold: for model 1 (less compact, i.e.~smaller
$A_{\rm ctr}$ than the threshold model) $\varphi$ peaks at the
centre $r=0$ and for model 2 (more compact with larger $A_{\rm
ctr}$) it peaks off centre.

\begin{figure}
    \includegraphics[width=0.5\textwidth]{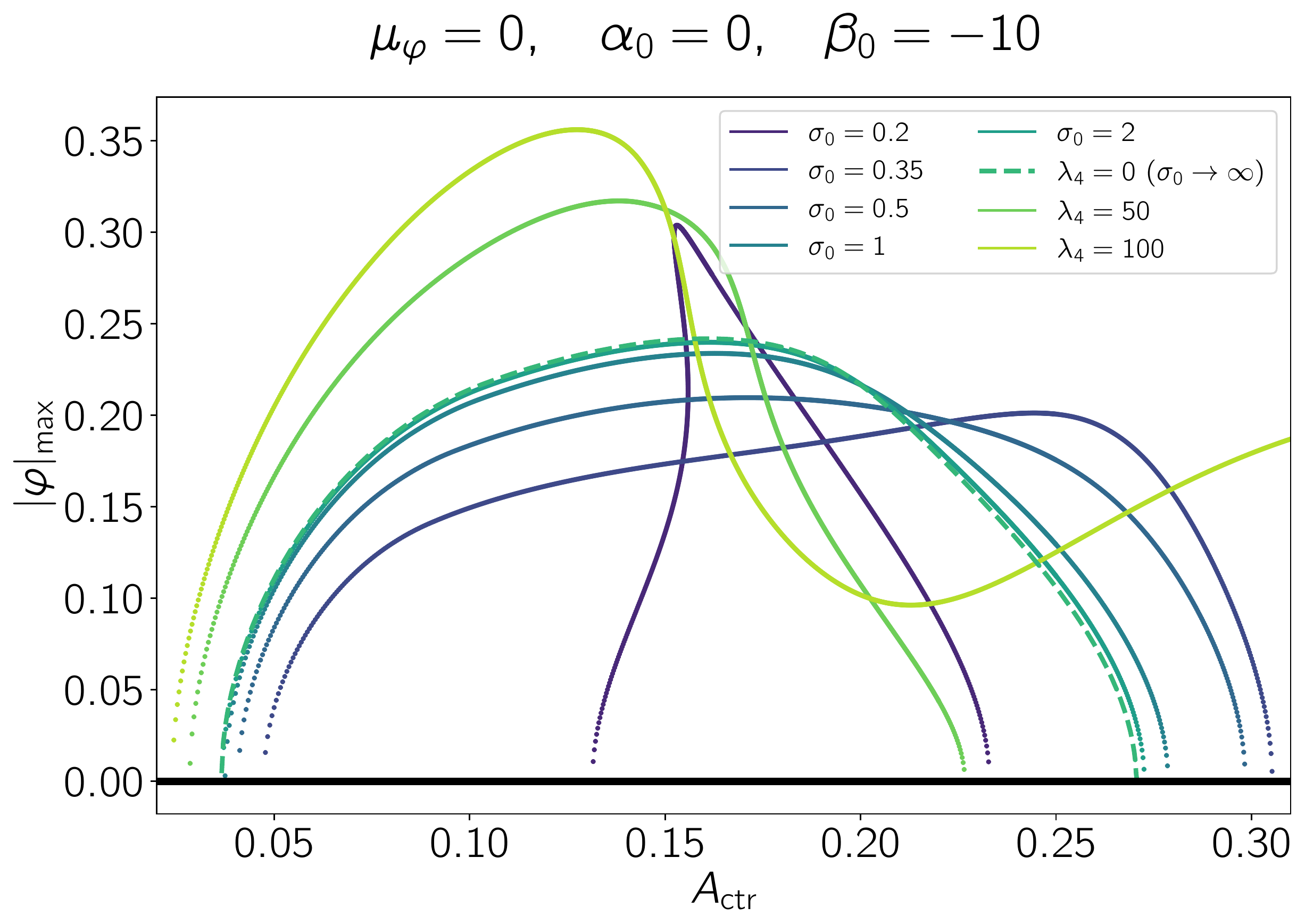}
    \caption{Maximum of the gravitational scalar field as a function
    of the central bosonic amplitude for various potentials considered
    in this work. We remark that repulsive potentials result in
    more strongly scalarized solution than any other potential
    considered here.
    }
    \label{fig:diffpotentials_beta10_alpha0}
\end{figure}

We can qualitatively understand this behaviour as follows: the shape
of the gravitational scalar field near the origin is controlled by
Eq.~(\ref{eq:allr}) for $\kappa$, which denotes a re-scaled version
of the radial derivative for the gravitational scalar field $\partial_r
\varphi$. The overall sign of $\partial_r \kappa$ at small radii
determines the concavity of $\varphi$ near the origin, and, thus,
whether it peaks off or at the centre. Without loss of generality,
we assume that $\varphi_{\rm{ctr}} > 0$, then the sign of $\partial_r
\kappa$ will be determined by the two terms: $2\alpha X F W_{,
\varphi^2} \varphi$ and $\left(\omega^2A^2 - 2\alpha^2 V - F^2
\theta^2 \right)$. The former is always positive or zero, whilst
the latter can be negative, provided the magnitudes of the potential
and the lapse (i.e.~$2 \alpha^2 V$) are large enough to counteract
the positive contribution of $\omega^2 A^2$.  The last term
$F^2\theta^2$ is negligible near the origin since $\theta(0)=0$ by
Eq.~(\ref{eq:BCs}).  The ingredients required to make $\partial_r
\kappa$ negative, and thus cause the gravitational scalar $\varphi$
to peak at the origin, are as follows:
\begin{enumerate}
    \item \textbf{Large potential values $V$}: for typical values
    of $A_{\rm ctr} \in [0.1, 0.2]$ in the regime of scalarization,
    the non-interacting and the repulsive potentials take on larger
    values compared to a solitonic potential with small $\sigma_0$;
    cf.~Fig.~\ref{fig:potentials}.
    \item \textbf{Small central amplitudes of the BS scalar}: models
    with small $A_{\rm{ctr}}$ result in less compact (or {\it
    fluffy}) stars. These fluffy configurations have shallower
    gravitational wells which manifest themselves in larger values
    of the lapse $\alpha$ near the origin\footnote{In our gauge,
    the lapse function $\alpha$ asymptotes to 1 and smaller values
    of the lapse function correspond to deeper gravitational wells
    and vice versa.}. Small $A_{\rm{ctr}}$ and the ensuing large
    $\alpha$ therefore are more likely to make the overall contribution
    to $\partial_r \kappa$ negative.  Empirically, we find that
    mini and repulsive potentials, but rarely solitonic with $\sigma_0
    \approx 0.2$, allow for scalarized BSs with small $A_{\rm ctr}$,
    i.e.~low compactness; cf.~Fig.~\ref{fig:diffpotentials_beta10_alpha0}.
\end{enumerate}

This behaviour of the gravitational scalar can also be explained
using the trace of the energy-momentum tensor which is obtained
from Eq.~(\ref{eq:em}),
\begin{equation} \label{eq:trace_em}
    T = g^{\alpha \beta} T_{\alpha \beta} = \frac{1}{F \alpha^2} \left(\omega^2 A^2 - 2\alpha^2 V - F^2 \theta^2 \right)\,.
\end{equation}
Up to an overall factor, this expression equals the second source
term of our above analysis that drives $\partial_r \kappa$ via
Eq.~\eqref{eq:allr}.  The sign of $T$ can therefore be directly
translated to the concavity of $\varphi$.  Recalling that the
tachyonic instability requires negative $T$, we expect strong
scalarization to occur in the region where $T<0$.
Fig.~\ref{fig:MofR_diff_pots} illustrates exactly this effect: for
less compact models with $T \lesssim 0$ at small $r$, $\varphi$
peaks at the origin, and for compact models with a central $T \gtrsim
0$, $\varphi$ peaks away from the origin; cf.~also
Fig.~\ref{fig:prototype_profiles}. The $\varphi(r)$ profile of
particularly compact scalarized models is therefore more
\textit{shell}-like.

\textit{The onset and degree of scalarization}: The nature of the
potential further determines the degree of scalarization of BS
solutions and the global properties of such models.  Figure
\ref{fig:MofR_diff_pots} illustrates one example of this for mini
and repulsive potentials in the mass-radius plane, where all of the
scalarized branch falls underneath the GR one for $\lambda_4=0$,
but only part of it for $\lambda_4=50$. Similarly, we see in
Fig.~\ref{fig:BSfamilies} in Appendix \ref{app:BSfamilies} that
many scalarized BSs for solitonic potentials with $\sigma \gtrsim
0.3$ have smaller masses than their GR counter parts with equal
radius, very much in contrast to the solitonic $\sigma_0=0.2$ case
of the prototypical example studied in Sec.~\ref{sec:prototypical_example}.

Furthermore, the threshold for $\beta_0$ at which scalarization
starts to occur varies with the potential; we summarize this
dependence in Table \ref{tab:massless_potentials_thr}. Overall, we
find that the repulsive potential requires less negative $\beta_0$
values for solutions to scalarize, which is then followed by the
mini and solitonic potentials. For massless ST theory, the thresholds
fall into a range $-7.5 \lesssim \beta_{0,{\rm thr}}\lesssim -4.96$,
a bit below the threshold of $-4.35$ typically found for neutron
stars.
\begin{table}[b]
\caption{Summary of threshold values of $\beta_0$ for the onset of
scalarization for different potentials in the massless case.}
\begin{tabular}{c c c c} 
\hline
\hline
~~~~~ Potential ~~~~~ & ~~~~~ Coupling ~~~~~ & ~~~~~ $\beta_{0,\rm{thr}}$ ~~~~~ \\
\hline
Solitonic & $\sigma_0 = 0.2$ & -6.079   \\
Solitonic & $\sigma_0 = 0.3$ & -7.282   \\
Solitonic & $\sigma_0 = 0.5$ & -6.641   \\
Solitonic & $\sigma_0 = 1$ & -6.219   \\
Solitonic & $\sigma_0 = 2$ & -6.128   \\
Mini & $\lambda_4 = 0$ & -6.098  \\
Repulsive & $\lambda_4 = 50$ & -5.272  \\
Repulsive & $\lambda_4 = 100$ & -4.961  \\
\hline
\end{tabular}
\label{tab:massless_potentials_thr}
\end{table}

Even though solutions are less strongly scalarized for the solitonic
potential with $\sigma_0 \gtrsim 0.3$, there are some subtle features
in the dependence of scalarization on the self-interaction term as
illustrated in Fig.~\ref{fig:diffpotentials_beta10_alpha0}.  First,
the degree of scalarization varies non-monotonically with $\sigma_0$.
Typically smaller values of $\sigma_0 \approx 0.2$ result in more
strongly scalarized solutions than models with  $\sigma_0 \approx
1$.  On the other hand, larger $\sigma_0$ lead to a wider range of
$A_{\rm ctr}$ for which scalarized solutions exist: in
Fig.~\ref{fig:diffpotentials_beta10_alpha0} the support of
$|\varphi|_{\rm{max}}(A_{\rm{ctr}})$ significantly increases for
models with $\sigma_0 \gtrsim 0.35$.  As $\sigma_0$ increases, the
scalarized branches slowly converge to the scalarized solutions of
mini BSs, which can be seen by a close overlap of the $\sigma_0 =
2$ and $\lambda_4 = 0$ curves. We present a more detailed illustration
of this transition in Appendix \ref{app:BSfamilies}, where we compute
families of solutions for a wider range of potentials. Finally, we
find that for a fixed value of $\beta_0$, the strongest scalarization
occurs in the case of the repulsive potential with our largest
$\lambda_4 = 100$.

\begin{table}[b]
\caption{The minimum value of $\sigma_0$ where we find BS models
with a gravitational scalar peaking at $r=0$ is listed for solitonic
BS families with $\alpha_0=0$ as well as the specified values of
$\mu_{\varphi}$ and $\beta_0$. A gravitational scalar $\varphi$
reaching its global extremum at the origin is easier to obtain for
smaller $\mu_{\varphi}$ and for larger $\beta_0$.
}
\begin{tabular}{ccc}
  \hline
  \hline
  $~~~~~\mu_{\varphi}$~~~~~ & ~~~~~$\beta_0$~~~~~ & ~~~~~$\sigma_{0,{\rm min}}~~~~~$
  \\
  \hline
  0 & -10 & 0.25
  \\
  0.1 & -10 & 0.34
  \\
  0.1 & -14 & 0.24
  \\
  0.3 & -14 & 0.55
  \\
  \hline
\end{tabular}
\label{tab:peak}
\end{table}
\begin{figure*}
    \includegraphics[width=18cm, valign=t]{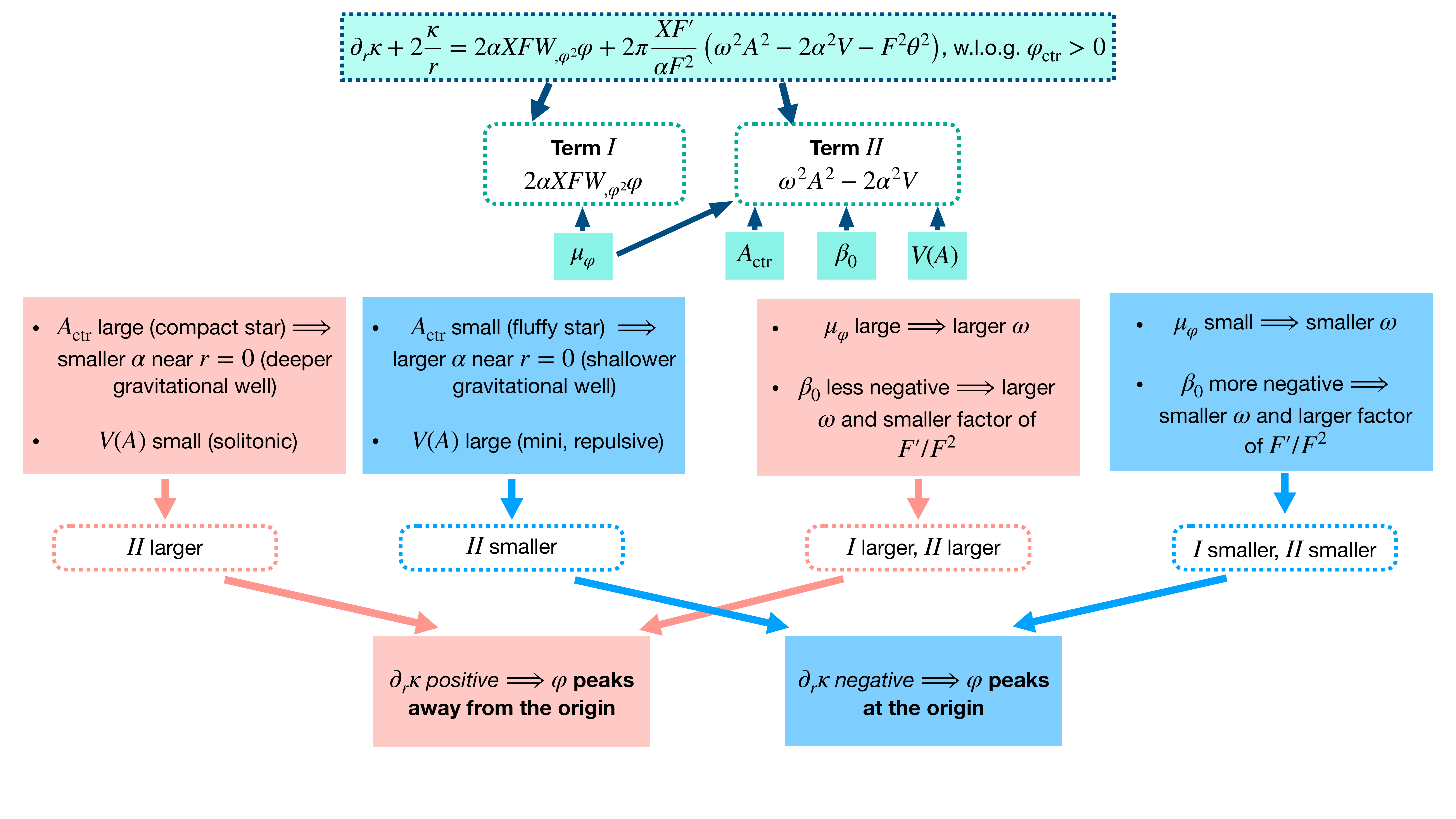}
    \caption{Flow-chart showing the dependence of the gravitational
    scalar profile $\varphi$ on the BS parameters, $A_{\rm{ctr}}$
    and $V(A)$, and the ST parameters, $\mu_{\varphi}$ and $\beta_0$.
    As discussed in the main text, the sign of $\partial_r \kappa$
    determines the behavior of $\varphi$ at the origin. If $\partial_r
    \kappa > 0$, $\varphi$ peaks away from $r=0$, whilst for
    $\partial_r \kappa < 0$, $\varphi$ peaks at $r=0$. The two
    source terms, denoted by $I$ and $II$ in the diagram, play the
    leading role in determining the sign of $\partial_r \kappa$ at
    $r=0$. By the boxes' colours we mark how the ST and BS parameters
    tend to impact the $\varphi$ profiles -- blue for peaking at
    the origin and pink for peaking off centre.
    }
    \label{fig:flow_chart}
\end{figure*}

\subsubsection{Massive gravitational scalar field}
\textit{Gravitational scalar profiles}: We next turn to analysing
the features of scalarized BS solutions with other bosonic potentials
in the massive case. A systematic investigation of the global maximum
of $|\varphi|$ across the parameter space confirms that for
$\mu_{\varphi}>0$, certain choices of the potential $V(A)$ still
result in the gravitational scalar profiles peaking at the origin.
However, we find that an increasing mass $\mu_{\varphi}$ tends to
push the extremum of $\varphi$ away from the origin. This effect
can be explained using the same reasoning as in the massless case
of Sec.~\ref{sec:potentials_massless} by assessing the sign of the
right-hand-side of Eq.~\eqref{eq:allr} for $\partial_r \kappa$.
Assuming that $\varphi_{\rm{ctr}} > 0$, we conclude that $\mu_{\varphi}
> 0$ increases the overall positive contribution from the $2\alpha
X F W_{, \varphi^2} \varphi$ term and empirically we find it also
increases $\omega$. Therefore, larger values of the gravitational
mass push $\partial_r \kappa$ towards positive values and the
gravitational scalar $\varphi$ tends to peak away from the origin.

It is possible however to push the gravitational scalar field maximum
towards $r=0$ by making $\beta_0$ negative enough. The impact of a
more negative $\beta_0$ parameter is two-fold:
\begin{enumerate}
    \item BS frequencies $\omega$ decrease,
    \item The overall magnitude of $\frac{2\pi XF'}{\alpha F^2}
    (\omega^2 A^2 - 2\alpha^2 V)$ increases due to the larger
    $F'/F^2$ factor.
\end{enumerate}
The combined effect of a more negative $\beta_0$ is therefore to
make the final term of the $\partial_r \kappa$ equation negative
and large enough to counteract the non-zero positive contribution
from $2\alpha X F W_{, \varphi^2} \varphi$ term. The sign of
$\partial_r \kappa$ becomes negative and $\varphi$ peaks at $r=0$.

We quantitatively illustrate the effect of $\mu_{\varphi}$ and
$\beta_0$ on the $\varphi$ profile in Table \ref{tab:peak} where
we list for several combinations of ($\mu_{\varphi}$, $\beta_0$)
the minimal $\sigma_0$ values at which we find models with $\varphi$
peaking at $r=0$. Below the respective $\sigma_{0,{\rm min}}$,
$\varphi$ peaks off center for all scalarized BS models. As the
table suggests, it is easier to find $\varphi$ profiles peaking at
the origin for smaller gravitational masses $\mu_{\varphi}$ and
more negative $\beta_0$. In Fig.~\ref{fig:flow_chart} we summarize
the dependence of the gravitational scalar profile on the BS and
ST parameters.

\begin{figure}%
    \includegraphics[width=0.45\textwidth,valign=t]{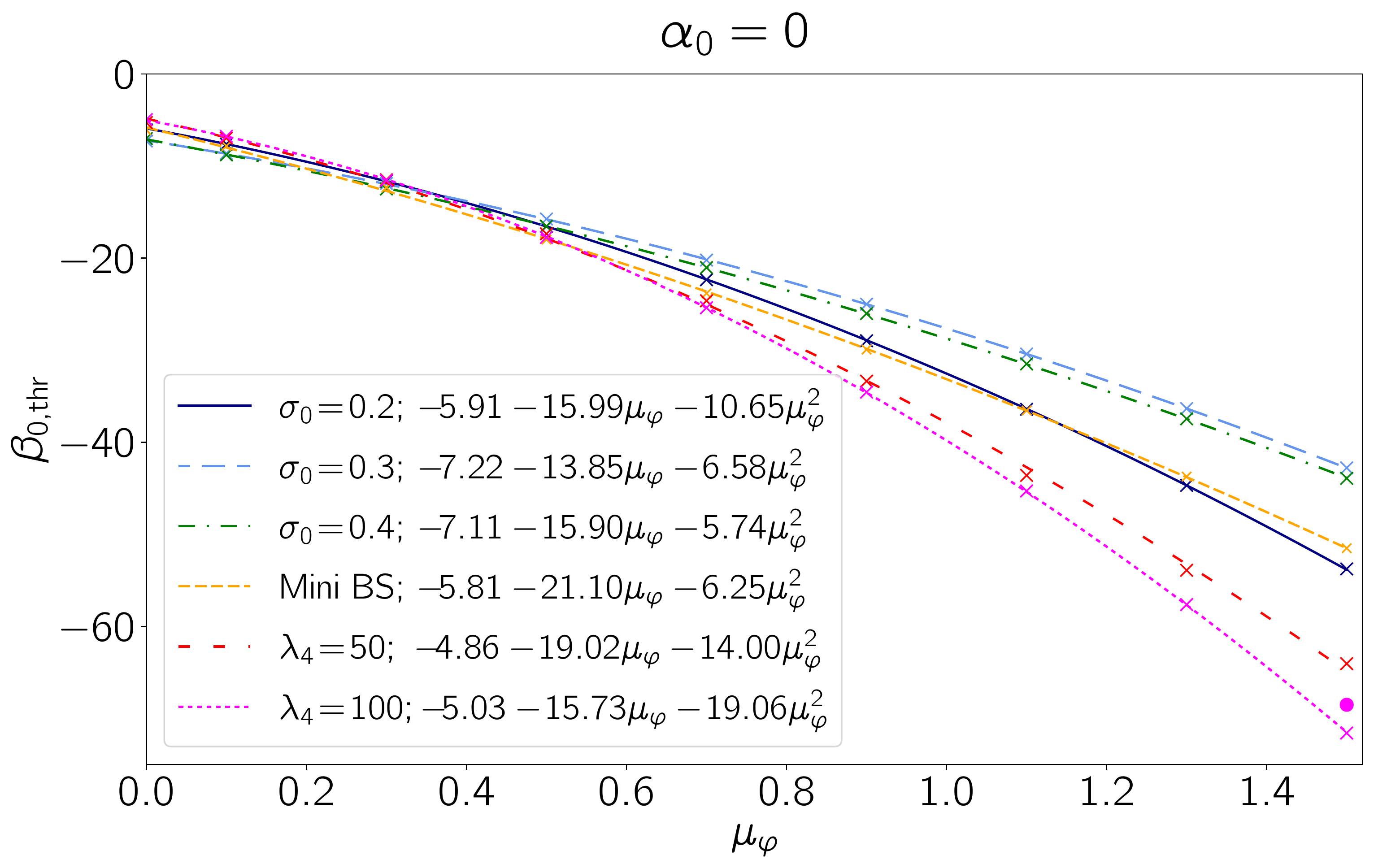}
    \caption{
    The threshold $\beta_{0,{\rm thr}}$ for the onset of scalarization
    as a function of $\mu_{\varphi}$ for different BS potentials
    $V(A)$ as given in Eq.~(\ref{eq:BSpotentials}) with the indicated
    parameter values $\sigma_0$ and $\lambda_4$. The lines represent
    quadratic fits with the given coefficients.  The case
    $\lambda_4=100$, $\mu_{\varphi}=1.5$ exhibits one minor anomaly:
    here, the $\times$ symbol marks the disappearance of stable
    scalarized models, but extremely compact unstable BSs are still
    found up to the $\beta_0$ value marked by the filled circle.
    Only the former ($\times$) data point is used in the fit.
    \label{fig:multionset}
    }
\end{figure}

\begin{figure}
  \includegraphics[width=0.5\textwidth]{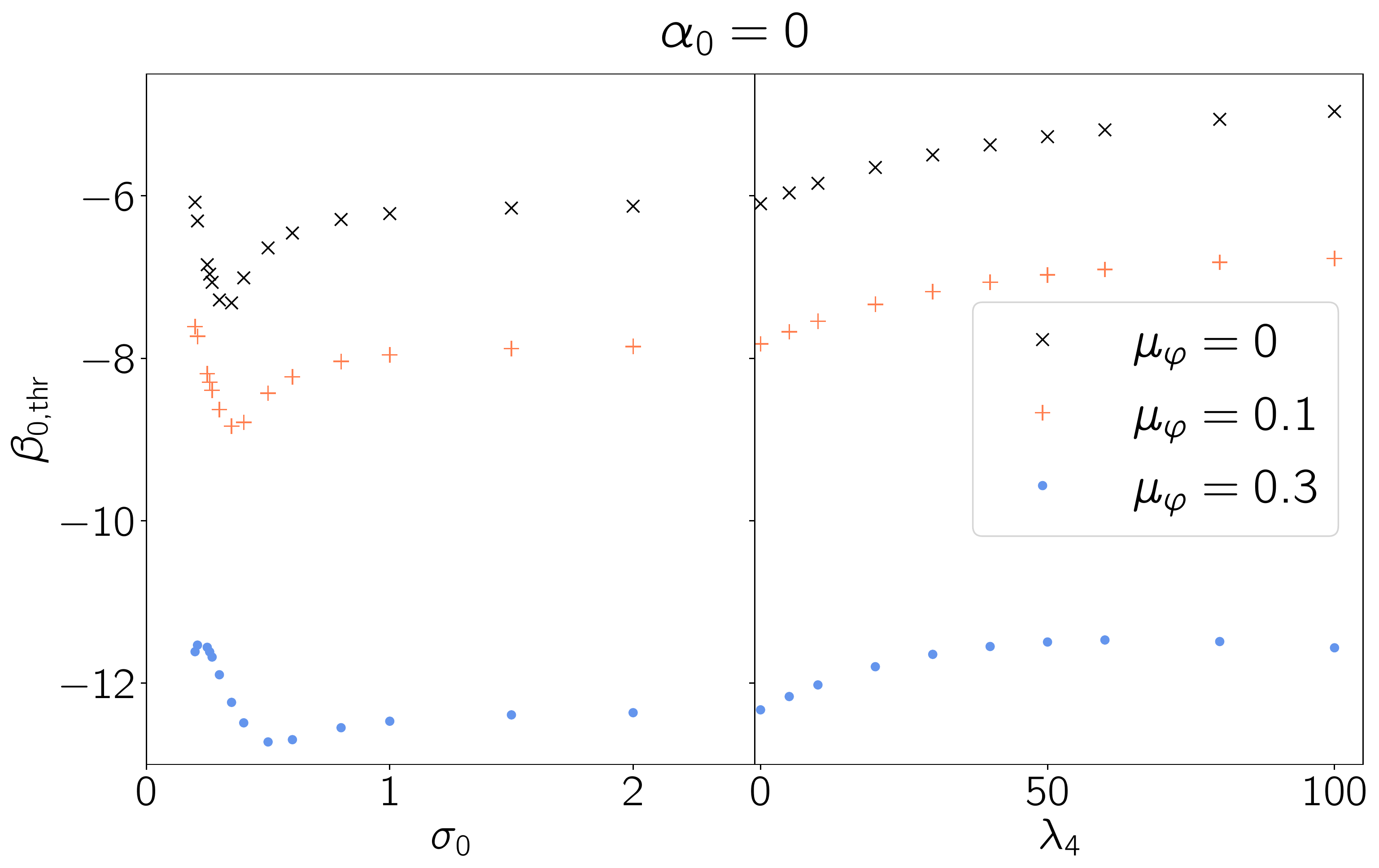}
  \caption{
  The threshold $\beta_{0,{\rm thr}}$ for the onset
  of strong scalarization is shown for the different
  types of potentials and three values of the gravitational
  scalar mass $\mu_{\varphi}=0,~0.1$ and $0.3$.
  The horizontal axis represents $\sigma_0$ for the
  solitonic potential on the left and $\lambda_4$ for
  the repulsive potential on the right. 
  \label{fig:onset_sigma0_lambda4}
  } 
\end{figure}

\textit{The onset and degree of scalarization}: Finally, we make a
few remarks on the effect of $\mu_{\varphi} > 0$ on the scalarization
of BS solutions.  As already seen in Fig.~\ref{fig:beta0thr}, the
threshold $\beta_{0,{\rm thr}}$ for the onset of strong scalarization
for solitonic BSs with $\sigma_0=0.2$ is well described by a quadratic
function of the mass parameter $\mu_{\varphi}$ for ST gravity with
$\alpha_0=0$.  In Fig.~\ref{fig:multionset} we plot the onset of
scalarization for several other potential functions with parameters
as listed in the legend. For all cases, the numerical data are in
excellent agreement with a monotonically decreasing quadratic fit
which confirms our general observation that an increasing $\mu_{\varphi}$
weakens scalarization. In order to assess in more detail how different
potential functions affect the onset of scalarization, we display
in Fig.~\ref{fig:onset_sigma0_lambda4} the threshold values
$\beta_{0,{\rm thr}}$ as a function of the parameters $\sigma_0$
and $\lambda_4$ for three fixed values of the mass $\mu_{\varphi}=0,~0.1$
and $0.3$. We recall that mini BSs are obtained in both limits,
$\sigma_0\rightarrow \infty$ and $\lambda_4=0$, and can be regarded
as the link between the solitonic and repulsive potential.  The
variation of $\beta_{0,{\rm thr}}$ in Figs.~\ref{fig:multionset}
and \ref{fig:onset_sigma0_lambda4} shows that for $\mu_{\varphi}
\lesssim 0.3$, similar to the massless case, BSs with a repulsive
potential are more susceptible to scalarization than solitonic and
mini BSs, although not by a huge amount.  For $\mu_{\varphi}\gtrsim
0.3$, however this trend is reversed and solitonic BSs scalarize
most easily.

%
\subsection{Dependence on $\alpha_0$}
\label{sec:alpha0ne0}
\begin{figure*}
    \includegraphics[width=8cm, valign=t]{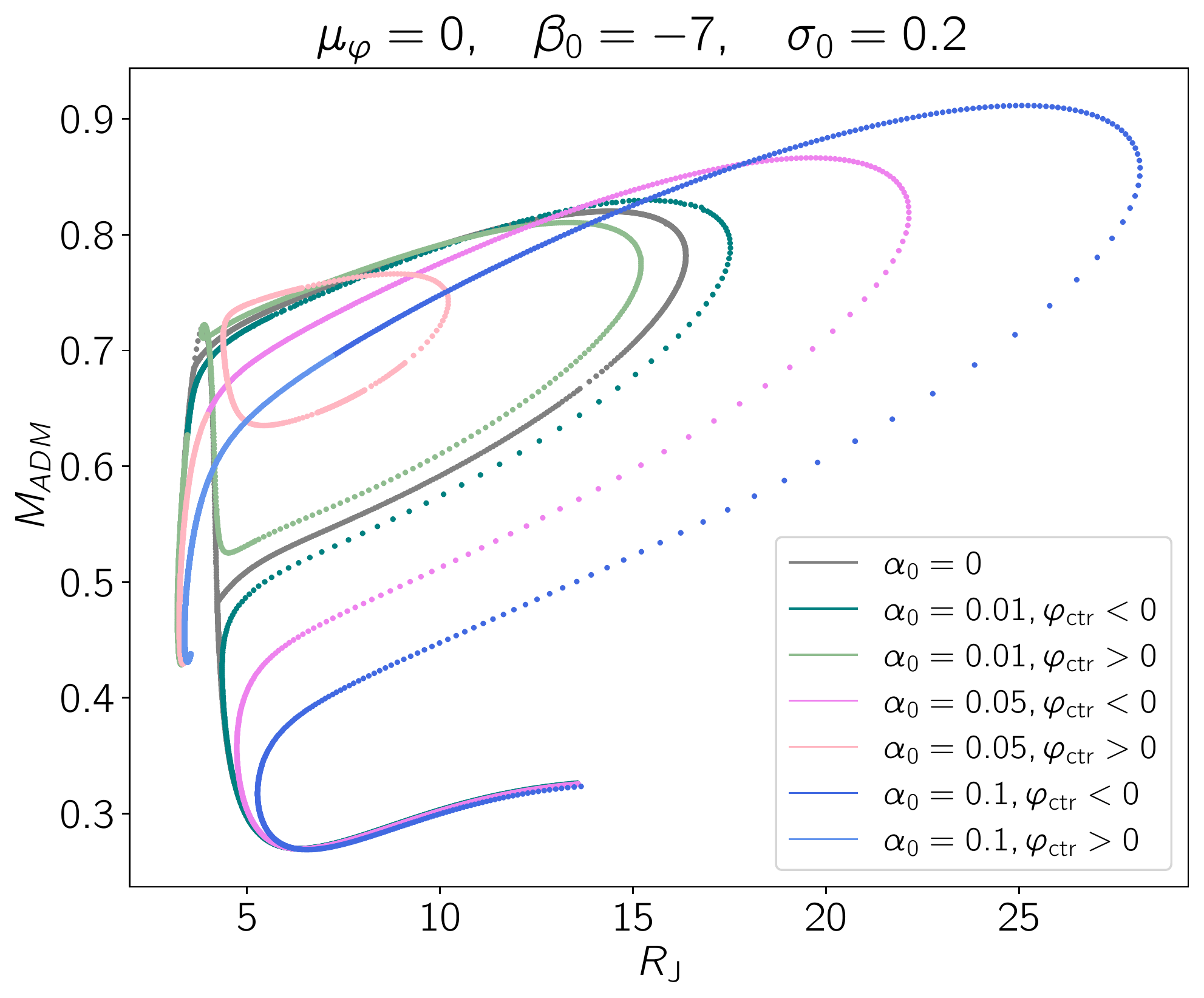}
    \quad
    \includegraphics[width=8cm, valign=t]{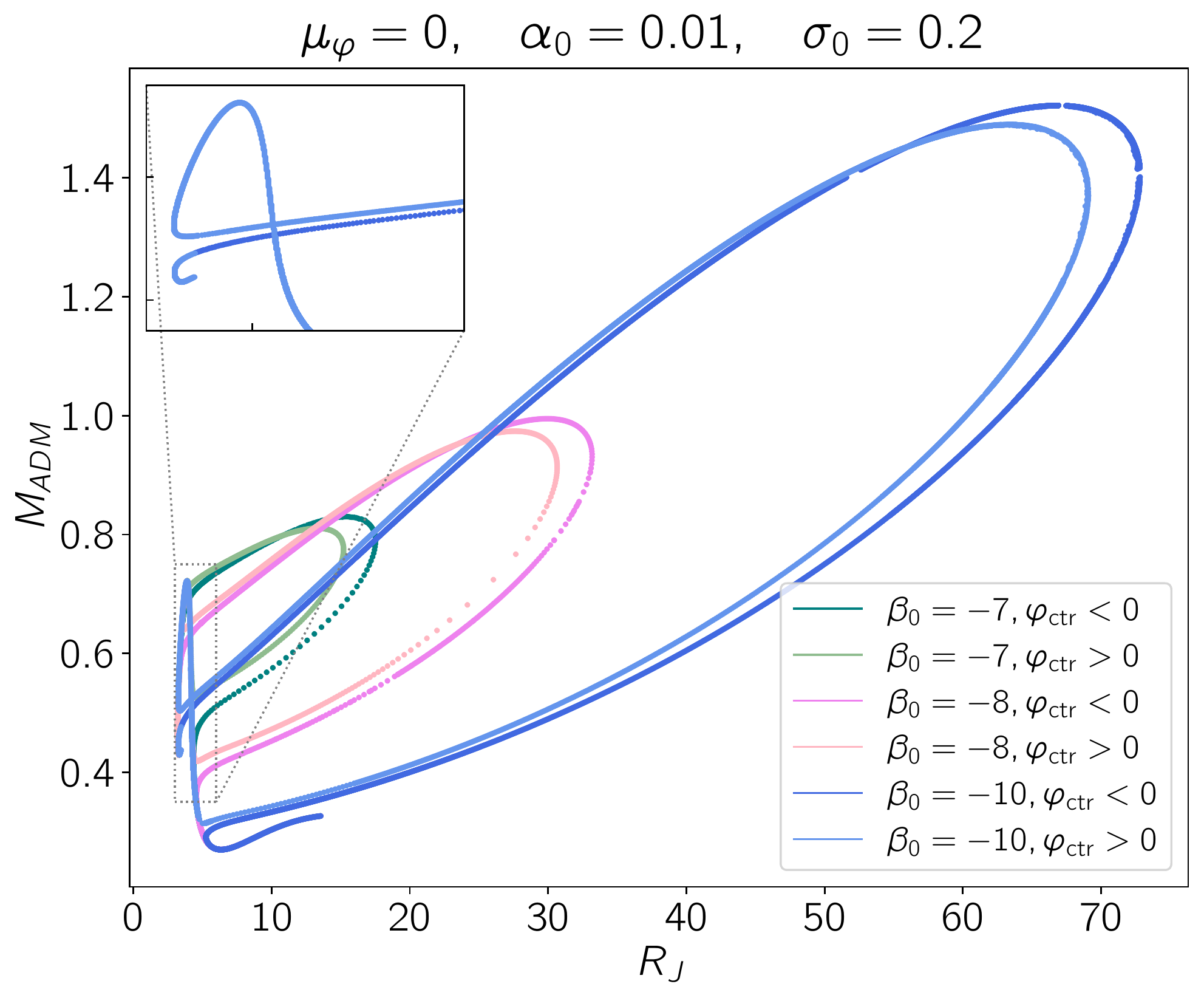}
    \caption{\textit{Left}: BS models in ST gravity with $\mu_{\varphi}
    = 0$, $\beta_0 = -7$ and varying $\alpha_0$ for a solitonic
    potential with $\sigma_0 = 0.2$. The branches obtained for each
    value of $\alpha_0$ are presented in different shades of one
    color to differentiate between positive and negative central
    gravitational field amplitudes. For $\alpha_0 = 0.01$, for
    instance, dark green denotes models where $\varphi_{\rm{ctr}}
    < 0$ and light green those with $\varphi_{\rm{ctr}} > 0$. For
    reference, we also plot the solutions for $\alpha_0 = 0$ which
    are shown in grey.  \textit{Right}: Dependence of the BS solutions
    on $\beta_0$ for $\alpha_0=0.01,~\mu_{\varphi}=0$ and a solitonic
    potential with $\sigma_0 = 0.2$. The inset highlights the
    self-intersecting $\varphi_{\rm{ctr}} > 0$ branch with $\infty$
    shape for the sequence of models with $\beta_0 = -10$.}
    \label{fig:MofR_alphavar} 
\end{figure*}

Previously we have considered ST theories with $\alpha_0 = 0$, where
the solutions are degenerate under the transformation $\varphi \to
-\varphi$. In the case, of $\alpha_0 \neq 0$ we acquire additional
structure through the splitting of the scalarized branches. More
specifically, we find\footnote{Recall that in our convention $\alpha_0
\geq 0$; for the opposite convention $\alpha_0 \leq 0$ the signs
of $\varphi$ would be reversed.}:
\begin{enumerate} 
    \item A larger branch (as viewed in the mass-radius plane),
    mostly with negative gravitational amplitudes $\varphi_{\rm{ctr}}$.
    At very high compactness (i.e.~large $A_{\rm{ctr}}$) these
    models start to have $\varphi_{\rm{ctr}} > 0$ and the $\varphi(r)$
    profiles have a zero-crossing.
    \item A smaller closed loop, where we exclusively observe
    positive $\varphi_{\rm{ctr}}$.
\end{enumerate}
These two branches contain strongly scalarized solutions, as well
as weakly scalarized ones, in which case the BS models have similar
mass and radius to their GR counterparts. This is illustrated in
Fig.~\ref{fig:MofR_alphavar}, where certain BS models of the $\alpha_0
> 0$ sequences closely resemble the GR solutions obtained for
$\alpha_0 = 0$.  For mild values of $\alpha_0$, the closed loop
intersects itself and acquires a shape similar to the $\infty$
symbol as in the case of $\alpha_0 = 0.01$. As we increase $\alpha_0$,
first the $\infty$-shape and then the entire loop slowly shrink and
disappear (see e.g.~$\alpha_0 = 0.1$). We note that certain values
of $\alpha_0$ we consider here are above the Cassini measurement
constraint. This choice is made for visualization purposes: for
smaller values of $\alpha_0$ the two branches lie more closely to
each other making their splitting less evident. Although here we
discuss solitonic BSs, for other potentials we observe a qualitatively
similar splitting of the solution branches.

The disappearance of one of the solution branches can be traced
back to the conformal coupling function
$F(\varphi)=\exp(-2\alpha_0\varphi-\beta_0\varphi^2)$.  When $\alpha_0
= 0$, the value $F$ is solely determined by the magnitude of
$\beta_0$: a more negative $\beta_0$ leads to stronger scalarization.
A non-zero $\alpha_0$, on the other hand, leads to a non-zero $-2
\alpha_0 \varphi$ contribution in the exponent of $F$. For $\varphi
> 0$, this curbs the effect of $\beta_0$ and subsequent scalarization
whilst for $\varphi <0$ the $\alpha_0$ term strengthens it: for
$\alpha_0>0$ we have $F(\varphi > 0) < F(\varphi < 0)$.  This effect
is further illustrated by the relative size of the scalarized
branches in Fig.~\ref{fig:MofR_alphavar}: the solution branches for
$\varphi_{\rm{ctr}} > 0$ are systematically smaller in size than
their $\varphi_{\rm{ctr}} < 0$ counterparts.  For sufficiently large
values of $\alpha_0$ with fixed $\beta_0$, the solutions with
$\varphi_{\rm ctr}>0$ cease to exist and the branch disappears.
Considering the case $\alpha_0 = 0.1$ and $\beta_0 = -7$, for
example, we can easily support this by a back-of-the-envelope
calculation: assuming $\varphi \sim 0.1$, the conformal coupling
function is $F(\varphi) \sim \rm{exp} (0.05)$, which is equivalent
to setting $\alpha_0 = 0$ and $\beta_0 = -5$. As discussed in Section
\ref{sec:mass_onset} (see also Fig.~\ref{fig:beta0thr} and Table \ref{tab:massless_potentials_thr}), however,
the onset of scalarization for $\mu_{\varphi}=0$ and a solitonic potential with $\sigma_0
= 0.2$ occurs for $\beta_0 \gtrsim -6.1$.  An effective $\beta_{0,{\rm
eff}} \approx -5$ falls out of this range and is therefore compatible
with the disappearance of positive solutions for $\alpha_0=0.1$.

Besides the branch splitting, BS solutions with $\alpha_0 > 0$
depend on $\beta_0$ and $\mu_{\varphi}$ in a way very similar to
the $\alpha_0 = 0$ case discussed previously. In particular, a more
negative $\beta_0$ results in stronger scalarization of solutions,
with larger mass $M_{\rm ADM}$ and radius $R_{\rm J}$ (cf.~the right
panel of Fig.~\ref{fig:MofR_alphavar}).  Increasing the mass parameter
$\mu_{\varphi}$, on the other hand, systematically reduces the
scalarization effect in the same way we have observed in
Sec.~\ref{sec:mass_onset} for $\alpha_0=0$.

\subsection{Thin-shell models}
\label{sec:thinshell}
In Ref.~\cite{Collodel:2022jly}, Collodel and Doneva computed an
intriguing type of BS solutions in GR dubbed {\it thin-shell} models (also see Ref.~\cite{Boskovic:2021nfs}).
The main characteristics of these BSs are (i) an approximately
constant scalar profile $A(r)$ extending out to rather large radius
and (ii) a low frequency $\omega$; see in particular their Fig.~6.
As a consequence of these two features, the derivatives $\partial_t
A$ and $\partial_r A$ become very small at small as well as very
large radii, resulting in a nearly vanishing energy density except
for a shell region where the scalar field eventually drops to zero.
In Fig.~\ref{fig:mvph005_a0m00_bm11_Soli020} for our prototypical
BS family, we can see that the scalarized stars tend to have lower
frequencies than their GR counterparts.  We have found this behaviour
systematically in our investigations of the BS and ST parameter
space (see Fig.~\ref{fig:families_omega}), suggesting that scalarization
can strengthen the thin-shell character of BSs.

\begin{figure}
    \includegraphics[width=0.45\textwidth,valign=t]{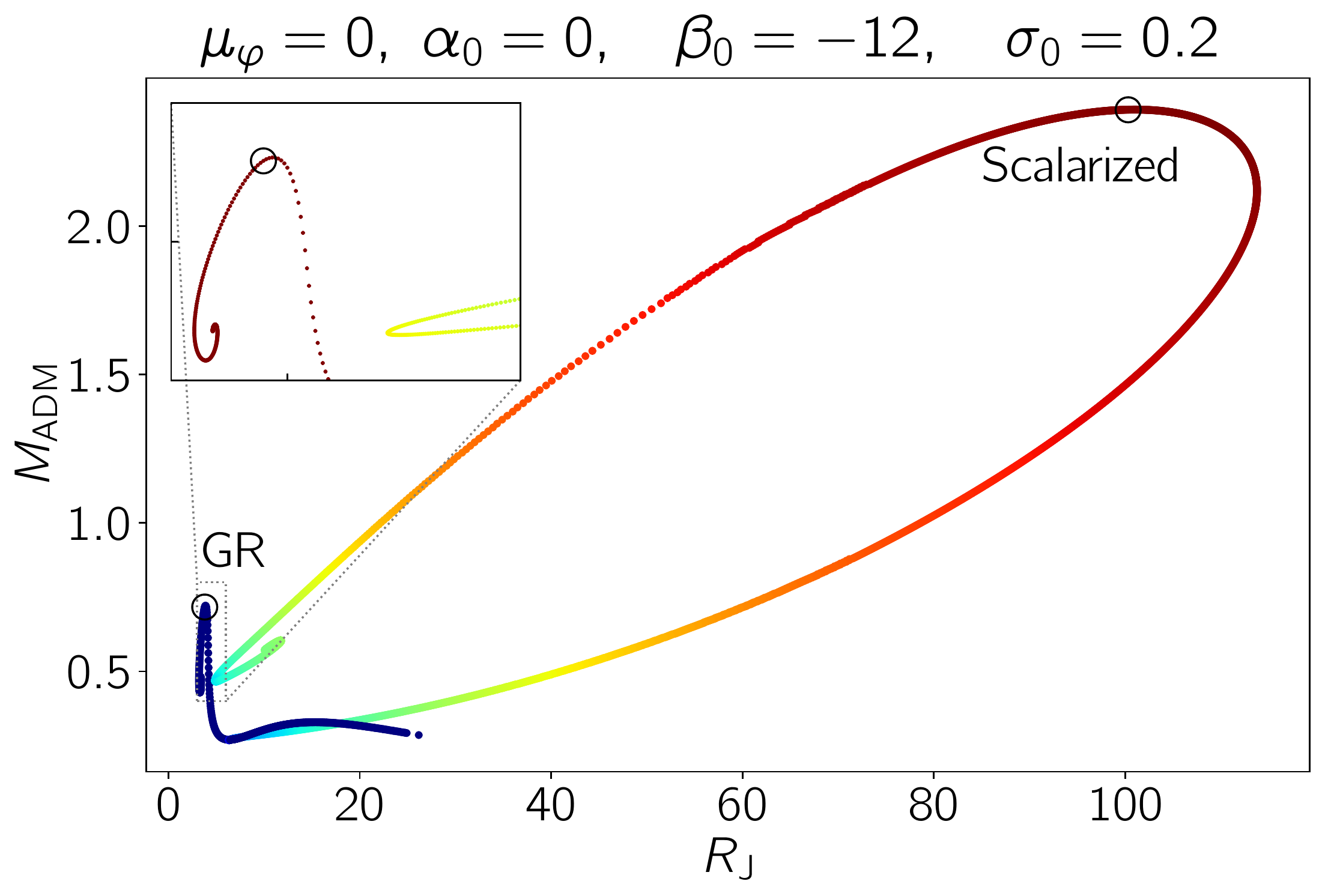}
    \includegraphics[width=0.45\textwidth,valign=t]{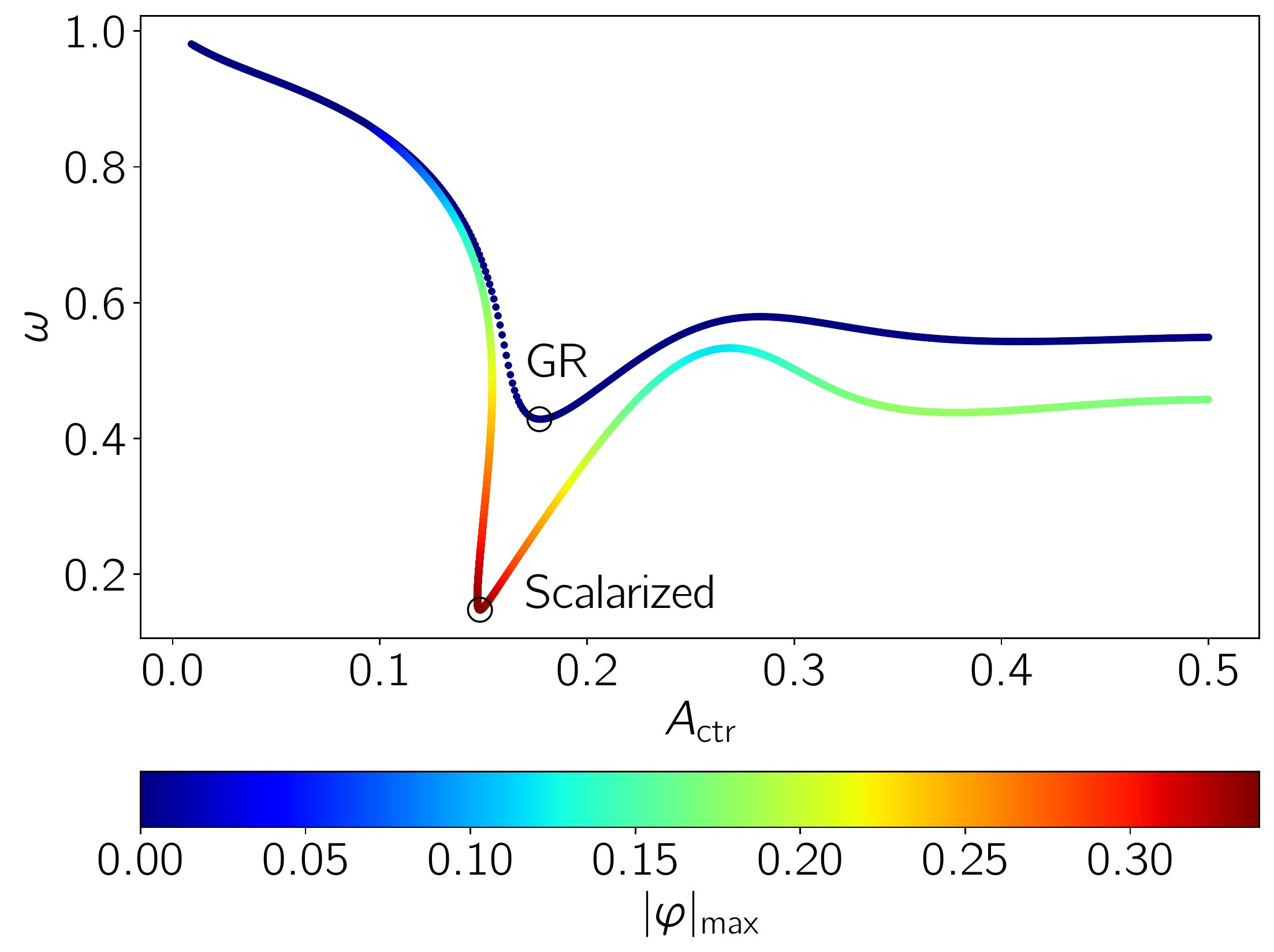}
    \caption{
    {\it Top}: Mass-radius diagram for solitonic BS models with
    $\mu_{\varphi}=0$, $\alpha_0=0$, $\beta_0=-12$ and $\sigma_0=0.2$.
    {\it Bottom:} Frequency $\omega$ as a function of the central
    BS scalar amplitude $A_{\rm ctr}$ for the same family of models.
    The color represents the maximal gravitational scalar in both
    panels.
    \label{fig:thinshellMR}
    }
\end{figure}
\begin{figure} [hbt!]
\includegraphics[width=0.45\textwidth,valign=t]{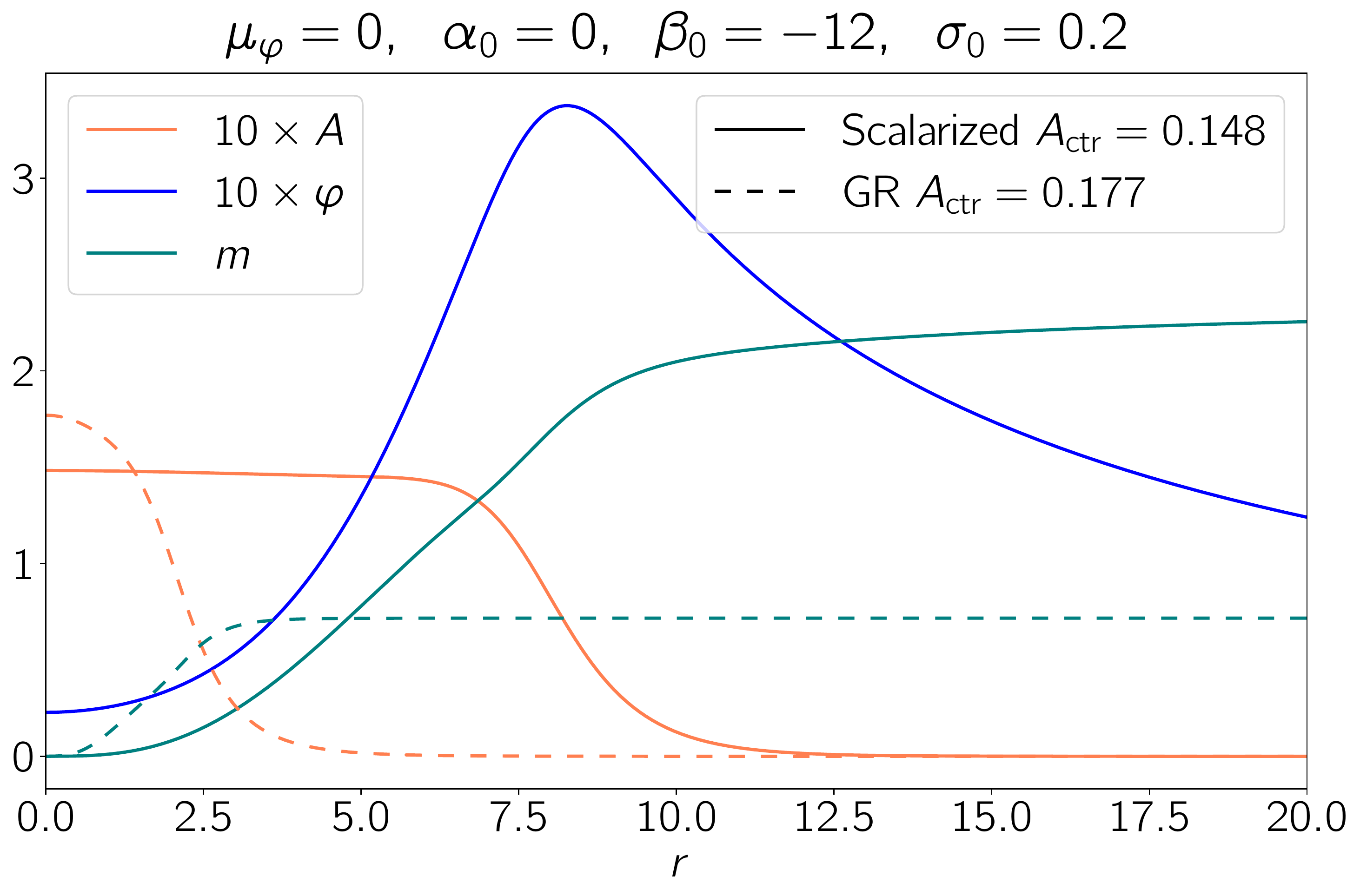}
    \caption{
    Radial profiles of the scalar functions $A(r),~\varphi(r)$ and
    the mass $m(r)$ for a scalarized BS with $A_{\rm ctr}=0.148$
    and a GR model with $A_{\rm ctr}=0.177$, corresponding to the
    minimal frequencies $\omega=0.1476$ and $\omega=0.4283$ and
    marked by circles in Fig.~\ref{fig:thinshellMR}. For presentation
    purposes, the scalar fields $A$ and $\varphi$ have been amplified
    by a factor $10$.
    \label{fig:thinshellAofr}
    }
\end{figure}

We study this feature in more detail for the specific case of
massless ST theory with $\alpha_0=0$ and $\beta_0=-12$ for the
solitonic potential with $\sigma_0=0.2$. Due to different conventions,
our $\sigma_0$ is related to the parameter $\sigma$ used by Collodel
and Doneva by $\sigma=\sqrt{2\pi}\sigma_0$, so that our scenario
corresponds to $\sigma \approx 0.5$ in Ref.~\cite{Collodel:2022jly},
well above the regime where they encounter thin-shell stars. We
likewise find no indication of shell-like structure on the GR branch
of our scenario which yields a minimal frequency of about $\omega=0.428$
in good agreement with Ref.~\cite{Collodel:2022jly}; this can be
seen by comparing the bottom panel of our Fig.~\ref{fig:thinshellMR}
with the left panel of their Fig.~2. Along the scalarized branch,
however, we obtain BS models with much lower frequencies $\omega
<0.2$.  As indicated in the upper panel of Fig.~\ref{fig:thinshellMR},
we also obtain significantly larger radii for these models, although
we emphasize that this is largely due to the reduced fall-off of
$A(r)$ rather than its behaviour in inner regions of the star.  The
radial profiles displayed in Fig.~\ref{fig:thinshellAofr} for the
strongly scalarized BS with minimal $\omega$ (marked by circles in
Fig.~\ref{fig:thinshellMR}), however, exhibit an approximately
constant $A(r)$ out to $r\approx 7$. In contrast, the GR model with
minimal $\omega$ displays a distinctly ``ordinary'' profile where
$A(r)$ rapidly drops away from the origin.
A similar investigation of neighbouring models along the GR and
scalarized branches yields very similar profiles and, as evident
in Fig.~\ref{fig:thinshellMR}, frequency values, and we observe
qualitatively the same behaviour for other values of $\sigma_0$.
Mathematically, we can explain this behaviour by considering
Eq.~(\ref{eq:allr}) for $\partial_r A$ and $\partial_r \theta$.
Thin-shell models can be regarded as a sequence of BSs that approach
a uniformly constant scalar-field solution $A=\mathrm{const}$
extending to larger and larger radii. In Eq.~(\ref{eq:allr}), this
limiting (Minkowski) solution is obtained for a zero right-hand
side of $\partial_r \theta$.  Ignoring the geometric term $-2\theta/r$,
which merely incorporates the $r^2$ behaviour of a 3D volume, the
key mathematical origin of thin-shell stars is a nearly vanishing
\begin{equation}
  \frac{X}{\alpha F}(\alpha^2 V_{,A^2}-\omega^2)\,.
  \label{eq:thinshellterm}
\end{equation}
For strongly scalarized BSs, on the other hand, $F(\varphi)$ is
dominated by the quadratic contribution in the exponential
(\ref{eq:ggbar}) and therefore $F>1$ which reduces the term
(\ref{eq:thinshellterm}) relative to its GR value and hence drives
BSs in the direction of thin-shell models.

\section{Stability} \label{sec:stability}

Depending on the ST parameters $(\mu_{\varphi}, \alpha_0, \beta_0)$
and the potential $V(A)$, there may exist up to seven (or even more
for non-zero $\alpha_0$) BS models with the same mass. This raises
the question which of these models is energetically favored.  We
assess this by using binding energy arguments for a wide range of
BS solutions. More specifically, we define the binding energy
\begin{equation}
    E_{\rm{b}} \defeq M_{\rm{ADM}}-Q_{\rm{BS}},
\end{equation}
where $Q_{\rm{BS}}$ denotes the conserved Noether charge, i.e.~the
number of bosonic particles making up the BS. It is defined as the
spatial integral of the time component of the Noether current,
\begin{equation}
J^{\alpha} = \frac{\rm{i}}{2} g^{\alpha \beta} \left(\psi^* \nabla_{\beta} \psi - \psi \nabla_{\beta} \psi^* \right),
\end{equation}
so that
\begin{equation} \label{eq:noether}
    Q_{\rm{BS}} = \int J^0 \sqrt{-g} \text{d} x^3 =  \int \frac{4 \pi A^2 \omega X} {F \alpha} r^2 \text{d} r.
\end{equation}
For each set of BSs with equal Noether charge \footnote{We  remark
that $Q_{\rm{BS}}$ in Eq.~\ref{eq:noether} is equal in both, Jordan
and Einstein frames, making it a universal diagnostic.}, the model
with the strongest binding energy $E_{\rm{b}}$, i.e.~the smallest
mass $M_{\rm ADM}$, is taken as the stable, energetically favored
configuration.  This does not necessarily imply that the other
models are {\it perturbatively} unstable, but under sufficiently strong
excitations, we expect them to either form a black hole, evaporate
or migrate to a stable BS; we therefore refer to these other
models as unstable in the following discussion.

\begin{figure}
  \includegraphics[width=0.45\textwidth]{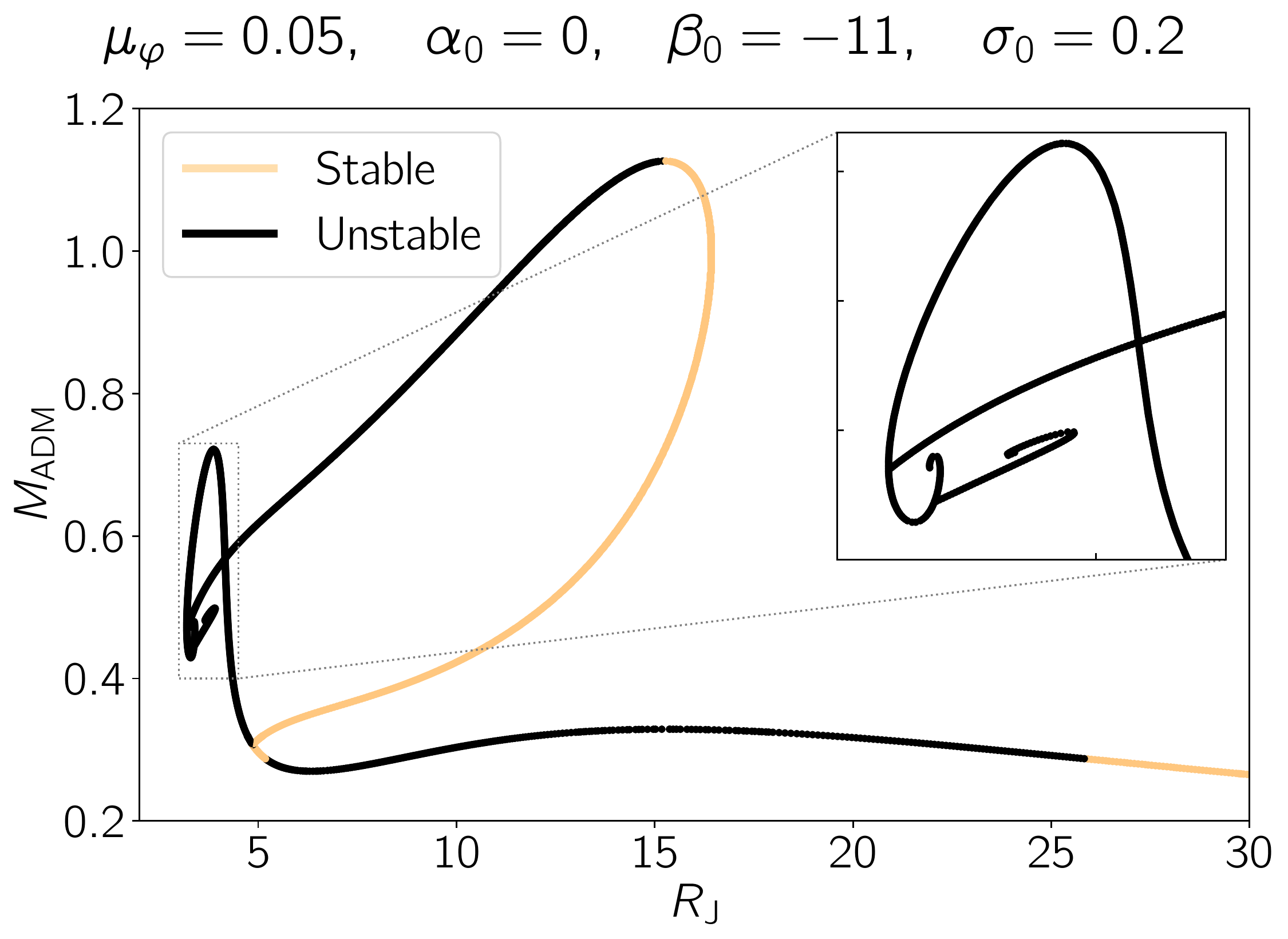}
  \caption{Mass-radius diagram for BS solutions for
  $\mu_{\varphi}=0.05,~\alpha_0=0,~\beta_0=-11$ and a solitonic
  potential with $\sigma_0=0.2$. Stable models are determined by
  identifying among the set of all stars with equal Noether charge
  $Q_{\rm BS}$ the BS with the strongest binding energy $E_{\rm
  b}=M_{\rm ADM}-Q_{\rm BS}$. These stable models are displayed in
  light (copper) color and the remaining unstable stars in dark (black).
  }
  \label{fig:stability_proto}
\end{figure}

We first apply this method to the prototypical example discussed
in Sec.~\ref{sec:prototypical_example}. This set of BS models is
obtained for ST parameters $\mu_{\varphi}=0.05,~\alpha_0=0,~\beta_0=-11$
and a solitonic potential with $\sigma_0=0.2$.  Figure
\ref{fig:stability_proto} shows the same mass-radius diagram as
Fig.~\ref{fig:mvph005_a0m00_bm11_Soli020} but now color codes
solutions in light (copper) for stable or dark (black) for unstable
stars.  We clearly see that the scalarized branch from the low-mass
bifurcation point all the way up to its maximal mass at about
$(R_{\rm J}\approx 15,\,M_{\rm ADM}\approx 1.1)$ contains energetically
favored models. This set of stable BSs is completed by a small piece
of the GR branch around $R_{\rm J}=5$ which contains low-mass but
comparatively compact BSs for whom no scalarized counterparts exist.

\begin{figure*}
  \includegraphics[width=\textwidth]{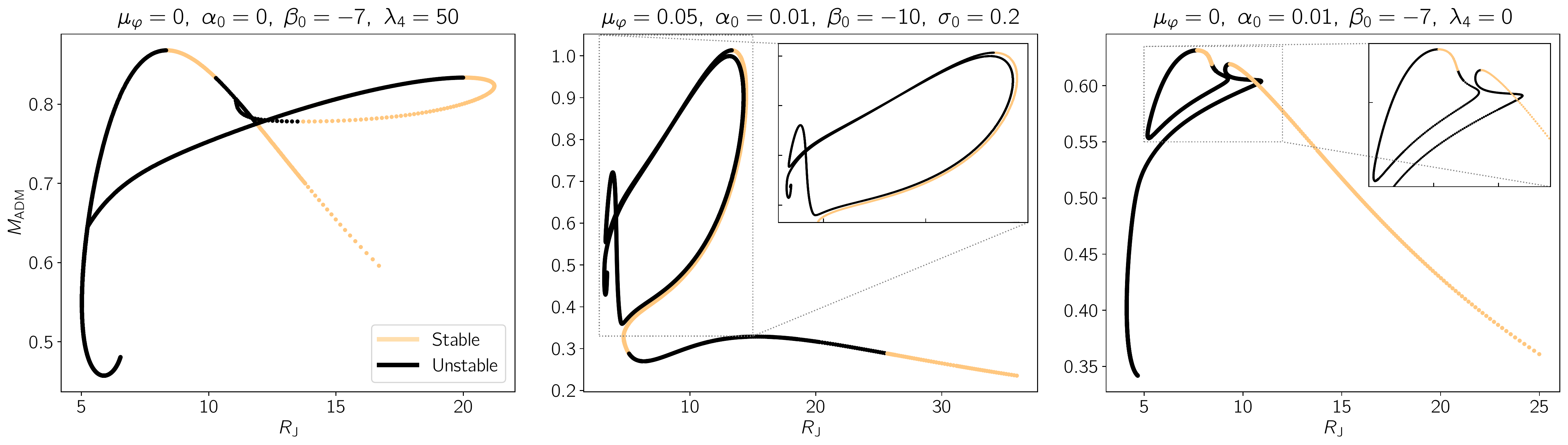}
  \caption{Stability of other BS models considered in this work.
  From left to right: (i) repulsive potential with $\lambda_4 =
  50$, massless $\varphi$, $\alpha_0 = 0$, $\beta_0 = -7$; (ii)
  solitonic potential with $\sigma_0 = 0.2$, massive $\varphi$ with
  $\mu_{\varphi} = 0.05$, $\alpha_0 = 0.01$, $\beta_0 = -10$; (iii)
  mini potential with $\lambda_4 = 0$, massless $\varphi$, $\alpha_0
  = 0$, $\beta_0 = -7$. As in Fig.~\ref{fig:stability_proto},
  light (copper) color denotes stable BS models and dark (black) unstable ones.  Note that
  for $\alpha_0 \ne 0$ (center and right panels) we always have two
  separate scalarized branches, either of which may self-intersect:
  one closed loop and one open branch with two loose ends. For the
  mini BS in the right panel, these two branches almost touch, yet
  remain separate as shown in the inset.}
  \label{fig:stability_others}
\end{figure*}

Repeating the same analysis in Fig.\ref{fig:stability_others} for
other bosonic potentials and ST parameters, unravels a very similar
pattern. For $\alpha_0 = 0$ scalarized BSs are commonly energetically
favored relative to their GR cousins with equal Noether charge. For
non-zero $\alpha_0$, where we commonly encounter multiple scalarized
branches, we observe that strongly scalarized BSs with {\it larger}
radius tend to be stable. We identify some exceptions to these
rules, however: scalarized BSs with relatively small mass or radius
may be energetically disfavored relative to other scalarized solutions
(see the left panel of Fig.~\ref{fig:stability_others}) and, in
some cases, even relative to GR models (see
Fig.~\ref{fig:families_stability} in Appendix \ref{app:BSfamilies},
where we explore a wider range of potentials $V$).  In general, we
observe that, similarly to BS models in GR, stable and unstable
models in ST gravity are separated by the maximum-mass model.

\section{Conclusions}\label{sec:conclusion}
\label{sec:conclusions}

In this work we have performed a systematic study of the structure
of strongly and weakly scalarized boson-star solutions in scalar-tensor
theory of gravity with Damour-Esposito-Far\`ese coupling. For this
purpose we compute spherically symmetric ground-state BS models
using shooting and relaxation algorithms. In order to control the
challenging exponential behaviour of the scalar field solutions,
we perform a comprehensive analysis of their asymptotic behaviour,
identifying incompatibilities that arise from using the flat-field
limit.

Our study suggests that there are many common features in the
dependence of scalarized solutions on the ST parameters $(\mu_{\varphi},
\alpha_0, \beta_0)$ as compared to the neutron star case. In
particular, we observe the following main features:
\begin{enumerate}
    \item The scalarized solutions form additional branches in the
    mass-radius plane. These models can have larger mass and radius
    compared to their GR counterparts, although exceptions to this
    rule exist, depending on the potential and ST parameters.
    \item A smaller gravitational scalar mass $\mu_{\varphi}$ and
    a more negative $\beta_0$ strengthen scalarization, pushing the
    scalarized branch to larger radii; cf.~Fig.~\ref{fig:mvph_MRAvphmax}
    and the right panel of Fig.~\ref{fig:MofR_alphavar}.
    \item A non-zero $\alpha_0$ splits the scalarized branches by
    breaking the $\varphi \rightarrow -\varphi$ degeneracy; cf.~the
    left panel of Fig.~\ref{fig:MofR_alphavar}. For sufficiently
    large values of $\alpha_0$ one of the branches disappears.
    \item Scalarized solutions are often energetically preferred
    (i.e.~stable) when compared to their GR counterparts with the
    same Noether charge. We similarly find that a larger BS radius
    tends to favour stability.
\end{enumerate}

Further to these main features, we summarize the specific traits
inherent to scalarized BSs and their variation over the parameter
space as follows.

{\it Onset of scalarization}: Compared with neutron stars, we require
more negative values of $\beta_0$ to obtain scalarized BSs.  Our
analysis suggests that BSs with repulsive potentials with large
$\lambda_4$ are most susceptible to scalarization in the massless
case: the onset of scalarization happens for $\beta_0 \lesssim
-4.96$, compared to $\beta_0 \lesssim -4.35$ for NSs. Mini and
solitonic BSs are less liable to scalarize and typically require
more negative $\beta_0$. This trend also holds in the massive case
up to $\mu_{\varphi} \approx 0.3$, beyond which solitonic BSs become
more receptive to scalarization. For all potentials, the threshold
for scalarization $\beta_{0,{\rm thr}}$ takes on a quadratic
dependence on the mass term $\mu_{\varphi}$; cf.~Fig.~\ref{fig:multionset}.

\textit{Degree of scalarization}: The repulsive potential for large
$\lambda_4$ results in the strongest scalarization, both in magnitude
and range of BS models.  For solitonic potentials, in contrast, we
observe a non-monotonic dependence of the degree of scalarization
on the self-interaction term $\sigma_0$. In general small $\sigma_0
\approx 0.2$ leads to stronger scalarization albeit over a narrower
range of BS models than larger values $\sigma_0 \gtrsim 1$ and
the mini BS limit $\sigma_0 \to \infty$;
cf.~Fig.~\ref{fig:diffpotentials_beta10_alpha0}.

\textit{Gravitational scalar behaviour}: The potential not only
determines the onset and degree of scalarization, but also the
resulting compactness of the star and the gravitational scalar
behaviour near the origin; cf.~Fig.~\ref{fig:flow_chart}. Generally,
we find that less compact stars support a gravitational scalar whose
radial profile peaks at the origin; this is most common for mini
and repulsive potentials that result in less compact stars. As
attractive self-interaction terms are included in the solitonic
potential, the scalarized solutions become more compact and the
gravitational scalar's extremum moves away from the origin. We also
relate this behavior to the trace of the energy-momentum tensor
$T$, which needs to be negative in the regions of strong scalarization.
We systematically observe that less compact models achieve negative
values of $T$ near the origin; cf. Fig.~\ref{fig:MofR_diff_pots}.
For more compact models, in contrast, $T$ only becomes negative
further away from $r=0$ resulting in $\textit{shell}$-like gravitational
scalar profiles.

\textit{Thin-shell like models}: Strong self-interactions in the
solitonic potential (small $\sigma_0$) paired with strong scalarization
(small gravitational scalar mass $\mu_{\varphi}$ and/or very negative
$\beta_0$) result in BS models akin to the \textit{thin-shell} BS
models found in GR \cite{Collodel:2022jly}. These models correspond
to highly massive BS solutions in their family and have a bosonic
field amplitude $A(r)$ approximating the shape of a Heaviside
function. In GR, typically small values of $\sigma_0 \lesssim 0.15$
are required to obtain such \text{thin-shell} like solutions. In
ST theory, however, we easily find such models already for $\sigma_0
= 0.2$.

Our results demonstrate the rich structure of BSs in ST theory of
gravity and open many opportunities for future work. A natural
extension consists in further exploration of more extreme regions
of the parameter space, as for example a thorough study of the onset
of scalarization for thin-shell models with $\sigma_0 < 0.2$.
Furthermore, dynamical evolutions of our models would provide a
robust picture of their stability complementary to our binding
energy estimates.  Finally, it will be intriguing to see how the
peculiarities identified in the structure of single BSs for different
ST parameters and/or BS potentials affect the gravitational-wave
emission of binary systems including the scalar or breathing
polarization mode.


\begin{acknowledgments}
T.E.~is supported by the Centre for Doctoral Training
(CDT) at the University of Cambridge funded through STFC.
R.R.M.~acknowledges support by the Deutsche Forschungsgemeinschaft
(DFG) under Grants No. 2176/7-1 and No. 406116891 within the Research
Training
Group RTG 2522/1.
This work has been supported by 
STFC Research Grant No. ST/V005669/1
``Probing Fundamental Physics with Gravitational-Wave Observations''.
We acknowledge support by the DiRAC project
ACTP284 from the Cambridge Service for Data Driven Discovery (CSD3)
system at the University of Cambridge
and Cosma7 and 8 of Durham University through STFC capital Grants
No.~ST/P002307/1 and No.~ST/R002452/1, and STFC operations Grant
No.~ST/R00689X/1. 
We also acknowledge support by the DiRAC project grant DiRAC Project
ACTP238 for use of Cosma7 and DiAL3.
This research project was conducted using computational resources
at the Maryland Advanced Research Computing Center (MARCC).
The authors acknowledge the Texas Advanced Computing Center (TACC)
at The University of Texas at Austin and the San Diego Supercomputer
Center for providing HPC resources that have contributed to the
research results reported within this paper through NSF grant
No.~PHY-090003. URLs: \url{http://www.tacc.utexas.edu},
\url{https://www.sdsc.edu/}.
\end{acknowledgments}


\bibliographystyle{apsrev4-1}
\bibliography{bibliography}

\begin{thebibliography}{108}%
\makeatletter
\providecommand \@ifxundefined [1]{%
 \@ifx{#1\undefined}
}%
\providecommand \@ifnum [1]{%
 \ifnum #1\expandafter \@firstoftwo
 \else \expandafter \@secondoftwo
 \fi
}%
\providecommand \@ifx [1]{%
 \ifx #1\expandafter \@firstoftwo
 \else \expandafter \@secondoftwo
 \fi
}%
\providecommand \natexlab [1]{#1}%
\providecommand \enquote  [1]{``#1''}%
\providecommand \bibnamefont  [1]{#1}%
\providecommand \bibfnamefont [1]{#1}%
\providecommand \citenamefont [1]{#1}%
\providecommand \href@noop [0]{\@secondoftwo}%
\providecommand \href [0]{\begingroup \@sanitize@url \@href}%
\providecommand \@href[1]{\@@startlink{#1}\@@href}%
\providecommand \@@href[1]{\endgroup#1\@@endlink}%
\providecommand \@sanitize@url [0]{\catcode `\\12\catcode `\$12\catcode
  `\&12\catcode `\#12\catcode `\^12\catcode `\_12\catcode `\%12\relax}%
\providecommand \@@startlink[1]{}%
\providecommand \@@endlink[0]{}%
\providecommand \url  [0]{\begingroup\@sanitize@url \@url }%
\providecommand \@url [1]{\endgroup\@href {#1}{\urlprefix }}%
\providecommand \urlprefix  [0]{URL }%
\providecommand \Eprint [0]{\href }%
\providecommand \doibase [0]{http://dx.doi.org/}%
\providecommand \selectlanguage [0]{\@gobble}%
\providecommand \bibinfo  [0]{\@secondoftwo}%
\providecommand \bibfield  [0]{\@secondoftwo}%
\providecommand \translation [1]{[#1]}%
\providecommand \BibitemOpen [0]{}%
\providecommand \bibitemStop [0]{}%
\providecommand \bibitemNoStop [0]{.\EOS\space}%
\providecommand \EOS [0]{\spacefactor3000\relax}%
\providecommand \BibitemShut  [1]{\csname bibitem#1\endcsname}%
\let\auto@bib@innerbib\@empty
\bibitem [{\citenamefont {Kaup}(1968)}]{Kaup:1968zz}%
  \BibitemOpen
  \bibfield  {author} {\bibinfo {author} {\bibfnamefont {D.~J.}\ \bibnamefont
  {Kaup}},\ }\href {\doibase 10.1103/PhysRev.172.1331} {\bibfield  {journal}
  {\bibinfo  {journal} {Phys. Rev.}\ }\textbf {\bibinfo {volume} {172}},\
  \bibinfo {pages} {1331} (\bibinfo {year} {1968})}\BibitemShut {NoStop}%
\bibitem [{\citenamefont {G\"uver}\ \emph {et~al.}(2014)\citenamefont
  {G\"uver}, \citenamefont {Erkoca}, \citenamefont {Hall~Reno},\ and\
  \citenamefont {Sarcevic}}]{Guver:2012ba}%
  \BibitemOpen
  \bibfield  {author} {\bibinfo {author} {\bibfnamefont {T.}~\bibnamefont
  {G\"uver}}, \bibinfo {author} {\bibfnamefont {A.~E.}\ \bibnamefont {Erkoca}},
  \bibinfo {author} {\bibfnamefont {M.}~\bibnamefont {Hall~Reno}}, \ and\
  \bibinfo {author} {\bibfnamefont {I.}~\bibnamefont {Sarcevic}},\ }\href
  {\doibase 10.1088/1475-7516/2014/05/013} {\bibfield  {journal} {\bibinfo
  {journal} {JCAP}\ }\textbf {\bibinfo {volume} {05}},\ \bibinfo {pages} {013}
  (\bibinfo {year} {2014})},\ \Eprint {http://arxiv.org/abs/1201.2400}
  {arXiv:1201.2400 [hep-ph]} \BibitemShut {NoStop}%
\bibitem [{\citenamefont {Alcubierre}\ \emph {et~al.}(2002)\citenamefont
  {Alcubierre}, \citenamefont {Guzman}, \citenamefont {Matos}, \citenamefont
  {Nunez}, \citenamefont {Urena-Lopez},\ and\ \citenamefont
  {Wiederhold}}]{Alcubierre:2001ea}%
  \BibitemOpen
  \bibfield  {author} {\bibinfo {author} {\bibfnamefont {M.}~\bibnamefont
  {Alcubierre}}, \bibinfo {author} {\bibfnamefont {F.~S.}\ \bibnamefont
  {Guzman}}, \bibinfo {author} {\bibfnamefont {T.}~\bibnamefont {Matos}},
  \bibinfo {author} {\bibfnamefont {D.}~\bibnamefont {Nunez}}, \bibinfo
  {author} {\bibfnamefont {L.~A.}\ \bibnamefont {Urena-Lopez}}, \ and\ \bibinfo
  {author} {\bibfnamefont {P.}~\bibnamefont {Wiederhold}},\ }\href {\doibase
  10.1088/0264-9381/19/19/314} {\bibfield  {journal} {\bibinfo  {journal}
  {Class. Quant. Grav.}\ }\textbf {\bibinfo {volume} {19}},\ \bibinfo {pages}
  {5017} (\bibinfo {year} {2002})},\ \Eprint
  {http://arxiv.org/abs/gr-qc/0110102} {arXiv:gr-qc/0110102} \BibitemShut
  {NoStop}%
\bibitem [{\citenamefont {Hu}\ \emph {et~al.}(2000)\citenamefont {Hu},
  \citenamefont {Barkana},\ and\ \citenamefont {Gruzinov}}]{Hu:2000ke}%
  \BibitemOpen
  \bibfield  {author} {\bibinfo {author} {\bibfnamefont {W.}~\bibnamefont
  {Hu}}, \bibinfo {author} {\bibfnamefont {R.}~\bibnamefont {Barkana}}, \ and\
  \bibinfo {author} {\bibfnamefont {A.}~\bibnamefont {Gruzinov}},\ }\href
  {\doibase 10.1103/PhysRevLett.85.1158} {\bibfield  {journal} {\bibinfo
  {journal} {Phys. Rev. Lett.}\ }\textbf {\bibinfo {volume} {85}},\ \bibinfo
  {pages} {1158} (\bibinfo {year} {2000})},\ \Eprint
  {http://arxiv.org/abs/astro-ph/0003365} {arXiv:astro-ph/0003365} \BibitemShut
  {NoStop}%
\bibitem [{\citenamefont {Macedo}\ \emph {et~al.}(2013)\citenamefont {Macedo},
  \citenamefont {Pani}, \citenamefont {Cardoso},\ and\ \citenamefont
  {Crispino}}]{Macedo:2013qea}%
  \BibitemOpen
  \bibfield  {author} {\bibinfo {author} {\bibfnamefont {C.~F.~B.}\
  \bibnamefont {Macedo}}, \bibinfo {author} {\bibfnamefont {P.}~\bibnamefont
  {Pani}}, \bibinfo {author} {\bibfnamefont {V.}~\bibnamefont {Cardoso}}, \
  and\ \bibinfo {author} {\bibfnamefont {L.~C.~B.}\ \bibnamefont {Crispino}},\
  }\href {\doibase 10.1088/0004-637X/774/1/48} {\bibfield  {journal} {\bibinfo
  {journal} {Astrophys. J.}\ }\textbf {\bibinfo {volume} {774}},\ \bibinfo
  {pages} {48} (\bibinfo {year} {2013})},\ \Eprint
  {http://arxiv.org/abs/1302.2646} {arXiv:1302.2646 [gr-qc]} \BibitemShut
  {NoStop}%
\bibitem [{\citenamefont {Sennett}\ \emph {et~al.}(2017)\citenamefont
  {Sennett}, \citenamefont {Hinderer}, \citenamefont {Steinhoff}, \citenamefont
  {Buonanno},\ and\ \citenamefont {Ossokine}}]{Sennett:2017etc}%
  \BibitemOpen
  \bibfield  {author} {\bibinfo {author} {\bibfnamefont {N.}~\bibnamefont
  {Sennett}}, \bibinfo {author} {\bibfnamefont {T.}~\bibnamefont {Hinderer}},
  \bibinfo {author} {\bibfnamefont {J.}~\bibnamefont {Steinhoff}}, \bibinfo
  {author} {\bibfnamefont {A.}~\bibnamefont {Buonanno}}, \ and\ \bibinfo
  {author} {\bibfnamefont {S.}~\bibnamefont {Ossokine}},\ }\href {\doibase
  10.1103/PhysRevD.96.024002} {\bibfield  {journal} {\bibinfo  {journal} {Phys.
  Rev. D}\ }\textbf {\bibinfo {volume} {96}},\ \bibinfo {pages} {024002}
  (\bibinfo {year} {2017})},\ \Eprint {http://arxiv.org/abs/1704.08651}
  {arXiv:1704.08651 [gr-qc]} \BibitemShut {NoStop}%
\bibitem [{\citenamefont {Herdeiro}\ \emph {et~al.}(2021)\citenamefont
  {Herdeiro}, \citenamefont {Pombo}, \citenamefont {Radu}, \citenamefont
  {Cunha},\ and\ \citenamefont {Sanchis-Gual}}]{Herdeiro:2021lwl}%
  \BibitemOpen
  \bibfield  {author} {\bibinfo {author} {\bibfnamefont {C.~A.~R.}\
  \bibnamefont {Herdeiro}}, \bibinfo {author} {\bibfnamefont {A.~M.}\
  \bibnamefont {Pombo}}, \bibinfo {author} {\bibfnamefont {E.}~\bibnamefont
  {Radu}}, \bibinfo {author} {\bibfnamefont {P.~V.~P.}\ \bibnamefont {Cunha}},
  \ and\ \bibinfo {author} {\bibfnamefont {N.}~\bibnamefont {Sanchis-Gual}},\
  }\href {\doibase 10.1088/1475-7516/2021/04/051} {\bibfield  {journal}
  {\bibinfo  {journal} {JCAP}\ }\textbf {\bibinfo {volume} {04}},\ \bibinfo
  {pages} {051} (\bibinfo {year} {2021})},\ \Eprint
  {http://arxiv.org/abs/2102.01703} {arXiv:2102.01703 [gr-qc]} \BibitemShut
  {NoStop}%
\bibitem [{\citenamefont {Rosa}\ and\ \citenamefont
  {Rubiera-Garcia}(2022)}]{Rosa:2022tfv}%
  \BibitemOpen
  \bibfield  {author} {\bibinfo {author} {\bibfnamefont {J.~a.~L.}\
  \bibnamefont {Rosa}}\ and\ \bibinfo {author} {\bibfnamefont {D.}~\bibnamefont
  {Rubiera-Garcia}},\ }\href {\doibase 10.1103/PhysRevD.106.084004} {\bibfield
  {journal} {\bibinfo  {journal} {Phys. Rev. D}\ }\textbf {\bibinfo {volume}
  {106}},\ \bibinfo {pages} {084004} (\bibinfo {year} {2022})},\ \Eprint
  {http://arxiv.org/abs/2204.12949} {arXiv:2204.12949 [gr-qc]} \BibitemShut
  {NoStop}%
\bibitem [{\citenamefont {Khlopov}\ \emph {et~al.}(1985)\citenamefont
  {Khlopov}, \citenamefont {Malomed},\ and\ \citenamefont
  {Zeldovich}}]{Khlopov:1985jw}%
  \BibitemOpen
  \bibfield  {author} {\bibinfo {author} {\bibfnamefont {M.}~\bibnamefont
  {Khlopov}}, \bibinfo {author} {\bibfnamefont {B.~A.}\ \bibnamefont
  {Malomed}}, \ and\ \bibinfo {author} {\bibfnamefont {I.~B.}\ \bibnamefont
  {Zeldovich}},\ }\href@noop {} {\bibfield  {journal} {\bibinfo  {journal}
  {Mon. Not. Roy. Astron. Soc.}\ }\textbf {\bibinfo {volume} {215}},\ \bibinfo
  {pages} {575} (\bibinfo {year} {1985})}\BibitemShut {NoStop}%
\bibitem [{\citenamefont {Brihaye}\ \emph {et~al.}(2020)\citenamefont
  {Brihaye}, \citenamefont {Ducobu},\ and\ \citenamefont
  {Hartmann}}]{Brihaye:2020klz}%
  \BibitemOpen
  \bibfield  {author} {\bibinfo {author} {\bibfnamefont {Y.}~\bibnamefont
  {Brihaye}}, \bibinfo {author} {\bibfnamefont {L.}~\bibnamefont {Ducobu}}, \
  and\ \bibinfo {author} {\bibfnamefont {B.}~\bibnamefont {Hartmann}},\ }\href
  {\doibase 10.1016/j.physletb.2020.135906} {\bibfield  {journal} {\bibinfo
  {journal} {Phys. Lett. B}\ }\textbf {\bibinfo {volume} {811}},\ \bibinfo
  {pages} {135906} (\bibinfo {year} {2020})},\ \Eprint
  {http://arxiv.org/abs/2004.08292} {arXiv:2004.08292 [gr-qc]} \BibitemShut
  {NoStop}%
\bibitem [{\citenamefont {Rosa}\ \emph {et~al.}(2023)\citenamefont {Rosa},
  \citenamefont {Macedo},\ and\ \citenamefont {Rubiera-Garcia}}]{Rosa:2023qcv}%
  \BibitemOpen
  \bibfield  {author} {\bibinfo {author} {\bibfnamefont {J.~a.~L.}\
  \bibnamefont {Rosa}}, \bibinfo {author} {\bibfnamefont {C.~F.~B.}\
  \bibnamefont {Macedo}}, \ and\ \bibinfo {author} {\bibfnamefont
  {D.}~\bibnamefont {Rubiera-Garcia}},\ }\href {\doibase
  10.1103/PhysRevD.108.044021} {\bibfield  {journal} {\bibinfo  {journal}
  {Phys. Rev. D}\ }\textbf {\bibinfo {volume} {108}},\ \bibinfo {pages}
  {044021} (\bibinfo {year} {2023})},\ \Eprint
  {http://arxiv.org/abs/2303.17296} {arXiv:2303.17296 [gr-qc]} \BibitemShut
  {NoStop}%
\bibitem [{\citenamefont {Rosa}\ \emph {et~al.}(2022)\citenamefont {Rosa},
  \citenamefont {Garcia}, \citenamefont {Vincent},\ and\ \citenamefont
  {Cardoso}}]{Rosa:2022toh}%
  \BibitemOpen
  \bibfield  {author} {\bibinfo {author} {\bibfnamefont {J.~a.~L.}\
  \bibnamefont {Rosa}}, \bibinfo {author} {\bibfnamefont {P.}~\bibnamefont
  {Garcia}}, \bibinfo {author} {\bibfnamefont {F.~H.}\ \bibnamefont {Vincent}},
  \ and\ \bibinfo {author} {\bibfnamefont {V.}~\bibnamefont {Cardoso}},\ }\href
  {\doibase 10.1103/PhysRevD.106.044031} {\bibfield  {journal} {\bibinfo
  {journal} {Phys. Rev. D}\ }\textbf {\bibinfo {volume} {106}},\ \bibinfo
  {pages} {044031} (\bibinfo {year} {2022})},\ \Eprint
  {http://arxiv.org/abs/2205.11541} {arXiv:2205.11541 [gr-qc]} \BibitemShut
  {NoStop}%
\bibitem [{\citenamefont {Colpi}\ \emph {et~al.}(1986)\citenamefont {Colpi},
  \citenamefont {Shapiro},\ and\ \citenamefont {Wasserman}}]{Colpi:1986ye}%
  \BibitemOpen
  \bibfield  {author} {\bibinfo {author} {\bibfnamefont {M.}~\bibnamefont
  {Colpi}}, \bibinfo {author} {\bibfnamefont {S.~L.}\ \bibnamefont {Shapiro}},
  \ and\ \bibinfo {author} {\bibfnamefont {I.}~\bibnamefont {Wasserman}},\
  }\href {\doibase 10.1103/PhysRevLett.57.2485} {\bibfield  {journal} {\bibinfo
   {journal} {Phys. Rev. Lett.}\ }\textbf {\bibinfo {volume} {57}},\ \bibinfo
  {pages} {2485} (\bibinfo {year} {1986})}\BibitemShut {NoStop}%
\bibitem [{\citenamefont {Seidel}\ and\ \citenamefont
  {Suen}(1990)}]{Seidel:1990jh}%
  \BibitemOpen
  \bibfield  {author} {\bibinfo {author} {\bibfnamefont {E.}~\bibnamefont
  {Seidel}}\ and\ \bibinfo {author} {\bibfnamefont {W.-M.}\ \bibnamefont
  {Suen}},\ }\href {\doibase 10.1103/PhysRevD.42.384} {\bibfield  {journal}
  {\bibinfo  {journal} {Phys. Rev. D}\ }\textbf {\bibinfo {volume} {42}},\
  \bibinfo {pages} {384} (\bibinfo {year} {1990})}\BibitemShut {NoStop}%
\bibitem [{\citenamefont {Kobayashi}\ \emph {et~al.}(1994)\citenamefont
  {Kobayashi}, \citenamefont {Kasai},\ and\ \citenamefont
  {Futamase}}]{Kobayashi:1994qi}%
  \BibitemOpen
  \bibfield  {author} {\bibinfo {author} {\bibfnamefont {Y.}~\bibnamefont
  {Kobayashi}}, \bibinfo {author} {\bibfnamefont {M.}~\bibnamefont {Kasai}}, \
  and\ \bibinfo {author} {\bibfnamefont {T.}~\bibnamefont {Futamase}},\ }\href
  {\doibase 10.1103/PhysRevD.50.7721} {\bibfield  {journal} {\bibinfo
  {journal} {Phys. Rev. D}\ }\textbf {\bibinfo {volume} {50}},\ \bibinfo
  {pages} {7721} (\bibinfo {year} {1994})}\BibitemShut {NoStop}%
\bibitem [{\citenamefont {Ryan}(1997)}]{Ryan:1996nk}%
  \BibitemOpen
  \bibfield  {author} {\bibinfo {author} {\bibfnamefont {F.~D.}\ \bibnamefont
  {Ryan}},\ }\href {\doibase 10.1103/PhysRevD.55.6081} {\bibfield  {journal}
  {\bibinfo  {journal} {Phys. Rev. D}\ }\textbf {\bibinfo {volume} {55}},\
  \bibinfo {pages} {6081} (\bibinfo {year} {1997})}\BibitemShut {NoStop}%
\bibitem [{\citenamefont {Schunck}\ and\ \citenamefont
  {Mielke}(1998)}]{Schunck:1996he}%
  \BibitemOpen
  \bibfield  {author} {\bibinfo {author} {\bibfnamefont {F.~E.}\ \bibnamefont
  {Schunck}}\ and\ \bibinfo {author} {\bibfnamefont {E.~W.}\ \bibnamefont
  {Mielke}},\ }\href {\doibase 10.1016/S0375-9601(98)00778-6} {\bibfield
  {journal} {\bibinfo  {journal} {Phys. Lett.}\ }\textbf {\bibinfo {volume}
  {A249}},\ \bibinfo {pages} {389} (\bibinfo {year} {1998})}\BibitemShut
  {NoStop}%
\bibitem [{\citenamefont {Balakrishna}\ \emph {et~al.}(1998)\citenamefont
  {Balakrishna}, \citenamefont {Seidel},\ and\ \citenamefont
  {Suen}}]{Balakrishna:1997ej}%
  \BibitemOpen
  \bibfield  {author} {\bibinfo {author} {\bibfnamefont {J.}~\bibnamefont
  {Balakrishna}}, \bibinfo {author} {\bibfnamefont {E.}~\bibnamefont {Seidel}},
  \ and\ \bibinfo {author} {\bibfnamefont {W.-M.}\ \bibnamefont {Suen}},\
  }\href {\doibase 10.1103/PhysRevD.58.104004} {\bibfield  {journal} {\bibinfo
  {journal} {Phys. Rev. D}\ }\textbf {\bibinfo {volume} {58}},\ \bibinfo
  {pages} {104004} (\bibinfo {year} {1998})},\ \Eprint
  {http://arxiv.org/abs/gr-qc/9712064} {gr-qc/9712064} \BibitemShut {NoStop}%
\bibitem [{\citenamefont {Yoshida}\ and\ \citenamefont
  {Eriguchi}(1997)}]{Yoshida:1997qf}%
  \BibitemOpen
  \bibfield  {author} {\bibinfo {author} {\bibfnamefont {S.}~\bibnamefont
  {Yoshida}}\ and\ \bibinfo {author} {\bibfnamefont {Y.}~\bibnamefont
  {Eriguchi}},\ }\href {\doibase 10.1103/PhysRevD.56.762} {\bibfield  {journal}
  {\bibinfo  {journal} {Phys. Rev. D}\ }\textbf {\bibinfo {volume} {56}},\
  \bibinfo {pages} {762} (\bibinfo {year} {1997})}\BibitemShut {NoStop}%
\bibitem [{\citenamefont {Schunck}\ and\ \citenamefont
  {Torres}(2000)}]{Schunck:1999zu}%
  \BibitemOpen
  \bibfield  {author} {\bibinfo {author} {\bibfnamefont {F.~E.}\ \bibnamefont
  {Schunck}}\ and\ \bibinfo {author} {\bibfnamefont {D.~F.}\ \bibnamefont
  {Torres}},\ }\href {\doibase 10.1142/S0218271800000608} {\bibfield  {journal}
  {\bibinfo  {journal} {Int. J. Mod. Phys. D}\ }\textbf {\bibinfo {volume}
  {9}},\ \bibinfo {pages} {601} (\bibinfo {year} {2000})},\ \Eprint
  {http://arxiv.org/abs/gr-qc/9911038} {arXiv:gr-qc/9911038} \BibitemShut
  {NoStop}%
\bibitem [{\citenamefont {Schunck}\ and\ \citenamefont
  {Mielke}(2003)}]{Schunck:2003kk}%
  \BibitemOpen
  \bibfield  {author} {\bibinfo {author} {\bibfnamefont {F.~E.}\ \bibnamefont
  {Schunck}}\ and\ \bibinfo {author} {\bibfnamefont {E.~W.}\ \bibnamefont
  {Mielke}},\ }\href {\doibase 10.1088/0264-9381/20/20/201} {\bibfield
  {journal} {\bibinfo  {journal} {Class. Quant. Grav.}\ }\textbf {\bibinfo
  {volume} {20}},\ \bibinfo {pages} {R301} (\bibinfo {year} {2003})},\ \Eprint
  {http://arxiv.org/abs/0801.0307} {arXiv:0801.0307 [astro-ph]} \BibitemShut
  {NoStop}%
\bibitem [{\citenamefont {Balakrishna}\ \emph {et~al.}(2006)\citenamefont
  {Balakrishna}, \citenamefont {Bondarescu}, \citenamefont {Daues},
  \citenamefont {Siddhartha~G.},\ and\ \citenamefont
  {Seidel}}]{Balakrishna:2006ru}%
  \BibitemOpen
  \bibfield  {author} {\bibinfo {author} {\bibfnamefont {J.}~\bibnamefont
  {Balakrishna}}, \bibinfo {author} {\bibfnamefont {R.}~\bibnamefont
  {Bondarescu}}, \bibinfo {author} {\bibfnamefont {G.}~\bibnamefont {Daues}},
  \bibinfo {author} {\bibfnamefont {F.}~\bibnamefont {Siddhartha~G.}}, \ and\
  \bibinfo {author} {\bibfnamefont {E.}~\bibnamefont {Seidel}},\ }\href
  {\doibase 10.1088/0264-9381/23/7/024} {\bibfield  {journal} {\bibinfo
  {journal} {Class. Quant. Grav.}\ }\textbf {\bibinfo {volume} {23}},\ \bibinfo
  {pages} {2631} (\bibinfo {year} {2006})},\ \Eprint
  {http://arxiv.org/abs/gr-qc/0602078} {arXiv:gr-qc/0602078} \BibitemShut
  {NoStop}%
\bibitem [{\citenamefont {Balakrishna}\ \emph {et~al.}(2008)\citenamefont
  {Balakrishna}, \citenamefont {Bondarescu}, \citenamefont {Daues},\ and\
  \citenamefont {Bondarescu}}]{Balakrishna:2007mr}%
  \BibitemOpen
  \bibfield  {author} {\bibinfo {author} {\bibfnamefont {J.}~\bibnamefont
  {Balakrishna}}, \bibinfo {author} {\bibfnamefont {R.}~\bibnamefont
  {Bondarescu}}, \bibinfo {author} {\bibfnamefont {G.}~\bibnamefont {Daues}}, \
  and\ \bibinfo {author} {\bibfnamefont {M.}~\bibnamefont {Bondarescu}},\
  }\href {\doibase 10.1103/PhysRevD.77.024028} {\bibfield  {journal} {\bibinfo
  {journal} {Phys. Rev. D}\ }\textbf {\bibinfo {volume} {77}},\ \bibinfo
  {pages} {024028} (\bibinfo {year} {2008})},\ \Eprint
  {http://arxiv.org/abs/0710.4131} {arXiv:0710.4131 [gr-qc]} \BibitemShut
  {NoStop}%
\bibitem [{\citenamefont {Hartmann}\ \emph {et~al.}(2012)\citenamefont
  {Hartmann}, \citenamefont {Kleihaus}, \citenamefont {Kunz},\ and\
  \citenamefont {Schaffer}}]{Hartmann:2012da}%
  \BibitemOpen
  \bibfield  {author} {\bibinfo {author} {\bibfnamefont {B.}~\bibnamefont
  {Hartmann}}, \bibinfo {author} {\bibfnamefont {B.}~\bibnamefont {Kleihaus}},
  \bibinfo {author} {\bibfnamefont {J.}~\bibnamefont {Kunz}}, \ and\ \bibinfo
  {author} {\bibfnamefont {I.}~\bibnamefont {Schaffer}},\ }\href {\doibase
  10.1016/j.physletb.2012.06.067} {\bibfield  {journal} {\bibinfo  {journal}
  {Phys. Lett. B}\ }\textbf {\bibinfo {volume} {714}},\ \bibinfo {pages} {120}
  (\bibinfo {year} {2012})},\ \Eprint {http://arxiv.org/abs/1205.0899}
  {arXiv:1205.0899 [gr-qc]} \BibitemShut {NoStop}%
\bibitem [{\citenamefont {Siemonsen}\ and\ \citenamefont
  {East}(2021)}]{Siemonsen:2020hcg}%
  \BibitemOpen
  \bibfield  {author} {\bibinfo {author} {\bibfnamefont {N.}~\bibnamefont
  {Siemonsen}}\ and\ \bibinfo {author} {\bibfnamefont {W.~E.}\ \bibnamefont
  {East}},\ }\href {\doibase 10.1103/PhysRevD.103.044022} {\bibfield  {journal}
  {\bibinfo  {journal} {Phys. Rev. D}\ }\textbf {\bibinfo {volume} {103}},\
  \bibinfo {pages} {044022} (\bibinfo {year} {2021})},\ \Eprint
  {http://arxiv.org/abs/2011.08247} {arXiv:2011.08247 [gr-qc]} \BibitemShut
  {NoStop}%
\bibitem [{\citenamefont {Schunck}\ and\ \citenamefont
  {Mielke}(1999)}]{Schunck:1999pm}%
  \BibitemOpen
  \bibfield  {author} {\bibinfo {author} {\bibfnamefont {F.~E.}\ \bibnamefont
  {Schunck}}\ and\ \bibinfo {author} {\bibfnamefont {E.~W.}\ \bibnamefont
  {Mielke}},\ }\href {\doibase 10.1023/A:1026673918588} {\bibfield  {journal}
  {\bibinfo  {journal} {Gen. Rel. Grav.}\ }\textbf {\bibinfo {volume} {31}},\
  \bibinfo {pages} {787} (\bibinfo {year} {1999})}\BibitemShut {NoStop}%
\bibitem [{\citenamefont {Sanchis-Gual}\ \emph {et~al.}(2019)\citenamefont
  {Sanchis-Gual}, \citenamefont {Di~Giovanni}, \citenamefont {Zilh\~ao},
  \citenamefont {Herdeiro}, \citenamefont {Cerd\'a-Dur\'an}, \citenamefont
  {Font},\ and\ \citenamefont {Radu}}]{Sanchis-Gual:2019ljs}%
  \BibitemOpen
  \bibfield  {author} {\bibinfo {author} {\bibfnamefont {N.}~\bibnamefont
  {Sanchis-Gual}}, \bibinfo {author} {\bibfnamefont {F.}~\bibnamefont
  {Di~Giovanni}}, \bibinfo {author} {\bibfnamefont {M.}~\bibnamefont
  {Zilh\~ao}}, \bibinfo {author} {\bibfnamefont {C.}~\bibnamefont {Herdeiro}},
  \bibinfo {author} {\bibfnamefont {P.}~\bibnamefont {Cerd\'a-Dur\'an}},
  \bibinfo {author} {\bibfnamefont {J.~A.}\ \bibnamefont {Font}}, \ and\
  \bibinfo {author} {\bibfnamefont {E.}~\bibnamefont {Radu}},\ }\href {\doibase
  10.1103/PhysRevLett.123.221101} {\bibfield  {journal} {\bibinfo  {journal}
  {Phys. Rev. Lett.}\ }\textbf {\bibinfo {volume} {123}},\ \bibinfo {pages}
  {221101} (\bibinfo {year} {2019})},\ \Eprint
  {http://arxiv.org/abs/1907.12565} {arXiv:1907.12565 [gr-qc]} \BibitemShut
  {NoStop}%
\bibitem [{\citenamefont {Siemonsen}\ and\ \citenamefont
  {East}(2023{\natexlab{a}})}]{Siemonsen:2023hko}%
  \BibitemOpen
  \bibfield  {author} {\bibinfo {author} {\bibfnamefont {N.}~\bibnamefont
  {Siemonsen}}\ and\ \bibinfo {author} {\bibfnamefont {W.~E.}\ \bibnamefont
  {East}},\ }\href {\doibase 10.1103/PhysRevD.107.124018} {\bibfield  {journal}
  {\bibinfo  {journal} {Phys. Rev. D}\ }\textbf {\bibinfo {volume} {107}},\
  \bibinfo {pages} {124018} (\bibinfo {year} {2023}{\natexlab{a}})},\ \Eprint
  {http://arxiv.org/abs/2302.06627} {arXiv:2302.06627 [gr-qc]} \BibitemShut
  {NoStop}%
\bibitem [{\citenamefont {Palenzuela}\ \emph {et~al.}(2007)\citenamefont
  {Palenzuela}, \citenamefont {Olabarrieta}, \citenamefont {Lehner},\ and\
  \citenamefont {Liebling}}]{Palenzuela:2006wp}%
  \BibitemOpen
  \bibfield  {author} {\bibinfo {author} {\bibfnamefont {C.}~\bibnamefont
  {Palenzuela}}, \bibinfo {author} {\bibfnamefont {I.}~\bibnamefont
  {Olabarrieta}}, \bibinfo {author} {\bibfnamefont {L.}~\bibnamefont {Lehner}},
  \ and\ \bibinfo {author} {\bibfnamefont {S.~L.}\ \bibnamefont {Liebling}},\
  }\href {\doibase 10.1103/PhysRevD.75.064005} {\bibfield  {journal} {\bibinfo
  {journal} {Phys. Rev. D}\ }\textbf {\bibinfo {volume} {75}},\ \bibinfo
  {pages} {064005} (\bibinfo {year} {2007})},\ \Eprint
  {http://arxiv.org/abs/gr-qc/0612067} {gr-qc/0612067} \BibitemShut {NoStop}%
\bibitem [{\citenamefont {Palenzuela}\ \emph {et~al.}(2008)\citenamefont
  {Palenzuela}, \citenamefont {Lehner},\ and\ \citenamefont
  {Liebling}}]{Palenzuela:2007dm}%
  \BibitemOpen
  \bibfield  {author} {\bibinfo {author} {\bibfnamefont {C.}~\bibnamefont
  {Palenzuela}}, \bibinfo {author} {\bibfnamefont {L.}~\bibnamefont {Lehner}},
  \ and\ \bibinfo {author} {\bibfnamefont {S.~L.}\ \bibnamefont {Liebling}},\
  }\href {\doibase 10.1103/PhysRevD.77.044036} {\bibfield  {journal} {\bibinfo
  {journal} {Phys. Rev. D}\ }\textbf {\bibinfo {volume} {77}},\ \bibinfo
  {pages} {044036} (\bibinfo {year} {2008})},\ \Eprint
  {http://arxiv.org/abs/arXiv:0706.2435 [gr-qc]} {arXiv:0706.2435 [gr-qc]}
  \BibitemShut {NoStop}%
\bibitem [{\citenamefont {Palenzuela}\ \emph {et~al.}(2017)\citenamefont
  {Palenzuela}, \citenamefont {Pani}, \citenamefont {Bezares}, \citenamefont
  {Cardoso}, \citenamefont {Lehner},\ and\ \citenamefont
  {Liebling}}]{Palenzuela:2017kcg}%
  \BibitemOpen
  \bibfield  {author} {\bibinfo {author} {\bibfnamefont {C.}~\bibnamefont
  {Palenzuela}}, \bibinfo {author} {\bibfnamefont {P.}~\bibnamefont {Pani}},
  \bibinfo {author} {\bibfnamefont {M.}~\bibnamefont {Bezares}}, \bibinfo
  {author} {\bibfnamefont {V.}~\bibnamefont {Cardoso}}, \bibinfo {author}
  {\bibfnamefont {L.}~\bibnamefont {Lehner}}, \ and\ \bibinfo {author}
  {\bibfnamefont {S.}~\bibnamefont {Liebling}},\ }\href {\doibase
  10.1103/PhysRevD.96.104058} {\bibfield  {journal} {\bibinfo  {journal} {Phys.
  Rev. D}\ }\textbf {\bibinfo {volume} {96}},\ \bibinfo {pages} {104058}
  (\bibinfo {year} {2017})},\ \Eprint {http://arxiv.org/abs/arXiv:1710.09432
  [gr-qc]} {arXiv:1710.09432 [gr-qc]} \BibitemShut {NoStop}%
\bibitem [{\citenamefont {Helfer}\ \emph {et~al.}(2022)\citenamefont {Helfer},
  \citenamefont {Sperhake}, \citenamefont {Croft}, \citenamefont {Radia},
  \citenamefont {Ge},\ and\ \citenamefont {Lim}}]{Helfer:2021brt}%
  \BibitemOpen
  \bibfield  {author} {\bibinfo {author} {\bibfnamefont {T.}~\bibnamefont
  {Helfer}}, \bibinfo {author} {\bibfnamefont {U.}~\bibnamefont {Sperhake}},
  \bibinfo {author} {\bibfnamefont {R.}~\bibnamefont {Croft}}, \bibinfo
  {author} {\bibfnamefont {M.}~\bibnamefont {Radia}}, \bibinfo {author}
  {\bibfnamefont {B.-X.}\ \bibnamefont {Ge}}, \ and\ \bibinfo {author}
  {\bibfnamefont {E.~A.}\ \bibnamefont {Lim}},\ }\href {\doibase
  10.1088/1361-6382/ac53b7} {\bibfield  {journal} {\bibinfo  {journal} {Class.
  Quant. Grav.}\ }\textbf {\bibinfo {volume} {39}},\ \bibinfo {pages} {074001}
  (\bibinfo {year} {2022})},\ \Eprint {http://arxiv.org/abs/2108.11995}
  {arXiv:2108.11995 [gr-qc]} \BibitemShut {NoStop}%
\bibitem [{\citenamefont {Sanchis-Gual}\ \emph {et~al.}(2020)\citenamefont
  {Sanchis-Gual}, \citenamefont {Zilh\~ao}, \citenamefont {Herdeiro},
  \citenamefont {Di~Giovanni}, \citenamefont {Font},\ and\ \citenamefont
  {Radu}}]{Sanchis-Gual:2020mzb}%
  \BibitemOpen
  \bibfield  {author} {\bibinfo {author} {\bibfnamefont {N.}~\bibnamefont
  {Sanchis-Gual}}, \bibinfo {author} {\bibfnamefont {M.}~\bibnamefont
  {Zilh\~ao}}, \bibinfo {author} {\bibfnamefont {C.}~\bibnamefont {Herdeiro}},
  \bibinfo {author} {\bibfnamefont {F.}~\bibnamefont {Di~Giovanni}}, \bibinfo
  {author} {\bibfnamefont {J.~A.}\ \bibnamefont {Font}}, \ and\ \bibinfo
  {author} {\bibfnamefont {E.}~\bibnamefont {Radu}},\ }\href {\doibase
  10.1103/PhysRevD.102.101504} {\bibfield  {journal} {\bibinfo  {journal}
  {Phys. Rev. D}\ }\textbf {\bibinfo {volume} {102}},\ \bibinfo {pages}
  {101504} (\bibinfo {year} {2020})},\ \Eprint
  {http://arxiv.org/abs/2007.11584} {arXiv:2007.11584 [gr-qc]} \BibitemShut
  {NoStop}%
\bibitem [{\citenamefont {Bezares}\ \emph {et~al.}(2022)\citenamefont
  {Bezares}, \citenamefont {Bo\v{s}kovi\'c}, \citenamefont {Liebling},
  \citenamefont {Palenzuela}, \citenamefont {Pani},\ and\ \citenamefont
  {Barausse}}]{Bezares:2022obu}%
  \BibitemOpen
  \bibfield  {author} {\bibinfo {author} {\bibfnamefont {M.}~\bibnamefont
  {Bezares}}, \bibinfo {author} {\bibfnamefont {M.}~\bibnamefont
  {Bo\v{s}kovi\'c}}, \bibinfo {author} {\bibfnamefont {S.}~\bibnamefont
  {Liebling}}, \bibinfo {author} {\bibfnamefont {C.}~\bibnamefont
  {Palenzuela}}, \bibinfo {author} {\bibfnamefont {P.}~\bibnamefont {Pani}}, \
  and\ \bibinfo {author} {\bibfnamefont {E.}~\bibnamefont {Barausse}},\ }\href
  {\doibase 10.1103/PhysRevD.105.064067} {\bibfield  {journal} {\bibinfo
  {journal} {Phys. Rev. D}\ }\textbf {\bibinfo {volume} {105}},\ \bibinfo
  {pages} {064067} (\bibinfo {year} {2022})},\ \Eprint
  {http://arxiv.org/abs/2201.06113} {arXiv:2201.06113 [gr-qc]} \BibitemShut
  {NoStop}%
\bibitem [{\citenamefont {Cardoso}\ \emph {et~al.}(2022)\citenamefont
  {Cardoso}, \citenamefont {Ikeda}, \citenamefont {Zhong},\ and\ \citenamefont
  {Zilh\~ao}}]{Cardoso:2022vpj}%
  \BibitemOpen
  \bibfield  {author} {\bibinfo {author} {\bibfnamefont {V.}~\bibnamefont
  {Cardoso}}, \bibinfo {author} {\bibfnamefont {T.}~\bibnamefont {Ikeda}},
  \bibinfo {author} {\bibfnamefont {Z.}~\bibnamefont {Zhong}}, \ and\ \bibinfo
  {author} {\bibfnamefont {M.}~\bibnamefont {Zilh\~ao}},\ }\href {\doibase
  10.1103/PhysRevD.106.044030} {\bibfield  {journal} {\bibinfo  {journal}
  {Phys. Rev. D}\ }\textbf {\bibinfo {volume} {106}},\ \bibinfo {pages}
  {044030} (\bibinfo {year} {2022})},\ \Eprint
  {http://arxiv.org/abs/2206.00021} {arXiv:2206.00021 [gr-qc]} \BibitemShut
  {NoStop}%
\bibitem [{\citenamefont {Croft}\ \emph {et~al.}(2023)\citenamefont {Croft},
  \citenamefont {Helfer}, \citenamefont {Ge}, \citenamefont {Radia},
  \citenamefont {Evstafyeva}, \citenamefont {Lim}, \citenamefont {Sperhake},\
  and\ \citenamefont {Clough}}]{Croft:2022bxq}%
  \BibitemOpen
  \bibfield  {author} {\bibinfo {author} {\bibfnamefont {R.}~\bibnamefont
  {Croft}}, \bibinfo {author} {\bibfnamefont {T.}~\bibnamefont {Helfer}},
  \bibinfo {author} {\bibfnamefont {B.-X.}\ \bibnamefont {Ge}}, \bibinfo
  {author} {\bibfnamefont {M.}~\bibnamefont {Radia}}, \bibinfo {author}
  {\bibfnamefont {T.}~\bibnamefont {Evstafyeva}}, \bibinfo {author}
  {\bibfnamefont {E.~A.}\ \bibnamefont {Lim}}, \bibinfo {author} {\bibfnamefont
  {U.}~\bibnamefont {Sperhake}}, \ and\ \bibinfo {author} {\bibfnamefont
  {K.}~\bibnamefont {Clough}},\ }\href {\doibase 10.1088/1361-6382/acace4}
  {\bibfield  {journal} {\bibinfo  {journal} {Class. Quant. Grav.}\ }\textbf
  {\bibinfo {volume} {40}},\ \bibinfo {pages} {065001} (\bibinfo {year}
  {2023})},\ \Eprint {http://arxiv.org/abs/2207.05690} {arXiv:2207.05690
  [gr-qc]} \BibitemShut {NoStop}%
\bibitem [{\citenamefont {Sanchis-Gual}\ \emph {et~al.}(2022)\citenamefont
  {Sanchis-Gual}, \citenamefont {Zilh\~ao},\ and\ \citenamefont
  {Cardoso}}]{Sanchis-Gual:2022zsr}%
  \BibitemOpen
  \bibfield  {author} {\bibinfo {author} {\bibfnamefont {N.}~\bibnamefont
  {Sanchis-Gual}}, \bibinfo {author} {\bibfnamefont {M.}~\bibnamefont
  {Zilh\~ao}}, \ and\ \bibinfo {author} {\bibfnamefont {V.}~\bibnamefont
  {Cardoso}},\ }\href {\doibase 10.1103/PhysRevD.106.064034} {\bibfield
  {journal} {\bibinfo  {journal} {Phys. Rev. D}\ }\textbf {\bibinfo {volume}
  {106}},\ \bibinfo {pages} {064034} (\bibinfo {year} {2022})},\ \Eprint
  {http://arxiv.org/abs/2207.05494} {arXiv:2207.05494 [gr-qc]} \BibitemShut
  {NoStop}%
\bibitem [{\citenamefont {Evstafyeva}\ \emph {et~al.}(2023)\citenamefont
  {Evstafyeva}, \citenamefont {Sperhake}, \citenamefont {Helfer}, \citenamefont
  {Croft}, \citenamefont {Radia}, \citenamefont {Ge},\ and\ \citenamefont
  {Lim}}]{Evstafyeva:2022bpr}%
  \BibitemOpen
  \bibfield  {author} {\bibinfo {author} {\bibfnamefont {T.}~\bibnamefont
  {Evstafyeva}}, \bibinfo {author} {\bibfnamefont {U.}~\bibnamefont
  {Sperhake}}, \bibinfo {author} {\bibfnamefont {T.}~\bibnamefont {Helfer}},
  \bibinfo {author} {\bibfnamefont {R.}~\bibnamefont {Croft}}, \bibinfo
  {author} {\bibfnamefont {M.}~\bibnamefont {Radia}}, \bibinfo {author}
  {\bibfnamefont {B.-X.}\ \bibnamefont {Ge}}, \ and\ \bibinfo {author}
  {\bibfnamefont {E.~A.}\ \bibnamefont {Lim}},\ }\href {\doibase
  10.1088/1361-6382/acc2a8} {\bibfield  {journal} {\bibinfo  {journal} {Class.
  Quant. Grav.}\ }\textbf {\bibinfo {volume} {40}},\ \bibinfo {pages} {085009}
  (\bibinfo {year} {2023})},\ \Eprint {http://arxiv.org/abs/2212.08023}
  {arXiv:2212.08023 [gr-qc]} \BibitemShut {NoStop}%
\bibitem [{\citenamefont {Siemonsen}\ and\ \citenamefont
  {East}(2023{\natexlab{b}})}]{Siemonsen:2023age}%
  \BibitemOpen
  \bibfield  {author} {\bibinfo {author} {\bibfnamefont {N.}~\bibnamefont
  {Siemonsen}}\ and\ \bibinfo {author} {\bibfnamefont {W.~E.}\ \bibnamefont
  {East}},\ }\href@noop {} {\  (\bibinfo {year} {2023}{\natexlab{b}})},\
  \Eprint {http://arxiv.org/abs/2306.17265} {arXiv:2306.17265 [gr-qc]}
  \BibitemShut {NoStop}%
\bibitem [{\citenamefont {Jetzer}(1992)}]{Jetzer:1991jr}%
  \BibitemOpen
  \bibfield  {author} {\bibinfo {author} {\bibfnamefont {P.}~\bibnamefont
  {Jetzer}},\ }\href {\doibase 10.1016/0370-1573(92)90123-H} {\bibfield
  {journal} {\bibinfo  {journal} {Phys. Rept.}\ }\textbf {\bibinfo {volume}
  {220}},\ \bibinfo {pages} {163} (\bibinfo {year} {1992})}\BibitemShut
  {NoStop}%
\bibitem [{\citenamefont {Mielke}\ and\ \citenamefont
  {Schunck}(1997)}]{Mielke:1997re}%
  \BibitemOpen
  \bibfield  {author} {\bibinfo {author} {\bibfnamefont {E.~W.}\ \bibnamefont
  {Mielke}}\ and\ \bibinfo {author} {\bibfnamefont {F.~E.}\ \bibnamefont
  {Schunck}},\ }in\ \href@noop {} {\emph {\bibinfo {booktitle} {{Recent
  developments in theoretical and experimental general relativity, gravitation,
  and relativistic field theories. Proceedings, 8th Marcel Grossmann meeting,
  MG8, Jerusalem, Israel, June+ 22-27, 1997. Pts. A, B}}}}\ (\bibinfo {year}
  {1997})\ pp.\ \bibinfo {pages} {1607--1626},\ \Eprint
  {http://arxiv.org/abs/gr-qc/9801063} {gr-qc/9801063} \BibitemShut {NoStop}%
\bibitem [{\citenamefont {Liebling}\ and\ \citenamefont
  {Palenzuela}(2012)}]{Liebling:2012fv}%
  \BibitemOpen
  \bibfield  {author} {\bibinfo {author} {\bibfnamefont {S.~L.}\ \bibnamefont
  {Liebling}}\ and\ \bibinfo {author} {\bibfnamefont {C.}~\bibnamefont
  {Palenzuela}},\ }\href {\doibase 10.12942/lrr-2012-6} {\bibfield  {journal}
  {\bibinfo  {journal} {Living Rev. Rel.}\ }\textbf {\bibinfo {volume} {15}},\
  \bibinfo {pages} {6} (\bibinfo {year} {2012})},\ \Eprint
  {http://arxiv.org/abs/arXiv:1202.5809 [gr-qc]} {arXiv:1202.5809 [gr-qc]}
  \BibitemShut {NoStop}%
\bibitem [{\citenamefont {Visinelli}(2021)}]{Visinelli:2021uve}%
  \BibitemOpen
  \bibfield  {author} {\bibinfo {author} {\bibfnamefont {L.}~\bibnamefont
  {Visinelli}},\ }\href@noop {} {\  (\bibinfo {year} {2021})},\ \Eprint
  {http://arxiv.org/abs/2109.05481} {arXiv:2109.05481 [gr-qc]} \BibitemShut
  {NoStop}%
\bibitem [{\citenamefont {Torres}(1997)}]{Torres:1997np}%
  \BibitemOpen
  \bibfield  {author} {\bibinfo {author} {\bibfnamefont {D.~F.}\ \bibnamefont
  {Torres}},\ }\href {\doibase 10.1103/PhysRevD.56.3478} {\bibfield  {journal}
  {\bibinfo  {journal} {Phys. Rev. D}\ }\textbf {\bibinfo {volume} {56}},\
  \bibinfo {pages} {3478} (\bibinfo {year} {1997})},\ \Eprint
  {http://arxiv.org/abs/gr-qc/9704006} {arXiv:gr-qc/9704006} \BibitemShut
  {NoStop}%
\bibitem [{\citenamefont {Balakrishna}\ and\ \citenamefont
  {Shinkai}(1998)}]{Balakrishna:1997ek}%
  \BibitemOpen
  \bibfield  {author} {\bibinfo {author} {\bibfnamefont {J.}~\bibnamefont
  {Balakrishna}}\ and\ \bibinfo {author} {\bibfnamefont {H.-a.}\ \bibnamefont
  {Shinkai}},\ }\href {\doibase 10.1103/PhysRevD.58.044016} {\bibfield
  {journal} {\bibinfo  {journal} {Phys. Rev. D}\ }\textbf {\bibinfo {volume}
  {58}},\ \bibinfo {pages} {044016} (\bibinfo {year} {1998})},\ \Eprint
  {http://arxiv.org/abs/gr-qc/9712065} {arXiv:gr-qc/9712065} \BibitemShut
  {NoStop}%
\bibitem [{\citenamefont {Comer}\ and\ \citenamefont
  {Shinkai}(1998)}]{Comer:1997ns}%
  \BibitemOpen
  \bibfield  {author} {\bibinfo {author} {\bibfnamefont {G.~L.}\ \bibnamefont
  {Comer}}\ and\ \bibinfo {author} {\bibfnamefont {H.-a.}\ \bibnamefont
  {Shinkai}},\ }\href {\doibase 10.1088/0264-9381/15/3/016} {\bibfield
  {journal} {\bibinfo  {journal} {Class. Quant. Grav.}\ }\textbf {\bibinfo
  {volume} {15}},\ \bibinfo {pages} {669} (\bibinfo {year} {1998})},\ \Eprint
  {http://arxiv.org/abs/gr-qc/9708071} {arXiv:gr-qc/9708071} \BibitemShut
  {NoStop}%
\bibitem [{\citenamefont {Mas\'o-Ferrando}\ \emph {et~al.}(2021)\citenamefont
  {Mas\'o-Ferrando}, \citenamefont {Sanchis-Gual}, \citenamefont {Font},\ and\
  \citenamefont {Olmo}}]{Maso-Ferrando:2021ngp}%
  \BibitemOpen
  \bibfield  {author} {\bibinfo {author} {\bibfnamefont {A.}~\bibnamefont
  {Mas\'o-Ferrando}}, \bibinfo {author} {\bibfnamefont {N.}~\bibnamefont
  {Sanchis-Gual}}, \bibinfo {author} {\bibfnamefont {J.~A.}\ \bibnamefont
  {Font}}, \ and\ \bibinfo {author} {\bibfnamefont {G.~J.}\ \bibnamefont
  {Olmo}},\ }\href {\doibase 10.1088/1361-6382/ac1fd0} {\bibfield  {journal}
  {\bibinfo  {journal} {Class. Quant. Grav.}\ }\textbf {\bibinfo {volume}
  {38}},\ \bibinfo {pages} {194003} (\bibinfo {year} {2021})},\ \Eprint
  {http://arxiv.org/abs/2103.15705} {arXiv:2103.15705 [gr-qc]} \BibitemShut
  {NoStop}%
\bibitem [{\citenamefont {Mas\'o-Ferrando}\ \emph {et~al.}(2023)\citenamefont
  {Mas\'o-Ferrando}, \citenamefont {Sanchis-Gual}, \citenamefont {Font},\ and\
  \citenamefont {Olmo}}]{Maso-Ferrando:2023nju}%
  \BibitemOpen
  \bibfield  {author} {\bibinfo {author} {\bibfnamefont {A.}~\bibnamefont
  {Mas\'o-Ferrando}}, \bibinfo {author} {\bibfnamefont {N.}~\bibnamefont
  {Sanchis-Gual}}, \bibinfo {author} {\bibfnamefont {J.~A.}\ \bibnamefont
  {Font}}, \ and\ \bibinfo {author} {\bibfnamefont {G.~J.}\ \bibnamefont
  {Olmo}},\ }\href {\doibase 10.1088/1475-7516/2023/06/028} {\bibfield
  {journal} {\bibinfo  {journal} {JCAP}\ }\textbf {\bibinfo {volume} {06}},\
  \bibinfo {pages} {028} (\bibinfo {year} {2023})},\ \Eprint
  {http://arxiv.org/abs/2304.12018} {arXiv:2304.12018 [gr-qc]} \BibitemShut
  {NoStop}%
\bibitem [{\citenamefont {Iliji\'c}\ and\ \citenamefont
  {Sossich}(2020)}]{Ilijic:2020vzu}%
  \BibitemOpen
  \bibfield  {author} {\bibinfo {author} {\bibfnamefont {S.}~\bibnamefont
  {Iliji\'c}}\ and\ \bibinfo {author} {\bibfnamefont {M.}~\bibnamefont
  {Sossich}},\ }\href {\doibase 10.1103/PhysRevD.102.084019} {\bibfield
  {journal} {\bibinfo  {journal} {Phys. Rev. D}\ }\textbf {\bibinfo {volume}
  {102}},\ \bibinfo {pages} {084019} (\bibinfo {year} {2020})},\ \Eprint
  {http://arxiv.org/abs/2007.12451} {arXiv:2007.12451 [gr-qc]} \BibitemShut
  {NoStop}%
\bibitem [{\citenamefont {Will}(1993)}]{Will:1993}%
  \BibitemOpen
  \bibfield  {author} {\bibinfo {author} {\bibfnamefont {C.~M.}\ \bibnamefont
  {Will}},\ }\href@noop {} {\emph {\bibinfo {title} {{Theory and Experiment in
  Gravitational Physics}}}}\ (\bibinfo  {publisher} {Cambridge University
  Press, Cambridge},\ \bibinfo {year} {1993})\BibitemShut {NoStop}%
\bibitem [{\citenamefont {Taylor}\ and\ \citenamefont
  {Veneziano}(1988)}]{Taylor:1988nw}%
  \BibitemOpen
  \bibfield  {author} {\bibinfo {author} {\bibfnamefont {T.~R.}\ \bibnamefont
  {Taylor}}\ and\ \bibinfo {author} {\bibfnamefont {G.}~\bibnamefont
  {Veneziano}},\ }\href {\doibase 10.1016/0370-2693(88)91290-7} {\bibfield
  {journal} {\bibinfo  {journal} {Phys. Lett. B}\ }\textbf {\bibinfo {volume}
  {213}},\ \bibinfo {pages} {450} (\bibinfo {year} {1988})}\BibitemShut
  {NoStop}%
\bibitem [{\citenamefont {{Maeda}}(1988)}]{1988MPLA....3..243M}%
  \BibitemOpen
  \bibfield  {author} {\bibinfo {author} {\bibfnamefont {K.-I.}\ \bibnamefont
  {{Maeda}}},\ }\href {\doibase 10.1142/S0217732388000295} {\bibfield
  {journal} {\bibinfo  {journal} {Modern Physics Letters A}\ }\textbf {\bibinfo
  {volume} {3}},\ \bibinfo {pages} {243} (\bibinfo {year} {1988})}\BibitemShut
  {NoStop}%
\bibitem [{\citenamefont {Damour}\ and\ \citenamefont
  {Polyakov}(1994)}]{Damour:1994zq}%
  \BibitemOpen
  \bibfield  {author} {\bibinfo {author} {\bibfnamefont {T.}~\bibnamefont
  {Damour}}\ and\ \bibinfo {author} {\bibfnamefont {A.~M.}\ \bibnamefont
  {Polyakov}},\ }\href {\doibase 10.1016/0550-3213(94)90143-0} {\bibfield
  {journal} {\bibinfo  {journal} {Nucl. Phys. B}\ }\textbf {\bibinfo {volume}
  {423}},\ \bibinfo {pages} {532} (\bibinfo {year} {1994})},\ \Eprint
  {http://arxiv.org/abs/hep-th/9401069} {arXiv:hep-th/9401069} \BibitemShut
  {NoStop}%
\bibitem [{\citenamefont {Riazuelo}\ and\ \citenamefont
  {Uzan}(2002)}]{Riazuelo:2001mg}%
  \BibitemOpen
  \bibfield  {author} {\bibinfo {author} {\bibfnamefont {A.}~\bibnamefont
  {Riazuelo}}\ and\ \bibinfo {author} {\bibfnamefont {J.-P.}\ \bibnamefont
  {Uzan}},\ }\href {\doibase 10.1103/PhysRevD.66.023525} {\bibfield  {journal}
  {\bibinfo  {journal} {Phys. Rev. D}\ }\textbf {\bibinfo {volume} {66}},\
  \bibinfo {pages} {023525} (\bibinfo {year} {2002})},\ \Eprint
  {http://arxiv.org/abs/astro-ph/0107386} {arXiv:astro-ph/0107386} \BibitemShut
  {NoStop}%
\bibitem [{\citenamefont {Schimd}\ \emph {et~al.}(2005)\citenamefont {Schimd},
  \citenamefont {Uzan},\ and\ \citenamefont {Riazuelo}}]{Schimd:2004nq}%
  \BibitemOpen
  \bibfield  {author} {\bibinfo {author} {\bibfnamefont {C.}~\bibnamefont
  {Schimd}}, \bibinfo {author} {\bibfnamefont {J.-P.}\ \bibnamefont {Uzan}}, \
  and\ \bibinfo {author} {\bibfnamefont {A.}~\bibnamefont {Riazuelo}},\ }\href
  {\doibase 10.1103/PhysRevD.71.083512} {\bibfield  {journal} {\bibinfo
  {journal} {Phys. Rev. D}\ }\textbf {\bibinfo {volume} {71}},\ \bibinfo
  {pages} {083512} (\bibinfo {year} {2005})},\ \Eprint
  {http://arxiv.org/abs/astro-ph/0412120} {arXiv:astro-ph/0412120} \BibitemShut
  {NoStop}%
\bibitem [{\citenamefont {Berti}\ \emph {et~al.}(2015)\citenamefont {Berti},
  \citenamefont {Barausse}, \citenamefont {Cardoso}, \citenamefont {Gualtieri},
  \citenamefont {Pani}, \citenamefont {Sperhake}, \citenamefont {Stein},
  \citenamefont {Wex}, \citenamefont {Yagi}, \citenamefont {Baker},\ and\
  \citenamefont {et~al.}}]{6d97dcf5be1a4f88aed4ff8b4d1635b7}%
  \BibitemOpen
  \bibfield  {author} {\bibinfo {author} {\bibfnamefont {E.}~\bibnamefont
  {Berti}}, \bibinfo {author} {\bibfnamefont {E.}~\bibnamefont {Barausse}},
  \bibinfo {author} {\bibfnamefont {V.}~\bibnamefont {Cardoso}}, \bibinfo
  {author} {\bibfnamefont {L.}~\bibnamefont {Gualtieri}}, \bibinfo {author}
  {\bibfnamefont {P.}~\bibnamefont {Pani}}, \bibinfo {author} {\bibfnamefont
  {U.}~\bibnamefont {Sperhake}}, \bibinfo {author} {\bibfnamefont
  {L.}~\bibnamefont {Stein}}, \bibinfo {author} {\bibfnamefont
  {N.}~\bibnamefont {Wex}}, \bibinfo {author} {\bibfnamefont {K.}~\bibnamefont
  {Yagi}}, \bibinfo {author} {\bibfnamefont {T.}~\bibnamefont {Baker}}, \ and\
  \bibinfo {author} {\bibnamefont {et~al.}},\ }\href {\doibase
  10.1088/0264-9381/32/24/243001} {\bibfield  {journal} {\bibinfo  {journal}
  {Classical and Quantum Gravity}\ }\textbf {\bibinfo {volume} {32}} (\bibinfo
  {year} {2015}),\ 10.1088/0264-9381/32/24/243001}\BibitemShut {NoStop}%
\bibitem [{\citenamefont {Doneva}\ \emph {et~al.}(2022)\citenamefont {Doneva},
  \citenamefont {Ramazano\u{g}lu}, \citenamefont {Silva}, \citenamefont
  {Sotiriou},\ and\ \citenamefont {Yazadjiev}}]{Doneva:2022ewd}%
  \BibitemOpen
  \bibfield  {author} {\bibinfo {author} {\bibfnamefont {D.~D.}\ \bibnamefont
  {Doneva}}, \bibinfo {author} {\bibfnamefont {F.~M.}\ \bibnamefont
  {Ramazano\u{g}lu}}, \bibinfo {author} {\bibfnamefont {H.~O.}\ \bibnamefont
  {Silva}}, \bibinfo {author} {\bibfnamefont {T.~P.}\ \bibnamefont {Sotiriou}},
  \ and\ \bibinfo {author} {\bibfnamefont {S.~S.}\ \bibnamefont {Yazadjiev}},\
  }\href@noop {} {\  (\bibinfo {year} {2022})},\ \Eprint
  {http://arxiv.org/abs/2211.01766} {arXiv:2211.01766 [gr-qc]} \BibitemShut
  {NoStop}%
\bibitem [{\citenamefont {Damour}\ and\ \citenamefont
  {Esposito-Far{\'e}se}(1992)}]{Damour:1992we}%
  \BibitemOpen
  \bibfield  {author} {\bibinfo {author} {\bibfnamefont {T.}~\bibnamefont
  {Damour}}\ and\ \bibinfo {author} {\bibfnamefont {G.}~\bibnamefont
  {Esposito-Far{\'e}se}},\ }\href {\doibase 10.1088/0264-9381/9/9/015}
  {\bibfield  {journal} {\bibinfo  {journal} {Class. Quant. Grav.}\ }\textbf
  {\bibinfo {volume} {9}},\ \bibinfo {pages} {2093} (\bibinfo {year}
  {1992})}\BibitemShut {NoStop}%
\bibitem [{\citenamefont {Damour}\ and\ \citenamefont
  {Esposito-Far{\`e}se}(1993)}]{Damour:1993hw}%
  \BibitemOpen
  \bibfield  {author} {\bibinfo {author} {\bibfnamefont {T.}~\bibnamefont
  {Damour}}\ and\ \bibinfo {author} {\bibfnamefont {G.}~\bibnamefont
  {Esposito-Far{\`e}se}},\ }\href {\doibase 10.1103/PhysRevLett.70.2220}
  {\bibfield  {journal} {\bibinfo  {journal} {Phys. Rev. Lett.}\ }\textbf
  {\bibinfo {volume} {70}},\ \bibinfo {pages} {2220} (\bibinfo {year}
  {1993})}\BibitemShut {NoStop}%
\bibitem [{\citenamefont {Damour}\ and\ \citenamefont
  {Esposito-Far{\'e}se}(1996)}]{Damour:1996ke}%
  \BibitemOpen
  \bibfield  {author} {\bibinfo {author} {\bibfnamefont {T.}~\bibnamefont
  {Damour}}\ and\ \bibinfo {author} {\bibfnamefont {G.}~\bibnamefont
  {Esposito-Far{\'e}se}},\ }\href {\doibase 10.1103/PhysRevD.54.1474}
  {\bibfield  {journal} {\bibinfo  {journal} {Phys. Rev. D}\ }\textbf {\bibinfo
  {volume} {54}},\ \bibinfo {pages} {1474} (\bibinfo {year} {1996})},\ \Eprint
  {http://arxiv.org/abs/gr-qc/9602056} {arXiv:gr-qc/9602056} \BibitemShut
  {NoStop}%
\bibitem [{\citenamefont {{Bertotti}}\ \emph {et~al.}(2003)\citenamefont
  {{Bertotti}}, \citenamefont {{Iess}},\ and\ \citenamefont
  {{Tortora}}}]{2003Natur.425..374B}%
  \BibitemOpen
  \bibfield  {author} {\bibinfo {author} {\bibfnamefont {B.}~\bibnamefont
  {{Bertotti}}}, \bibinfo {author} {\bibfnamefont {L.}~\bibnamefont {{Iess}}},
  \ and\ \bibinfo {author} {\bibfnamefont {P.}~\bibnamefont {{Tortora}}},\
  }\href {\doibase 10.1038/nature01997} {\bibfield  {journal} {\bibinfo
  {journal} {\nat}\ }\textbf {\bibinfo {volume} {425}},\ \bibinfo {pages} {374}
  (\bibinfo {year} {2003})}\BibitemShut {NoStop}%
\bibitem [{\citenamefont {Freire}\ \emph {et~al.}(2012)\citenamefont {Freire},
  \citenamefont {Wex}, \citenamefont {Esposito-Farese}, \citenamefont
  {Verbiest}, \citenamefont {Bailes}, \citenamefont {Jacoby}, \citenamefont
  {Kramer}, \citenamefont {Stairs}, \citenamefont {Antoniadis},\ and\
  \citenamefont {Janssen}}]{Freire:2012mg}%
  \BibitemOpen
  \bibfield  {author} {\bibinfo {author} {\bibfnamefont {P.~C.~C.}\
  \bibnamefont {Freire}}, \bibinfo {author} {\bibfnamefont {N.}~\bibnamefont
  {Wex}}, \bibinfo {author} {\bibfnamefont {G.}~\bibnamefont
  {Esposito-Farese}}, \bibinfo {author} {\bibfnamefont {J.~P.~W.}\ \bibnamefont
  {Verbiest}}, \bibinfo {author} {\bibfnamefont {M.}~\bibnamefont {Bailes}},
  \bibinfo {author} {\bibfnamefont {B.~A.}\ \bibnamefont {Jacoby}}, \bibinfo
  {author} {\bibfnamefont {M.}~\bibnamefont {Kramer}}, \bibinfo {author}
  {\bibfnamefont {I.~H.}\ \bibnamefont {Stairs}}, \bibinfo {author}
  {\bibfnamefont {J.}~\bibnamefont {Antoniadis}}, \ and\ \bibinfo {author}
  {\bibfnamefont {G.~H.}\ \bibnamefont {Janssen}},\ }\href {\doibase
  10.1111/j.1365-2966.2012.21253.x} {\bibfield  {journal} {\bibinfo  {journal}
  {Mon. Not. Roy. Astron. Soc.}\ }\textbf {\bibinfo {volume} {423}},\ \bibinfo
  {pages} {3328} (\bibinfo {year} {2012})},\ \Eprint
  {http://arxiv.org/abs/1205.1450} {arXiv:1205.1450 [astro-ph.GA]} \BibitemShut
  {NoStop}%
\bibitem [{\citenamefont {Antoniadis}\ \emph {et~al.}(2013)\citenamefont
  {Antoniadis} \emph {et~al.}}]{Antoniadis:2013pzd}%
  \BibitemOpen
  \bibfield  {author} {\bibinfo {author} {\bibfnamefont {J.}~\bibnamefont
  {Antoniadis}} \emph {et~al.},\ }\href {\doibase 10.1126/science.1233232}
  {\bibfield  {journal} {\bibinfo  {journal} {Science}\ }\textbf {\bibinfo
  {volume} {340}},\ \bibinfo {pages} {6131} (\bibinfo {year} {2013})},\ \Eprint
  {http://arxiv.org/abs/1304.6875} {arXiv:1304.6875 [astro-ph.HE]} \BibitemShut
  {NoStop}%
\bibitem [{\citenamefont {Yunes}\ \emph {et~al.}(2016)\citenamefont {Yunes},
  \citenamefont {Yagi},\ and\ \citenamefont {Pretorius}}]{Yunes:2016jcc}%
  \BibitemOpen
  \bibfield  {author} {\bibinfo {author} {\bibfnamefont {N.}~\bibnamefont
  {Yunes}}, \bibinfo {author} {\bibfnamefont {K.}~\bibnamefont {Yagi}}, \ and\
  \bibinfo {author} {\bibfnamefont {F.}~\bibnamefont {Pretorius}},\ }\href
  {\doibase 10.1103/PhysRevD.94.084002} {\bibfield  {journal} {\bibinfo
  {journal} {Phys. Rev. D}\ }\textbf {\bibinfo {volume} {94}},\ \bibinfo
  {pages} {084002} (\bibinfo {year} {2016})},\ \Eprint
  {http://arxiv.org/abs/1603.08955} {arXiv:1603.08955 [gr-qc]} \BibitemShut
  {NoStop}%
\bibitem [{\citenamefont {Ramazanoglu}\ and\ \citenamefont
  {Pretorius}(2016)}]{Ramazanoglu:2016kul}%
  \BibitemOpen
  \bibfield  {author} {\bibinfo {author} {\bibfnamefont {F.~M.}\ \bibnamefont
  {Ramazanoglu}}\ and\ \bibinfo {author} {\bibfnamefont {F.}~\bibnamefont
  {Pretorius}},\ }\href {\doibase 10.1103/PhysRevD.93.064005} {\bibfield
  {journal} {\bibinfo  {journal} {Phys. Rev. D}\ }\textbf {\bibinfo {volume}
  {93}},\ \bibinfo {pages} {064005} (\bibinfo {year} {2016})},\ \Eprint
  {http://arxiv.org/abs/1601.07475} {arXiv:1601.07475 [gr-qc]} \BibitemShut
  {NoStop}%
\bibitem [{\citenamefont {{Sperhake}}\ \emph {et~al.}(2017)\citenamefont
  {{Sperhake}}, \citenamefont {{Moore}}, \citenamefont {{Rosca}}, \citenamefont
  {{Agathos}}, \citenamefont {{Gerosa}},\ and\ \citenamefont
  {{Ott}}}]{2017PhRvL.119t1103S}%
  \BibitemOpen
  \bibfield  {author} {\bibinfo {author} {\bibfnamefont {U.}~\bibnamefont
  {{Sperhake}}}, \bibinfo {author} {\bibfnamefont {C.~J.}\ \bibnamefont
  {{Moore}}}, \bibinfo {author} {\bibfnamefont {R.}~\bibnamefont {{Rosca}}},
  \bibinfo {author} {\bibfnamefont {M.}~\bibnamefont {{Agathos}}}, \bibinfo
  {author} {\bibfnamefont {D.}~\bibnamefont {{Gerosa}}}, \ and\ \bibinfo
  {author} {\bibfnamefont {C.~D.}\ \bibnamefont {{Ott}}},\ }\href {\doibase
  10.1103/PhysRevLett.119.201103} {\bibfield  {journal} {\bibinfo  {journal}
  {Phys.~Rev.~Lett.}\ }\textbf {\bibinfo {volume} {119}},\ \bibinfo {eid}
  {201103} (\bibinfo {year} {2017})},\ \Eprint
  {http://arxiv.org/abs/1708.03651} {arXiv:1708.03651 [gr-qc]} \BibitemShut
  {NoStop}%
\bibitem [{\citenamefont {{Rosca-Mead}}\ \emph {et~al.}(2020)\citenamefont
  {{Rosca-Mead}}, \citenamefont {{Sperhake}}, \citenamefont {{Moore}},
  \citenamefont {{Agathos}}, \citenamefont {{Gerosa}},\ and\ \citenamefont
  {{Ott}}}]{2020PhRvD.102d4010R}%
  \BibitemOpen
  \bibfield  {author} {\bibinfo {author} {\bibfnamefont {R.}~\bibnamefont
  {{Rosca-Mead}}}, \bibinfo {author} {\bibfnamefont {U.}~\bibnamefont
  {{Sperhake}}}, \bibinfo {author} {\bibfnamefont {C.~J.}\ \bibnamefont
  {{Moore}}}, \bibinfo {author} {\bibfnamefont {M.}~\bibnamefont {{Agathos}}},
  \bibinfo {author} {\bibfnamefont {D.}~\bibnamefont {{Gerosa}}}, \ and\
  \bibinfo {author} {\bibfnamefont {C.~D.}\ \bibnamefont {{Ott}}},\ }\href
  {\doibase 10.1103/PhysRevD.102.044010} {\bibfield  {journal} {\bibinfo
  {journal} {\prd}\ }\textbf {\bibinfo {volume} {102}},\ \bibinfo {eid}
  {044010} (\bibinfo {year} {2020})},\ \Eprint
  {http://arxiv.org/abs/2005.09728} {arXiv:2005.09728 [gr-qc]} \BibitemShut
  {NoStop}%
\bibitem [{\citenamefont {Aurrekoetxea}\ \emph {et~al.}(2022)\citenamefont
  {Aurrekoetxea}, \citenamefont {Ferreira}, \citenamefont {Clough},
  \citenamefont {Lim},\ and\ \citenamefont
  {Tattersall}}]{Aurrekoetxea:2022ika}%
  \BibitemOpen
  \bibfield  {author} {\bibinfo {author} {\bibfnamefont {J.~C.}\ \bibnamefont
  {Aurrekoetxea}}, \bibinfo {author} {\bibfnamefont {P.}~\bibnamefont
  {Ferreira}}, \bibinfo {author} {\bibfnamefont {K.}~\bibnamefont {Clough}},
  \bibinfo {author} {\bibfnamefont {E.~A.}\ \bibnamefont {Lim}}, \ and\
  \bibinfo {author} {\bibfnamefont {O.~J.}\ \bibnamefont {Tattersall}},\ }\href
  {\doibase 10.1103/PhysRevD.106.104002} {\bibfield  {journal} {\bibinfo
  {journal} {Phys. Rev. D}\ }\textbf {\bibinfo {volume} {106}},\ \bibinfo
  {pages} {104002} (\bibinfo {year} {2022})},\ \Eprint
  {http://arxiv.org/abs/2205.15878} {arXiv:2205.15878 [gr-qc]} \BibitemShut
  {NoStop}%
\bibitem [{\citenamefont {Bl\'azquez-Salcedo}\ \emph
  {et~al.}(2021)\citenamefont {Bl\'azquez-Salcedo}, \citenamefont {Khoo},
  \citenamefont {Kunz},\ and\ \citenamefont
  {Preut}}]{Blazquez-Salcedo:2021exm}%
  \BibitemOpen
  \bibfield  {author} {\bibinfo {author} {\bibfnamefont {J.~L.}\ \bibnamefont
  {Bl\'azquez-Salcedo}}, \bibinfo {author} {\bibfnamefont {F.~S.}\ \bibnamefont
  {Khoo}}, \bibinfo {author} {\bibfnamefont {J.}~\bibnamefont {Kunz}}, \ and\
  \bibinfo {author} {\bibfnamefont {V.}~\bibnamefont {Preut}},\ }\href
  {\doibase 10.3389/fphy.2021.741427} {\bibfield  {journal} {\bibinfo
  {journal} {Front. in Phys.}\ }\textbf {\bibinfo {volume} {9}},\ \bibinfo
  {pages} {741427} (\bibinfo {year} {2021})},\ \Eprint
  {http://arxiv.org/abs/2107.06726} {arXiv:2107.06726 [gr-qc]} \BibitemShut
  {NoStop}%
\bibitem [{\citenamefont {Kuan}\ \emph {et~al.}(2022)\citenamefont {Kuan},
  \citenamefont {Suvorov}, \citenamefont {Doneva},\ and\ \citenamefont
  {Yazadjiev}}]{Kuan:2022oxs}%
  \BibitemOpen
  \bibfield  {author} {\bibinfo {author} {\bibfnamefont {H.-J.}\ \bibnamefont
  {Kuan}}, \bibinfo {author} {\bibfnamefont {A.~G.}\ \bibnamefont {Suvorov}},
  \bibinfo {author} {\bibfnamefont {D.~D.}\ \bibnamefont {Doneva}}, \ and\
  \bibinfo {author} {\bibfnamefont {S.~S.}\ \bibnamefont {Yazadjiev}},\ }\href
  {\doibase 10.1103/PhysRevLett.129.121104} {\bibfield  {journal} {\bibinfo
  {journal} {Phys. Rev. Lett.}\ }\textbf {\bibinfo {volume} {129}},\ \bibinfo
  {pages} {121104} (\bibinfo {year} {2022})},\ \Eprint
  {http://arxiv.org/abs/2203.03672} {arXiv:2203.03672 [gr-qc]} \BibitemShut
  {NoStop}%
\bibitem [{\citenamefont {Rosca-Mead}\ \emph {et~al.}(2023)\citenamefont
  {Rosca-Mead}, \citenamefont {Agathos}, \citenamefont {Moore},\ and\
  \citenamefont {Sperhake}}]{Rosca-Mead:2023tdc}%
  \BibitemOpen
  \bibfield  {author} {\bibinfo {author} {\bibfnamefont {R.}~\bibnamefont
  {Rosca-Mead}}, \bibinfo {author} {\bibfnamefont {M.}~\bibnamefont {Agathos}},
  \bibinfo {author} {\bibfnamefont {C.~J.}\ \bibnamefont {Moore}}, \ and\
  \bibinfo {author} {\bibfnamefont {U.}~\bibnamefont {Sperhake}},\ }\href
  {\doibase 10.1103/PhysRevD.107.124040} {\bibfield  {journal} {\bibinfo
  {journal} {Phys. Rev. D}\ }\textbf {\bibinfo {volume} {107}},\ \bibinfo
  {pages} {124040} (\bibinfo {year} {2023})},\ \Eprint
  {http://arxiv.org/abs/2302.04995} {arXiv:2302.04995 [gr-qc]} \BibitemShut
  {NoStop}%
\bibitem [{\citenamefont {Kuroda}\ and\ \citenamefont
  {Shibata}(2023)}]{Kuroda:2023zbz}%
  \BibitemOpen
  \bibfield  {author} {\bibinfo {author} {\bibfnamefont {T.}~\bibnamefont
  {Kuroda}}\ and\ \bibinfo {author} {\bibfnamefont {M.}~\bibnamefont
  {Shibata}},\ }\href {\doibase 10.1103/PhysRevD.107.103025} {\bibfield
  {journal} {\bibinfo  {journal} {Phys. Rev. D}\ }\textbf {\bibinfo {volume}
  {107}},\ \bibinfo {pages} {103025} (\bibinfo {year} {2023})},\ \Eprint
  {http://arxiv.org/abs/2302.09853} {arXiv:2302.09853 [astro-ph.HE]}
  \BibitemShut {NoStop}%
\bibitem [{\citenamefont {Yazadjiev}\ \emph {et~al.}(2016)\citenamefont
  {Yazadjiev}, \citenamefont {Doneva},\ and\ \citenamefont
  {Popchev}}]{Yazadjiev:2016pcb}%
  \BibitemOpen
  \bibfield  {author} {\bibinfo {author} {\bibfnamefont {S.~S.}\ \bibnamefont
  {Yazadjiev}}, \bibinfo {author} {\bibfnamefont {D.~D.}\ \bibnamefont
  {Doneva}}, \ and\ \bibinfo {author} {\bibfnamefont {D.}~\bibnamefont
  {Popchev}},\ }\href {\doibase 10.1103/PhysRevD.93.084038} {\bibfield
  {journal} {\bibinfo  {journal} {Phys. Rev. D}\ }\textbf {\bibinfo {volume}
  {93}},\ \bibinfo {pages} {084038} (\bibinfo {year} {2016})},\ \Eprint
  {http://arxiv.org/abs/1602.04766} {arXiv:1602.04766 [gr-qc]} \BibitemShut
  {NoStop}%
\bibitem [{\citenamefont {Morisaki}\ and\ \citenamefont
  {Suyama}(2017)}]{Morisaki:2017nit}%
  \BibitemOpen
  \bibfield  {author} {\bibinfo {author} {\bibfnamefont {S.}~\bibnamefont
  {Morisaki}}\ and\ \bibinfo {author} {\bibfnamefont {T.}~\bibnamefont
  {Suyama}},\ }\href {\doibase 10.1103/PhysRevD.96.084026} {\bibfield
  {journal} {\bibinfo  {journal} {Phys. Rev. D}\ }\textbf {\bibinfo {volume}
  {96}},\ \bibinfo {pages} {084026} (\bibinfo {year} {2017})},\ \Eprint
  {http://arxiv.org/abs/1707.02809} {arXiv:1707.02809 [gr-qc]} \BibitemShut
  {NoStop}%
\bibitem [{\citenamefont {Ramazano\ifmmode~\breve{g}\else \u{g}\fi{}lu}\ and\
  \citenamefont {Pretorius}(2016)}]{PhysRevD.93.064005}%
  \BibitemOpen
  \bibfield  {author} {\bibinfo {author} {\bibfnamefont {F.~M.}\ \bibnamefont
  {Ramazano\ifmmode~\breve{g}\else \u{g}\fi{}lu}}\ and\ \bibinfo {author}
  {\bibfnamefont {F.}~\bibnamefont {Pretorius}},\ }\href {\doibase
  10.1103/PhysRevD.93.064005} {\bibfield  {journal} {\bibinfo  {journal} {Phys.
  Rev. D}\ }\textbf {\bibinfo {volume} {93}},\ \bibinfo {pages} {064005}
  (\bibinfo {year} {2016})}\BibitemShut {NoStop}%
\bibitem [{\citenamefont {Rosca-Mead}\ \emph {et~al.}(2019)\citenamefont
  {Rosca-Mead}, \citenamefont {Moore}, \citenamefont {Agathos},\ and\
  \citenamefont {Sperhake}}]{Rosca-Mead:2019seq}%
  \BibitemOpen
  \bibfield  {author} {\bibinfo {author} {\bibfnamefont {R.}~\bibnamefont
  {Rosca-Mead}}, \bibinfo {author} {\bibfnamefont {C.~J.}\ \bibnamefont
  {Moore}}, \bibinfo {author} {\bibfnamefont {M.}~\bibnamefont {Agathos}}, \
  and\ \bibinfo {author} {\bibfnamefont {U.}~\bibnamefont {Sperhake}},\ }\href
  {\doibase 10.1088/1361-6382/ab256f} {\bibfield  {journal} {\bibinfo
  {journal} {Class. Quant. Grav.}\ }\textbf {\bibinfo {volume} {36}},\ \bibinfo
  {pages} {134003} (\bibinfo {year} {2019})},\ \Eprint
  {http://arxiv.org/abs/1903.09704} {arXiv:1903.09704 [gr-qc]} \BibitemShut
  {NoStop}%
\bibitem [{\citenamefont {Huang}\ \emph {et~al.}(2021)\citenamefont {Huang},
  \citenamefont {Geng},\ and\ \citenamefont {Kuan}}]{Huang:2021tpu}%
  \BibitemOpen
  \bibfield  {author} {\bibinfo {author} {\bibfnamefont {D.}~\bibnamefont
  {Huang}}, \bibinfo {author} {\bibfnamefont {C.-Q.}\ \bibnamefont {Geng}}, \
  and\ \bibinfo {author} {\bibfnamefont {H.-J.}\ \bibnamefont {Kuan}},\ }\href
  {\doibase 10.1088/1361-6382/ac35ab} {\bibfield  {journal} {\bibinfo
  {journal} {Class. Quant. Grav.}\ }\textbf {\bibinfo {volume} {38}},\ \bibinfo
  {pages} {245006} (\bibinfo {year} {2021})},\ \Eprint
  {http://arxiv.org/abs/2106.13065} {arXiv:2106.13065 [gr-qc]} \BibitemShut
  {NoStop}%
\bibitem [{\citenamefont {Asakawa}\ and\ \citenamefont
  {Sekiguchi}(2023)}]{Asakawa:2023obq}%
  \BibitemOpen
  \bibfield  {author} {\bibinfo {author} {\bibfnamefont {N.}~\bibnamefont
  {Asakawa}}\ and\ \bibinfo {author} {\bibfnamefont {Y.}~\bibnamefont
  {Sekiguchi}},\ }\href {\doibase 10.1103/PhysRevD.108.044060} {\bibfield
  {journal} {\bibinfo  {journal} {Phys. Rev. D}\ }\textbf {\bibinfo {volume}
  {108}},\ \bibinfo {pages} {044060} (\bibinfo {year} {2023})},\ \Eprint
  {http://arxiv.org/abs/2308.15052} {arXiv:2308.15052 [gr-qc]} \BibitemShut
  {NoStop}%
\bibitem [{\citenamefont {Ma}\ \emph {et~al.}(2023)\citenamefont {Ma},
  \citenamefont {Varma}, \citenamefont {Stein}, \citenamefont {Foucart},
  \citenamefont {Duez}, \citenamefont {Kidder}, \citenamefont {Pfeiffer},\ and\
  \citenamefont {Scheel}}]{Ma:2023sok}%
  \BibitemOpen
  \bibfield  {author} {\bibinfo {author} {\bibfnamefont {S.}~\bibnamefont
  {Ma}}, \bibinfo {author} {\bibfnamefont {V.}~\bibnamefont {Varma}}, \bibinfo
  {author} {\bibfnamefont {L.~C.}\ \bibnamefont {Stein}}, \bibinfo {author}
  {\bibfnamefont {F.}~\bibnamefont {Foucart}}, \bibinfo {author} {\bibfnamefont
  {M.~D.}\ \bibnamefont {Duez}}, \bibinfo {author} {\bibfnamefont {L.~E.}\
  \bibnamefont {Kidder}}, \bibinfo {author} {\bibfnamefont {H.~P.}\
  \bibnamefont {Pfeiffer}}, \ and\ \bibinfo {author} {\bibfnamefont {M.~A.}\
  \bibnamefont {Scheel}},\ }\href {\doibase 10.1103/PhysRevD.107.124051}
  {\bibfield  {journal} {\bibinfo  {journal} {Phys. Rev. D}\ }\textbf {\bibinfo
  {volume} {107}},\ \bibinfo {pages} {124051} (\bibinfo {year} {2023})},\
  \Eprint {http://arxiv.org/abs/2304.11836} {arXiv:2304.11836 [gr-qc]}
  \BibitemShut {NoStop}%
\bibitem [{\citenamefont {Yagi}\ and\ \citenamefont
  {Stepniczka}(2021)}]{Yagi:2021loe}%
  \BibitemOpen
  \bibfield  {author} {\bibinfo {author} {\bibfnamefont {K.}~\bibnamefont
  {Yagi}}\ and\ \bibinfo {author} {\bibfnamefont {M.}~\bibnamefont
  {Stepniczka}},\ }\href {\doibase 10.1103/PhysRevD.104.044017} {\bibfield
  {journal} {\bibinfo  {journal} {Phys. Rev. D}\ }\textbf {\bibinfo {volume}
  {104}},\ \bibinfo {pages} {044017} (\bibinfo {year} {2021})},\ \Eprint
  {http://arxiv.org/abs/2105.01614} {arXiv:2105.01614 [gr-qc]} \BibitemShut
  {NoStop}%
\bibitem [{\citenamefont {Barausse}\ \emph {et~al.}(2013)\citenamefont
  {Barausse}, \citenamefont {Palenzuela}, \citenamefont {Ponce},\ and\
  \citenamefont {Lehner}}]{Barausse:2012da}%
  \BibitemOpen
  \bibfield  {author} {\bibinfo {author} {\bibfnamefont {E.}~\bibnamefont
  {Barausse}}, \bibinfo {author} {\bibfnamefont {C.}~\bibnamefont
  {Palenzuela}}, \bibinfo {author} {\bibfnamefont {M.}~\bibnamefont {Ponce}}, \
  and\ \bibinfo {author} {\bibfnamefont {L.}~\bibnamefont {Lehner}},\ }\href
  {\doibase 10.1103/PhysRevD.87.081506} {\bibfield  {journal} {\bibinfo
  {journal} {Phys. Rev. D}\ }\textbf {\bibinfo {volume} {87}},\ \bibinfo
  {pages} {081506} (\bibinfo {year} {2013})},\ \Eprint
  {http://arxiv.org/abs/1212.5053} {arXiv:1212.5053 [gr-qc]} \BibitemShut
  {NoStop}%
\bibitem [{\citenamefont {Odintsov}\ and\ \citenamefont
  {Oikonomou}(2022)}]{Odintsov:2021nqa}%
  \BibitemOpen
  \bibfield  {author} {\bibinfo {author} {\bibfnamefont {S.~D.}\ \bibnamefont
  {Odintsov}}\ and\ \bibinfo {author} {\bibfnamefont {V.~K.}\ \bibnamefont
  {Oikonomou}},\ }\href {\doibase 10.1016/j.aop.2022.168839} {\bibfield
  {journal} {\bibinfo  {journal} {Annals Phys.}\ }\textbf {\bibinfo {volume}
  {440}},\ \bibinfo {pages} {168839} (\bibinfo {year} {2022})},\ \Eprint
  {http://arxiv.org/abs/2104.01982} {arXiv:2104.01982 [gr-qc]} \BibitemShut
  {NoStop}%
\bibitem [{\citenamefont {Whinnett}\ and\ \citenamefont
  {Torres}(1999)}]{Whinnett:1999ma}%
  \BibitemOpen
  \bibfield  {author} {\bibinfo {author} {\bibfnamefont {A.~W.}\ \bibnamefont
  {Whinnett}}\ and\ \bibinfo {author} {\bibfnamefont {D.~F.}\ \bibnamefont
  {Torres}},\ }\href {\doibase 10.1103/PhysRevD.60.104050} {\bibfield
  {journal} {\bibinfo  {journal} {Phys. Rev. D}\ }\textbf {\bibinfo {volume}
  {60}},\ \bibinfo {pages} {104050} (\bibinfo {year} {1999})},\ \Eprint
  {http://arxiv.org/abs/gr-qc/9905017} {arXiv:gr-qc/9905017} \BibitemShut
  {NoStop}%
\bibitem [{\citenamefont {Whinnett}(2000)}]{Whinnett:1999sc}%
  \BibitemOpen
  \bibfield  {author} {\bibinfo {author} {\bibfnamefont {A.~W.}\ \bibnamefont
  {Whinnett}},\ }\href {\doibase 10.1103/PhysRevD.61.124014} {\bibfield
  {journal} {\bibinfo  {journal} {Phys. Rev. D}\ }\textbf {\bibinfo {volume}
  {61}},\ \bibinfo {pages} {124014} (\bibinfo {year} {2000})},\ \Eprint
  {http://arxiv.org/abs/gr-qc/9911052} {arXiv:gr-qc/9911052} \BibitemShut
  {NoStop}%
\bibitem [{\citenamefont {Ruiz}\ \emph {et~al.}(2012)\citenamefont {Ruiz},
  \citenamefont {Degollado}, \citenamefont {Alcubierre}, \citenamefont
  {Nunez},\ and\ \citenamefont {Salgado}}]{Ruiz:2012jt}%
  \BibitemOpen
  \bibfield  {author} {\bibinfo {author} {\bibfnamefont {M.}~\bibnamefont
  {Ruiz}}, \bibinfo {author} {\bibfnamefont {J.~C.}\ \bibnamefont {Degollado}},
  \bibinfo {author} {\bibfnamefont {M.}~\bibnamefont {Alcubierre}}, \bibinfo
  {author} {\bibfnamefont {D.}~\bibnamefont {Nunez}}, \ and\ \bibinfo {author}
  {\bibfnamefont {M.}~\bibnamefont {Salgado}},\ }\href {\doibase
  10.1103/PhysRevD.86.104044} {\bibfield  {journal} {\bibinfo  {journal} {Phys.
  Rev. D}\ }\textbf {\bibinfo {volume} {86}},\ \bibinfo {pages} {104044}
  (\bibinfo {year} {2012})},\ \Eprint {http://arxiv.org/abs/1207.6142}
  {arXiv:1207.6142 [gr-qc]} \BibitemShut {NoStop}%
\bibitem [{\citenamefont {Alcubierre}\ \emph {et~al.}(2010)\citenamefont
  {Alcubierre}, \citenamefont {Degollado}, \citenamefont {Nunez}, \citenamefont
  {Ruiz},\ and\ \citenamefont {Salgado}}]{Alcubierre:2010ea}%
  \BibitemOpen
  \bibfield  {author} {\bibinfo {author} {\bibfnamefont {M.}~\bibnamefont
  {Alcubierre}}, \bibinfo {author} {\bibfnamefont {J.~C.}\ \bibnamefont
  {Degollado}}, \bibinfo {author} {\bibfnamefont {D.}~\bibnamefont {Nunez}},
  \bibinfo {author} {\bibfnamefont {M.}~\bibnamefont {Ruiz}}, \ and\ \bibinfo
  {author} {\bibfnamefont {M.}~\bibnamefont {Salgado}},\ }\href {\doibase
  10.1103/PhysRevD.81.124018} {\bibfield  {journal} {\bibinfo  {journal} {Phys.
  Rev. D}\ }\textbf {\bibinfo {volume} {81}},\ \bibinfo {pages} {124018}
  (\bibinfo {year} {2010})},\ \Eprint {http://arxiv.org/abs/1003.4767}
  {arXiv:1003.4767 [gr-qc]} \BibitemShut {NoStop}%
\bibitem [{\citenamefont {Brihaye}\ and\ \citenamefont
  {Hartmann}(2019)}]{Brihaye:2019puo}%
  \BibitemOpen
  \bibfield  {author} {\bibinfo {author} {\bibfnamefont {Y.}~\bibnamefont
  {Brihaye}}\ and\ \bibinfo {author} {\bibfnamefont {B.}~\bibnamefont
  {Hartmann}},\ }\href {\doibase 10.1007/JHEP09(2019)049} {\bibfield  {journal}
  {\bibinfo  {journal} {JHEP}\ }\textbf {\bibinfo {volume} {09}},\ \bibinfo
  {pages} {049} (\bibinfo {year} {2019})},\ \Eprint
  {http://arxiv.org/abs/1903.10471} {arXiv:1903.10471 [gr-qc]} \BibitemShut
  {NoStop}%
\bibitem [{\citenamefont {Damour}\ and\ \citenamefont
  {Esposito-Far\`ese}(1993)}]{PhysRevLett.70.2220}%
  \BibitemOpen
  \bibfield  {author} {\bibinfo {author} {\bibfnamefont {T.}~\bibnamefont
  {Damour}}\ and\ \bibinfo {author} {\bibfnamefont {G.}~\bibnamefont
  {Esposito-Far\`ese}},\ }\href {\doibase 10.1103/PhysRevLett.70.2220}
  {\bibfield  {journal} {\bibinfo  {journal} {Phys.~Rev.~Lett.}\ }\textbf
  {\bibinfo {volume} {70}},\ \bibinfo {pages} {2220} (\bibinfo {year}
  {1993})}\BibitemShut {NoStop}%
\bibitem [{\citenamefont {Salgado}(2006)}]{Salgado:2005hx}%
  \BibitemOpen
  \bibfield  {author} {\bibinfo {author} {\bibfnamefont {M.}~\bibnamefont
  {Salgado}},\ }\href {\doibase 10.1088/0264-9381/23/14/010} {\bibfield
  {journal} {\bibinfo  {journal} {Class. Quantum Grav.}\ }\textbf {\bibinfo
  {volume} {23}},\ \bibinfo {pages} {4719} (\bibinfo {year} {2006})},\ \Eprint
  {http://arxiv.org/abs/gr-qc/0509001} {arXiv:gr-qc/0509001} \BibitemShut
  {NoStop}%
\bibitem [{\citenamefont {Chiba}\ \emph {et~al.}(1997)\citenamefont {Chiba},
  \citenamefont {Harada},\ and\ \citenamefont {Nakao}}]{Chiba:1997ms}%
  \BibitemOpen
  \bibfield  {author} {\bibinfo {author} {\bibfnamefont {T.}~\bibnamefont
  {Chiba}}, \bibinfo {author} {\bibfnamefont {T.}~\bibnamefont {Harada}}, \
  and\ \bibinfo {author} {\bibfnamefont {K.-i.}\ \bibnamefont {Nakao}},\ }\href
  {\doibase 10.1143/PTPS.128.335} {\bibfield  {journal} {\bibinfo  {journal}
  {Prog. Theor. Phys. Suppl.}\ }\textbf {\bibinfo {volume} {128}},\ \bibinfo
  {pages} {335} (\bibinfo {year} {1997})}\BibitemShut {NoStop}%
\bibitem [{\citenamefont {Faraoni}(1923)}]{Eddington:1923}%
  \BibitemOpen
  \bibfield  {author} {\bibinfo {author} {\bibfnamefont {A.~S.}\ \bibnamefont
  {Faraoni}},\ }\href@noop {} {\emph {\bibinfo {title} {{The Mathematical
  Theory of Relativity}}}}\ (\bibinfo  {publisher} {Cambridge University Press,
  Cambridge},\ \bibinfo {year} {1923})\BibitemShut {NoStop}%
\bibitem [{\citenamefont {Gerosa}\ \emph {et~al.}(2016)\citenamefont {Gerosa},
  \citenamefont {Sperhake},\ and\ \citenamefont {Ott}}]{Gerosa:2016fri}%
  \BibitemOpen
  \bibfield  {author} {\bibinfo {author} {\bibfnamefont {D.}~\bibnamefont
  {Gerosa}}, \bibinfo {author} {\bibfnamefont {U.}~\bibnamefont {Sperhake}}, \
  and\ \bibinfo {author} {\bibfnamefont {C.~D.}\ \bibnamefont {Ott}},\ }\href
  {\doibase 10.1088/0264-9381/33/13/135002} {\bibfield  {journal} {\bibinfo
  {journal} {Class. Quant. Grav.}\ }\textbf {\bibinfo {volume} {33}},\ \bibinfo
  {pages} {135002} (\bibinfo {year} {2016})},\ \Eprint
  {http://arxiv.org/abs/1602.06952} {arXiv:1602.06952 [gr-qc]} \BibitemShut
  {NoStop}%
\bibitem [{\citenamefont {Brans}\ and\ \citenamefont
  {Dicke}(1961)}]{Brans:1961sx}%
  \BibitemOpen
  \bibfield  {author} {\bibinfo {author} {\bibfnamefont {C.}~\bibnamefont
  {Brans}}\ and\ \bibinfo {author} {\bibfnamefont {R.}~\bibnamefont {Dicke}},\
  }\href {\doibase 10.1103/PhysRev.124.925} {\bibfield  {journal} {\bibinfo
  {journal} {Phys.Rev.}\ }\textbf {\bibinfo {volume} {124}},\ \bibinfo {pages}
  {925} (\bibinfo {year} {1961})}\BibitemShut {NoStop}%
\bibitem [{\citenamefont {Kuan}\ \emph {et~al.}(2023)\citenamefont {Kuan},
  \citenamefont {Van~Aelst}, \citenamefont {Lam},\ and\ \citenamefont
  {Shibata}}]{Kuan:2023hrh}%
  \BibitemOpen
  \bibfield  {author} {\bibinfo {author} {\bibfnamefont {H.-J.}\ \bibnamefont
  {Kuan}}, \bibinfo {author} {\bibfnamefont {K.}~\bibnamefont {Van~Aelst}},
  \bibinfo {author} {\bibfnamefont {A.-L.}\ \bibnamefont {Lam}}, \ and\
  \bibinfo {author} {\bibfnamefont {M.}~\bibnamefont {Shibata}},\ }\href@noop
  {} {\  (\bibinfo {year} {2023})},\ \Eprint {http://arxiv.org/abs/2309.01709}
  {arXiv:2309.01709 [gr-qc]} \BibitemShut {NoStop}%
\bibitem [{\citenamefont {Narayan}\ and\ \citenamefont
  {McClintock}(2013)}]{Narayan:2013gca}%
  \BibitemOpen
  \bibfield  {author} {\bibinfo {author} {\bibfnamefont {R.}~\bibnamefont
  {Narayan}}\ and\ \bibinfo {author} {\bibfnamefont {J.~E.}\ \bibnamefont
  {McClintock}},\ }\href@noop {} {\  (\bibinfo {year} {2013})},\ \Eprint
  {http://arxiv.org/abs/1312.6698} {arXiv:1312.6698 [astro-ph.HE]} \BibitemShut
  {NoStop}%
\bibitem [{\citenamefont {Brito}\ \emph {et~al.}(2015)\citenamefont {Brito},
  \citenamefont {Cardoso},\ and\ \citenamefont {Pani}}]{Brito:2015oca}%
  \BibitemOpen
  \bibfield  {author} {\bibinfo {author} {\bibfnamefont {R.}~\bibnamefont
  {Brito}}, \bibinfo {author} {\bibfnamefont {V.}~\bibnamefont {Cardoso}}, \
  and\ \bibinfo {author} {\bibfnamefont {P.}~\bibnamefont {Pani}},\ }\href
  {\doibase 10.1007/978-3-319-19000-6} {\bibfield  {journal} {\bibinfo
  {journal} {Lect. Notes Phys.}\ }\textbf {\bibinfo {volume} {906}},\ \bibinfo
  {pages} {pp.1} (\bibinfo {year} {2015})},\ \Eprint
  {http://arxiv.org/abs/arXiv:1501.06570 [gr-qc]} {arXiv:1501.06570 [gr-qc]}
  \BibitemShut {NoStop}%
\bibitem [{\citenamefont {Arnowitt}\ \emph {et~al.}(1962)\citenamefont
  {Arnowitt}, \citenamefont {Deser},\ and\ \citenamefont
  {Misner}}]{Arnowitt:1962hi}%
  \BibitemOpen
  \bibfield  {author} {\bibinfo {author} {\bibfnamefont {R.}~\bibnamefont
  {Arnowitt}}, \bibinfo {author} {\bibfnamefont {S.}~\bibnamefont {Deser}}, \
  and\ \bibinfo {author} {\bibfnamefont {C.~W.}\ \bibnamefont {Misner}},\ }in\
  \href@noop {} {\emph {\bibinfo {booktitle} {Gravitation an introduction to
  current research}}},\ \bibinfo {editor} {edited by\ \bibinfo {editor}
  {\bibfnamefont {L.}~\bibnamefont {Witten}}}\ (\bibinfo  {publisher} {John
  Wiley, New York},\ \bibinfo {year} {1962})\ pp.\ \bibinfo {pages}
  {227--265},\ \Eprint {http://arxiv.org/abs/gr-qc/0405109}
  {arXiv:gr-qc/0405109} \BibitemShut {NoStop}%
\bibitem [{\citenamefont {Rosca-Mead}\ \emph {et~al.}(2020)\citenamefont
  {Rosca-Mead}, \citenamefont {Moore}, \citenamefont {Sperhake}, \citenamefont
  {Agathos},\ and\ \citenamefont {Gerosa}}]{Rosca-Mead:2020bzt}%
  \BibitemOpen
  \bibfield  {author} {\bibinfo {author} {\bibfnamefont {R.}~\bibnamefont
  {Rosca-Mead}}, \bibinfo {author} {\bibfnamefont {C.~J.}\ \bibnamefont
  {Moore}}, \bibinfo {author} {\bibfnamefont {U.}~\bibnamefont {Sperhake}},
  \bibinfo {author} {\bibfnamefont {M.}~\bibnamefont {Agathos}}, \ and\
  \bibinfo {author} {\bibfnamefont {D.}~\bibnamefont {Gerosa}},\ }\href
  {\doibase 10.3390/sym12091384} {\bibfield  {journal} {\bibinfo  {journal}
  {Symmetry}\ }\textbf {\bibinfo {volume} {12}},\ \bibinfo {pages} {9}
  (\bibinfo {year} {2020})},\ \Eprint {http://arxiv.org/abs/2007.14429}
  {arXiv:2007.14429 [gr-qc]} \BibitemShut {NoStop}%
\bibitem [{\citenamefont {Novak}(1998{\natexlab{a}})}]{Novak:1997hw}%
  \BibitemOpen
  \bibfield  {author} {\bibinfo {author} {\bibfnamefont {J.}~\bibnamefont
  {Novak}},\ }\href {\doibase 10.1103/PhysRevD.57.4789} {\bibfield  {journal}
  {\bibinfo  {journal} {Phys. Rev. D}\ }\textbf {\bibinfo {volume} {57}},\
  \bibinfo {pages} {4789} (\bibinfo {year} {1998}{\natexlab{a}})},\ \Eprint
  {http://arxiv.org/abs/gr-qc/9707041} {arXiv:gr-qc/9707041} \BibitemShut
  {NoStop}%
\bibitem [{\citenamefont {Novak}(1998{\natexlab{b}})}]{Novak:1998rk}%
  \BibitemOpen
  \bibfield  {author} {\bibinfo {author} {\bibfnamefont {J.}~\bibnamefont
  {Novak}},\ }\href {\doibase 10.1103/PhysRevD.58.064019} {\bibfield  {journal}
  {\bibinfo  {journal} {Phys. Rev. D}\ }\textbf {\bibinfo {volume} {58}},\
  \bibinfo {pages} {064019} (\bibinfo {year} {1998}{\natexlab{b}})},\ \Eprint
  {http://arxiv.org/abs/gr-qc/9806022} {arXiv:gr-qc/9806022} \BibitemShut
  {NoStop}%
\bibitem [{\citenamefont {Pani}\ and\ \citenamefont
  {Berti}(2014)}]{Pani:2014jra}%
  \BibitemOpen
  \bibfield  {author} {\bibinfo {author} {\bibfnamefont {P.}~\bibnamefont
  {Pani}}\ and\ \bibinfo {author} {\bibfnamefont {E.}~\bibnamefont {Berti}},\
  }\href {\doibase 10.1103/PhysRevD.90.024025} {\bibfield  {journal} {\bibinfo
  {journal} {Phys. Rev. D}\ }\textbf {\bibinfo {volume} {90}},\ \bibinfo
  {pages} {024025} (\bibinfo {year} {2014})},\ \Eprint
  {http://arxiv.org/abs/1405.4547} {arXiv:1405.4547 [gr-qc]} \BibitemShut
  {NoStop}%
\bibitem [{\citenamefont {Mendes}\ \emph {et~al.}(2014)\citenamefont {Mendes},
  \citenamefont {Matsas},\ and\ \citenamefont {Vanzella}}]{Mendes:2014vna}%
  \BibitemOpen
  \bibfield  {author} {\bibinfo {author} {\bibfnamefont {R.~F.~P.}\
  \bibnamefont {Mendes}}, \bibinfo {author} {\bibfnamefont {G.~E.~A.}\
  \bibnamefont {Matsas}}, \ and\ \bibinfo {author} {\bibfnamefont {D.~A.~T.}\
  \bibnamefont {Vanzella}},\ }\href {\doibase 10.1103/PhysRevD.90.044053}
  {\bibfield  {journal} {\bibinfo  {journal} {Phys. Rev. D}\ }\textbf {\bibinfo
  {volume} {90}},\ \bibinfo {pages} {044053} (\bibinfo {year} {2014})},\
  \Eprint {http://arxiv.org/abs/1407.6405} {arXiv:1407.6405 [gr-qc]}
  \BibitemShut {NoStop}%
\bibitem [{\citenamefont {Mendes}(2015)}]{Mendes:2014ufa}%
  \BibitemOpen
  \bibfield  {author} {\bibinfo {author} {\bibfnamefont {R.~F.~P.}\
  \bibnamefont {Mendes}},\ }\href {\doibase 10.1103/PhysRevD.91.064024}
  {\bibfield  {journal} {\bibinfo  {journal} {Phys. Rev. D}\ }\textbf {\bibinfo
  {volume} {91}},\ \bibinfo {pages} {064024} (\bibinfo {year} {2015})},\
  \Eprint {http://arxiv.org/abs/1412.6789} {arXiv:1412.6789 [gr-qc]}
  \BibitemShut {NoStop}%
\bibitem [{\citenamefont {Mendes}\ and\ \citenamefont
  {Ortiz}(2016)}]{Mendes:2016fby}%
  \BibitemOpen
  \bibfield  {author} {\bibinfo {author} {\bibfnamefont {R.~F.}\ \bibnamefont
  {Mendes}}\ and\ \bibinfo {author} {\bibfnamefont {N.}~\bibnamefont {Ortiz}},\
  }\href {\doibase 10.1103/PhysRevD.93.124035} {\bibfield  {journal} {\bibinfo
  {journal} {Phys. Rev. D}\ }\textbf {\bibinfo {volume} {93}},\ \bibinfo
  {pages} {124035} (\bibinfo {year} {2016})},\ \Eprint
  {http://arxiv.org/abs/1604.04175} {arXiv:1604.04175 [gr-qc]} \BibitemShut
  {NoStop}%
\bibitem [{\citenamefont {Ramazano\u{g}lu}(2017)}]{Ramazanoglu:2017xbl}%
  \BibitemOpen
  \bibfield  {author} {\bibinfo {author} {\bibfnamefont {F.~M.}\ \bibnamefont
  {Ramazano\u{g}lu}},\ }\href {\doibase 10.1103/PhysRevD.96.064009} {\bibfield
  {journal} {\bibinfo  {journal} {Phys. Rev. D}\ }\textbf {\bibinfo {volume}
  {96}},\ \bibinfo {pages} {064009} (\bibinfo {year} {2017})},\ \Eprint
  {http://arxiv.org/abs/1706.01056} {arXiv:1706.01056 [gr-qc]} \BibitemShut
  {NoStop}%
\bibitem [{\citenamefont {Collodel}\ and\ \citenamefont
  {Doneva}(2022)}]{Collodel:2022jly}%
  \BibitemOpen
  \bibfield  {author} {\bibinfo {author} {\bibfnamefont {L.~G.}\ \bibnamefont
  {Collodel}}\ and\ \bibinfo {author} {\bibfnamefont {D.~D.}\ \bibnamefont
  {Doneva}},\ }\href {\doibase 10.1103/PhysRevD.106.084057} {\bibfield
  {journal} {\bibinfo  {journal} {Phys. Rev. D}\ }\textbf {\bibinfo {volume}
  {106}},\ \bibinfo {pages} {084057} (\bibinfo {year} {2022})},\ \Eprint
  {http://arxiv.org/abs/2203.08203} {arXiv:2203.08203 [gr-qc]} \BibitemShut
  {NoStop}%
\bibitem [{\citenamefont {Bo\v{s}kovi\'c}\ and\ \citenamefont
  {Barausse}(2022)}]{Boskovic:2021nfs}%
  \BibitemOpen
  \bibfield  {author} {\bibinfo {author} {\bibfnamefont {M.}~\bibnamefont
  {Bo\v{s}kovi\'c}}\ and\ \bibinfo {author} {\bibfnamefont {E.}~\bibnamefont
  {Barausse}},\ }\href {\doibase 10.1088/1475-7516/2022/02/032} {\bibfield
  {journal} {\bibinfo  {journal} {JCAP}\ }\textbf {\bibinfo {volume} {02}},\
  \bibinfo {pages} {032} (\bibinfo {year} {2022})},\ \Eprint
  {http://arxiv.org/abs/2111.03870} {arXiv:2111.03870 [gr-qc]} \BibitemShut
  {NoStop}%
\bibitem [{\citenamefont {{Press, William H. and Teukolsky, Saul A. and
  Vetterling, William T. and Flannery, Brian P.}}(1992)}]{Press1992}%
  \BibitemOpen
  \bibfield  {author} {\bibinfo {author} {\bibnamefont {{Press, William H. and
  Teukolsky, Saul A. and Vetterling, William T. and Flannery, Brian P.}}},\
  }\href@noop {} {\emph {\bibinfo {title} {{Numerical recipes in C (2nd ed.):
  the art of scientific computing}}}}\ (\bibinfo  {publisher} {{Cambridge
  University Press}},\ \bibinfo {address} {{New York, NY, USA}},\ \bibinfo
  {year} {1992})\BibitemShut {NoStop}%
\end{thebibliography}%

\appendix
\section{Asymptotic Behaviour}\label{app:asymptotic_behaviour}
In this Appendix, we analyse the asymptotic behaviour of the BS and
gravitational scalar fields in ST theory of gravity at infinity.
It is convenient to work in the compactified coordinate $y \defeq
1/r$ and define re-scaled variables, separating the exponential
behaviour in the following way,
\begin{eqnarray} \label{eq:rescaled_vars} 
\zeta \defeq \frac{e^{k/y}}{y^{\epsilon}}A\,,
~~~~~&&~~~~~
\Pi\defeq -\frac{e^{k/y}}{y^{\epsilon}} \theta\,,
\nonumber \\
\varrho \defeq \frac{e^{h/y}}{y^{\delta}}\varphi\,,
~~~~~&&~~~~~
\Lambda \defeq -\frac{e^{h/y}}{y^{\delta}} \kappa\,.
\label{eq:matching}
\end{eqnarray}
Note that the exponents $\epsilon, k, h, \delta$ are as yet unspecified
parameters and the goal of this section will be to determine them.
As we will see below, their values will depend on the specifics of
the BS model and ST parameters in question. The key difference of
the asymptotic behaviour we assume here as compared to the flat-field
limit \eqref{eq:asymp_wrong} is the inclusion of the non-integer
$\epsilon$ and $\delta$ parameters, which are set to zero in the
flat-field limit. We use the term \textit{flat-field limit} for
this latter case since it becomes correct when $M=0$, i.e.~when we
regard the matter as a perturbation on a flat Minkowski metric
without backreaction on spacetime curvature.

The field equations for the re-scaled variables \eqref{eq:rescaled_vars}
are obtained by substituting them into Eqs.~\eqref{eq:allr}, which
leads to
\begin{widetext}
\begin{eqnarray}
    \partial_y \Phi
  &=&
  -\frac{FX^2-1}{2y}
  +\frac{y^{2\delta-3}}{e^{2h/y}} X^2
  \left(
  F\hat{W}\varrho^2 - \frac{\Lambda^2}{2\alpha^2}
  \right)
  +2\pi y^{2\epsilon-3}
  \frac{X^2}{F\alpha^2}
  e^{-2k/y}
  \left(
  \alpha^2\hat{V}\zeta^2 - \omega^2\zeta^2 - F^2\Pi^2
  \right)
  \nonumber \\
  \partial_y m
  &=&
  -\frac{y^{2\delta - 4}}{e^{2h/y}}
  \left(
  \hat{W}\varrho^2 + \frac{\Lambda^2}{2F\alpha^2}
  \right)
  -\frac{2\pi y^{2\epsilon -4}}{F^2\alpha^2}e^{-2k/y}
  \left(
  \omega^2 \zeta^2 + \alpha^2\hat{V}\zeta^2 + F^2\Pi^2
  \right)\,,
  \nonumber \\
  \partial_y \varrho
  &=&
  \frac{X\Lambda/\alpha - (y\delta + h)\varrho}{y^2}\,,
  \nonumber \\
  \partial_y \Lambda
  &=&
  \frac{(2-\delta)y-h}{y^2}\Lambda
  + 2\frac{\alpha XF}{y^2}W_{,\varphi^2} \varrho
  + \frac{2\pi}{y^{2+\delta-2\epsilon}}
  e^{(h-2k)/y}\frac{X}{\alpha}
  \frac{F'}{F^2}
  \left(
  \omega^2 \zeta^2-2\alpha^2\hat{V}\zeta^2-F^2\Pi^2
  \right)\,,
  \nonumber \\
  \partial_y \zeta
  &=&
  \frac{FX\Pi/\alpha - (\epsilon y+k)\zeta}{y^2}\,,
  \nonumber \\
  \partial_y \Pi
  &=&
  \frac{(2-\epsilon)y-k}{y^2}\Pi
  + \frac{X\zeta}{\alpha F y^2}
  \left(
  \alpha^2 V_{,A^2} - \omega^2
  \right)\,,
  \label{eq:allyf}
\end{eqnarray}
\end{widetext}
where we denoted $\hat{W} \defeq W/\varphi^2$ and $\hat{V} \defeq
V/A^2$. Note that the equation for $\partial_y \Lambda$ contains
an exponential factor $\exp[(h-2k)/y]$ which diverges when $h >
2k$, resulting in unphysical spacetimes. We are thus left with two
remaining possible scenarios: $h < 2k$ and $h=2k$. All BS families
studied in this work comfortably fall into the range $h<2k$ which
is the case we henceforth focus on.

Assuming that we recover a Schwarzschild metric at infinity, we
propose the following ans{\"a}tze around $y=0$
\begin{eqnarray} \label{eq:series_ansatze}
  \Phi(y) &=& \Phi_0+\Phi_1 y + \Phi_2 y^2 + \Phi_3y^3+\Phi_4y^4+\ldots \,,
  \nonumber \\
  m(y) &=& M + m_1 y + m_2y^2+m_3y^3+m_4y^4+\ldots\,,
  \nonumber \\
  \varrho(y) &=& \varrho_1 y+\varrho_2y^2+\varrho_3y^3+\varrho_4y^4+\ldots\,,
  \nonumber \\
  \Lambda(y) &=& \Lambda_1y + \Lambda_2 y^2+\Lambda_3y^3+\Lambda_4y^4+\ldots \,,
  \nonumber \\
  \zeta(y) &=& \zeta_1 y+\zeta_2 y^2+\zeta_3 y^3+ \zeta_4y^4+\ldots \,,
  \nonumber \\
  \Pi(y) &=& \Pi_1 y + \Pi_2 y^2+\Pi_3y^3+\Pi_4y^4+\ldots \,.
  \label{eq:seriesy}
\end{eqnarray}
Plugging these expansions into Eqs.~\eqref{eq:allyf} and considering
terms order by order, we obtain a set of equations for the coefficients
that leave\footnote{The coefficient $M$ turns out to equal the ADM
mass, $M_{\rm{ADM}}$; for simplicity we use $M$ in this discussion.}
$\Phi_0, M, \zeta_1, \varrho_1$ unspecified but uniquely determine
all other coefficients in terms of the free parameters. The latter
are eventually obtained by matching the exterior solution of
Eqs.~(\ref{eq:allyf}) to the interior solution of Eqs.~(\ref{eq:allr})
as described in more detail in Appendix \ref{app:numerics}.

Applying the series expansion for a massive gravitational scalar
field, we arrive at the following relations for the free parameters
in Eq.~\eqref{eq:rescaled_vars},
\begin{equation} \label{eq:asymptotics_massive}
k^2 = 1-\frac{\omega^2}{e^{2\Phi_0}}, \text{  }
\epsilon = M\frac{2k^2-1}{k}, \text{  }
\delta = Mh, \text{  } h^2 =\mu_{\varphi}^2.
\end{equation}
Let us now consider two limiting cases of these conditions. First,
we obtain the asymptotic behavior of BSs in GR by simply omitting
the $\varrho$ and $\Lambda$ coefficients intrinsic to ST gravity.
Second, by setting $M=0$, we recover the flat-field limit
(\ref{eq:wrongasymptotics}).

In the massless case $\mu_{\varphi} = 0$, we follow a similar
procedure, except now there is no exponential fall-off for the
gravitational scalar field, i.e. $h=0$ and $\delta = 0$.  Again,
applying the asymptotic ans\"{a}tze \eqref{eq:series_ansatze} and
considering terms order by order in Eqs.~\eqref{eq:allyf}, we arrive
at
\begin{eqnarray} \label{eq:asymptotics_massless}
k^2~=~1-\frac{\omega^2}{e^{2\Phi_0}}\,, \text{  }
\epsilon ~=~ M\frac{2k^2-1}{k} + \frac{\alpha_0\varrho_1}{k}.
\end{eqnarray}

In summary, the re-scaled variables in Eq.~\ref{eq:rescaled_vars}
together with the series expansions in Eq.~\ref{eq:series_ansatze}
imply that the bosonic and gravitational scalar fields have asymptotic
behaviours given by,
\begin{equation}
  \varphi \sim e^{\pm h / y}y^{1+\delta}\, ~~ \text{and} ~~
  A \sim e^{\pm k / y}y^{1+\epsilon}\,,
\end{equation}
where $h, k, \delta$ and $\epsilon$ are determined by
Eqs.~\eqref{eq:asymptotics_massless} and \eqref{eq:asymptotics_massive}
for massless and massive $\varphi$, respectively.

All frequencies we report in this work correspond to a unit lapse
function at infinity, i.e.~$\Phi_0=0$; even if a non-zero $\Phi_0$
is used in the numerical calculation, this is straightforwardly
achieved by an a-posteriori rescaling of $\omega$.  The frequencies
thus obtained ubiquitously fall into the range $0<\omega <1$. The
functional behaviour $k(\omega)=\sqrt{(1-\omega^2)}$ furthermore
pushes $k$ towards values rather close to unity in exactly the same
way the Lorentz factor $\gamma=1/\sqrt{1-v^2}$ reduces relativistic
effects for all velocities but those close to the speed of light.
This effect avoids extreme values of $\epsilon$ which we typically
find to be in the range $-2\lesssim \epsilon \lesssim 3$. An
unscalarized low-density mini BS with $M=0.345$ and $\omega=0.9785$,
for instance, has $\epsilon=-1.53$ whereas the strongly scalarized
thin-shell BS with $M=2.39$ and $\omega=0.148$ of the
$(\mu_{\varphi}=0,~\alpha_0=0,~\beta_0=-12,~\sigma_0=0.2)$ family
discussed in Sec.~\ref{sec:thinshell} has $\epsilon=2.31$.  The
exponential factor $h$ for the asymptotic behaviour of the gravitational
scalar, on the other hand, is directly given by the scalar mass
$\mu_{\varphi}$ and the exponent $\delta$ merely acquires an extra
factor of the BS mass $M$; $\delta$ vanishes in massless ST theory,
but can reach values up to order unity for the BS sequences studied
in this work.  We similarly obtain values $\varrho_1 =\mathcal{O}(1)$
for strongly scalarized stars.

Finally, let us illustrate where and how the flat-field asymptotics
go wrong, when used in the presence of gravity. Using the flat-field
asymptotic behaviour leads to rescaled scalar variables
\begin{eqnarray} \label{eq:rescaled_vars_wrong}
\zeta \defeq e^{k/y}A\,,
~~~~~&&~~~~~
\Pi\defeq -e^{k/y} \theta.
\end{eqnarray} 
At $\mathcal{O}(y^{-1})$, the equation for $\partial_y \Pi$ then
gives the first ingredient of our result in
Eq.~\eqref{eq:asymptotics_massive}, i.e. $k^2 = 1-\omega^2/e^{2\Phi_0}$.
However, as we proceed to consider the next leading contributions
at $\mathcal{O}(y^0)$ in the equations for $\partial_y \Pi$ and
$\partial_y \zeta$, we arrive at,
\begin{equation}
    2M\Pi_1(1-2\omega^2) = 0,
\end{equation}
which can only be satisfied when $M=0$ or when $\omega^2 = 1/2$
with $M \neq 0$. This analysis evidently results in a contradiction,
as it postulates that BSs do not exist for frequencies $\omega^2
\neq 1/2$. The inclusion of the non-trivial parameter $\epsilon$
in Eqs.~\eqref{eq:rescaled_vars} avoids this contradiction.
\label{page:asymptotics}
%

\section{Numerics}
\label{app:numerics}
We have written three different codes for the numerical solution
of the ODE system (\ref{eq:allr}) with the boundary conditions
(\ref{eq:BCs}).  This slightly unusual approach was motivated by
two main considerations. First, it was not entirely clear how the
challenging asymptotic behaviour of the two scalar fields might be
best controlled numerically. Second, in the absence of literature
results for comparison, we wished to verify our results with
independent means. Here, we briefly describe our schemes and assess
the uncertainty of our results.

{\it Numerical shooting:}
Our first code operates in a way similar to the BS shooting code
developed for GR in Ref.~\cite{Helfer:2021brt}.  Starting with
$\Phi=0$, $X=1/\sqrt{F(\varphi)}$, $A=A_{\rm ctr}$, $\varphi =
\varphi_{\rm ctr}$, $\kappa=0$, $\theta=0$ at $r=0$ as well as an
initial guess for the frequency eigenvalue $\omega$, the system
(\ref{eq:allr}) is integrated outwards using a fourth-order Runge
Kutta scheme. At some radius, similar to the GR case, the scalar
fields start to diverge and we deal with this divergence for each
of the fields individually in the following way:
\begin{enumerate}
    \item We find the radius where the scalar field modulus starts
    to exceed its central value by an order of magnitude (this
    usually happens sooner for $A$).
    \item We terminate the integration at that point and search for
    the last local extremum of the scalar field.
    \item As a buffer, we take $90 \%$ of the radius where that
    extremum occurs, which consequently determines our matching
    radius, $r_{\rm{m}}$.
    \item At $r_{\rm{m}}$, we match the integrated solution to the
    asymptotics presented in  Appendix \ref{app:asymptotic_behaviour}.
\end{enumerate}
In the massless case, we find that the gravitational scalar field
can be integrated out, without any matching, as long as the BS
scalar field is matched to its corresponding asymptotic expansion.
In the massive case, in contrast, both fields have to be matched
and this is done at radii $r_{{\rm m},A}$ and $r_{{\rm m},\varphi}$
which are not necessarily equal.  Once the full solution is
constructed, we assess the quality of the matching for both scalar
fields. In the massive case, we do so by computing smoothness
conditions at the matching radii:
\begin{align}
  C(\varphi) &= \left|\varphi + \frac{\varphi'(r_{\rm{m}, \varphi})}{1+\delta+r_{\rm{m}, \varphi}m_{\varphi}}r_{\rm{m}, \varphi} \right|\,, \\
  C(A) &= \left|A + \frac{A'(r_{\rm{m}, A})}{1+\epsilon + r_{\rm{m}, A}\sqrt{1-\omega^2}} r_{\rm{m}, A} \right|\,,
  \label{eq:code1crit}
\end{align}
where $\delta$ and $\epsilon$ are defined in
Eq.~\eqref{eq:asymptotics_massive}. These conditions are obtained
by translating the asymptotic expressions in Eq.~\eqref{eq:rescaled_vars}
into relations between the scalar variables and their derivatives.
In the massless case, we replace the $C(\varphi)$ criterion by an
analytic condition that relates the vacuum solution at the stellar
surface (shortly to be defined), $\varphi_{\rm s}$, to its asymptotic
value, $\varphi_0$, at $r=\infty$ \cite{Damour:1993hw},
\begin{equation} \label{eq:Damour_condition}
    \varphi_{\rm{s}} = \varphi_0 - \frac{X_{\rm{s}} \eta_{\rm{s}}}{\sqrt{(\partial_r \Phi_{\rm{s}})^2 + X_{\rm{s}} \eta_{\rm{s}}}} \rm{artanh} \frac{\sqrt{(\partial_r \Phi_{\rm{s}})^2 + X_{\rm{s}} \eta_{\rm{s}}}}{\partial_r \Phi_{\rm{s}} + 1/r_{\rm{s}}}.
\end{equation}
Here $\eta_s = \kappa_s/ \alpha_s$ and, in our convention,
$\varphi_0=0$. Since BSs do not have a surface, we compute the
gravitational scalar field at the outer edge of the grid,
$\varphi(r_{\rm{end}})$ (typically $r_{\rm{end}} = 1200$), and
evaluate the quality of the solution using a new criterion,
$C(\varphi_{\rm{massless}}) = \left|\varphi(r_{\rm{end}}) -
\varphi_{\rm{s}} \right|$.  Improved guesses for $\varphi_{\rm ctr}$
and $\omega$ are then constructed through a standard Newton-Raphson
method \cite{Press1992}, until the conditions $C(\varphi)$ and
$C(A)$ are satisfied at the required threshold level, chosen to be
$10^{-7}$ or smaller.

{\it Two-way shooting:} The second code operates with an inner
region $r\in [0,2]$ where Eqs.~(\ref{eq:allr}) are integrated outward
and an exterior region $y\in [0,\tfrac{1}{2}]$ where $y=1/r$ and
the BS equations (\ref{eq:allyf}) for the rescaled variables are
integrated inwards from infinity to $y=1/2$.  Whereas integrating
out of $r=0$ is straightforwardly achieved by using the regularity
of the variables, the integration out of $y=0$ can result in numerical
instabilities at very small $y$, likely due to the divisions by
positive powers of $y$ in several terms on the right-hand side of
Eq.~(\ref{eq:allyf}). We overcome these instabilities by using the
asymptotic expansion (\ref{eq:seriesy}) to a small but finite $y$,
typically $y=0.01$. We can reduce this value to arbitrarily small
level by increasing the number of grid points in the exterior and
employ this reduction in the uncertainty estimate through convergence
analysis.  A solution is obtained when the interior and exterior
solutions thus obtained satisfy the 6 matching conditions
(\ref{eq:matching}) and (\ref{eq:Phim}) at $r=2$ with a relative
precision of $10^{-6}$ or better. In order to start the integrations
at $r=0$ and $y=0$, we need to specify 7 parameters, $\varphi_{\rm
ctr}$, $A_{\rm ctr}$ and $\Phi_{\rm ctr}$ at the origin, $M$,
$\zeta_1$ and $\varrho_1$ at infinity as well as the frequency
$\omega$. With 7 free parameters for 6 matching conditions, we
obtain one-parameter families of solutions, as expected. In practice,
we choose one of the 7 free parameters to control the specific BS
model, initialize the other 6 parameters with initial guesses
typically obtained from neighbouring BSs, and solve the matching
conditons with a Newton-Raphson iteration.

{\it Relaxation:} The third code uses a Newton-Raphson based
relaxation scheme, implemented according to chapter 17.3 of
Ref.~\cite{Press1992}. Here, we again operate with an inner and an
outer region using the ODE system in the forms (\ref{eq:allr}) and
(\ref{eq:allyf}), respectively. The compactified grid and algorithm
allows us to impose exact boundary conditions at spatial infinity.
We also have the seven free parameters as in the two-way shooting
code which now impose 6 boundary conditions needed in the relaxation
scheme and leave one parameter free to control the specific model.
The initial guess for the iterative procedure is either obtained
from other BS solutions close by in parameter space, from the
shooting code discussed at the start of this subsection or from
outward integrated profiles starting with guesses for $A_{\rm ctr}$
and $\varphi_{\rm ctr}$ to which we append the exponentially decaying
behaviour on the outer grid.

We have assessed the numerical uncertainties of our codes by verifying
the respective convergence properties and by comparing the codes'
results directly. For all our diagnostics this results in a relative
error and agreement of $10^{-4}$ or better.

\vspace{1cm} 
\section{Boson-star families}
\label{app:BSfamilies}
\begin{figure*}
  \includegraphics[width=0.9\textwidth]{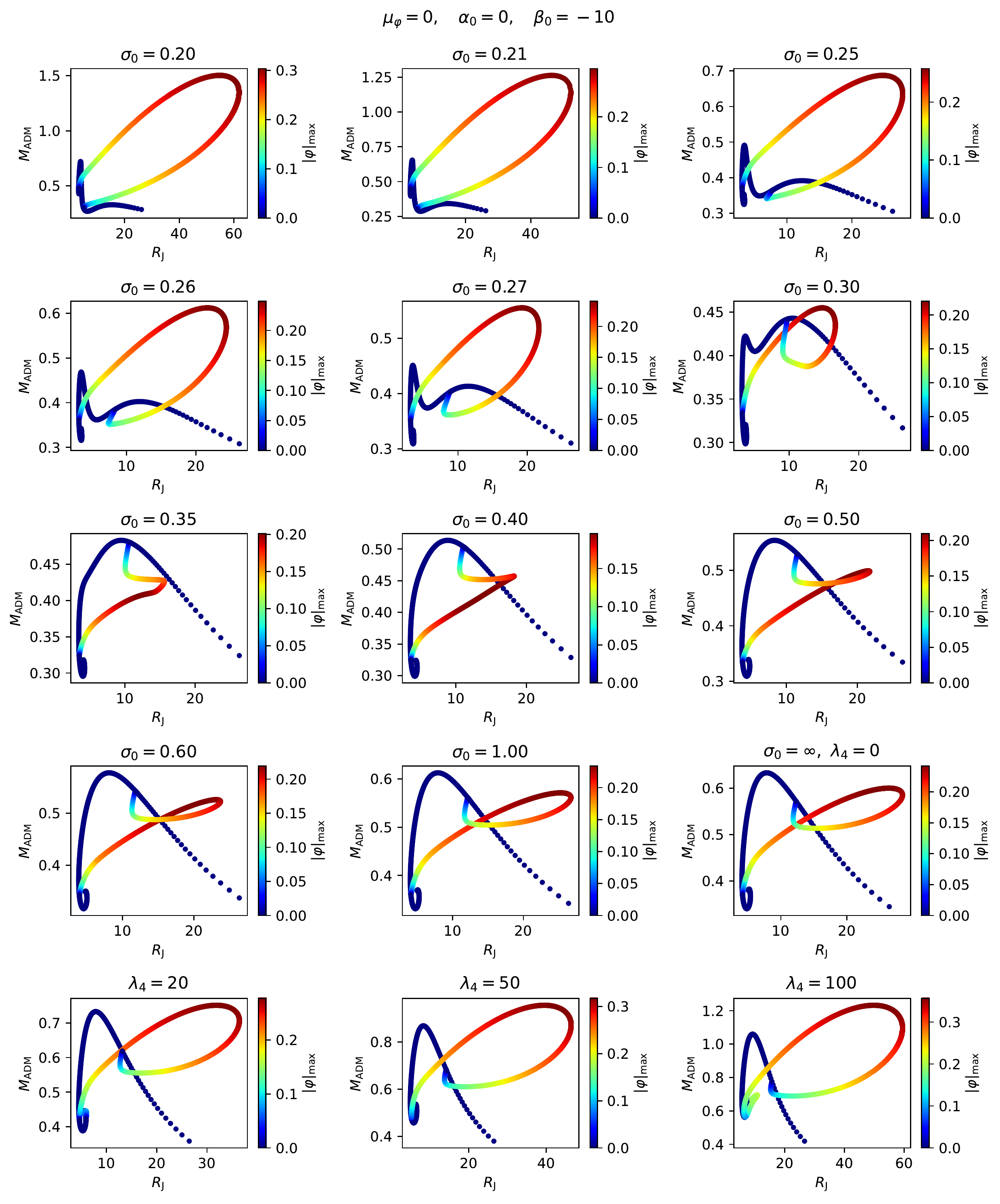}
  \caption{Boson-star families in the mass-radius plane for fixed
  ST parameters $\alpha_0=0$, $\beta_0=-10$, $\mu_{\varphi}=0$ and
  BS potential functions $V(A)$ as given in Eq.~(\ref{eq:BSpotentials}).
  The parameters $\sigma_0$ or $\lambda_4$ are specified in each
  panel for the corresponding BS family. The mini-BS case is recovered
  in either of the limits $\sigma_0 \rightarrow \infty$ or
  $\lambda_4=0$. The color bar marks the degree of scalarization
  in terms of the maximum of $|\varphi(r)|$.
  }
  \label{fig:BSfamilies}
\end{figure*}

In this appendix we illustrate the bulk properties of a wide range
of BS solutions in the form of mass-radius diagrams for various
potentials of the BS scalar. For this purpose, we have chosen a
massless gravitational scalar, $\mu_{\varphi}=0$, with vanishing
$\alpha_0$ (for non-zero $\alpha_0$ the branches would exhibit the
splitting discussed in Sec.~\ref{sec:alpha0ne0}), and $\beta_0=-10$
which results in strongly scalarized models for all potentials
considered. In each panel, the potential $V(A)$ is characterized
by the corresponding values of $\sigma_0$ for solitonic BSs and
$\lambda_4$ for massive BSs. The results shown in Fig.~\ref{fig:BSfamilies}
demonstrate a rather complex variation of the branches with $\sigma_0$
for solitonic BSs, including even one case, $\sigma_0=0.35$, where
all scalarized stars have a lower mass than their GR counterparts
with equal radius. The parameter $\lambda_4$ for massive BSs, in
constrast, has a relatively straightforward impact on the scalarized
BSs' bulk properties, leading to systematically larger radii and
masses as $\lambda_4$ is increased.

\begin{figure*}
  \includegraphics[width=0.9\textwidth]{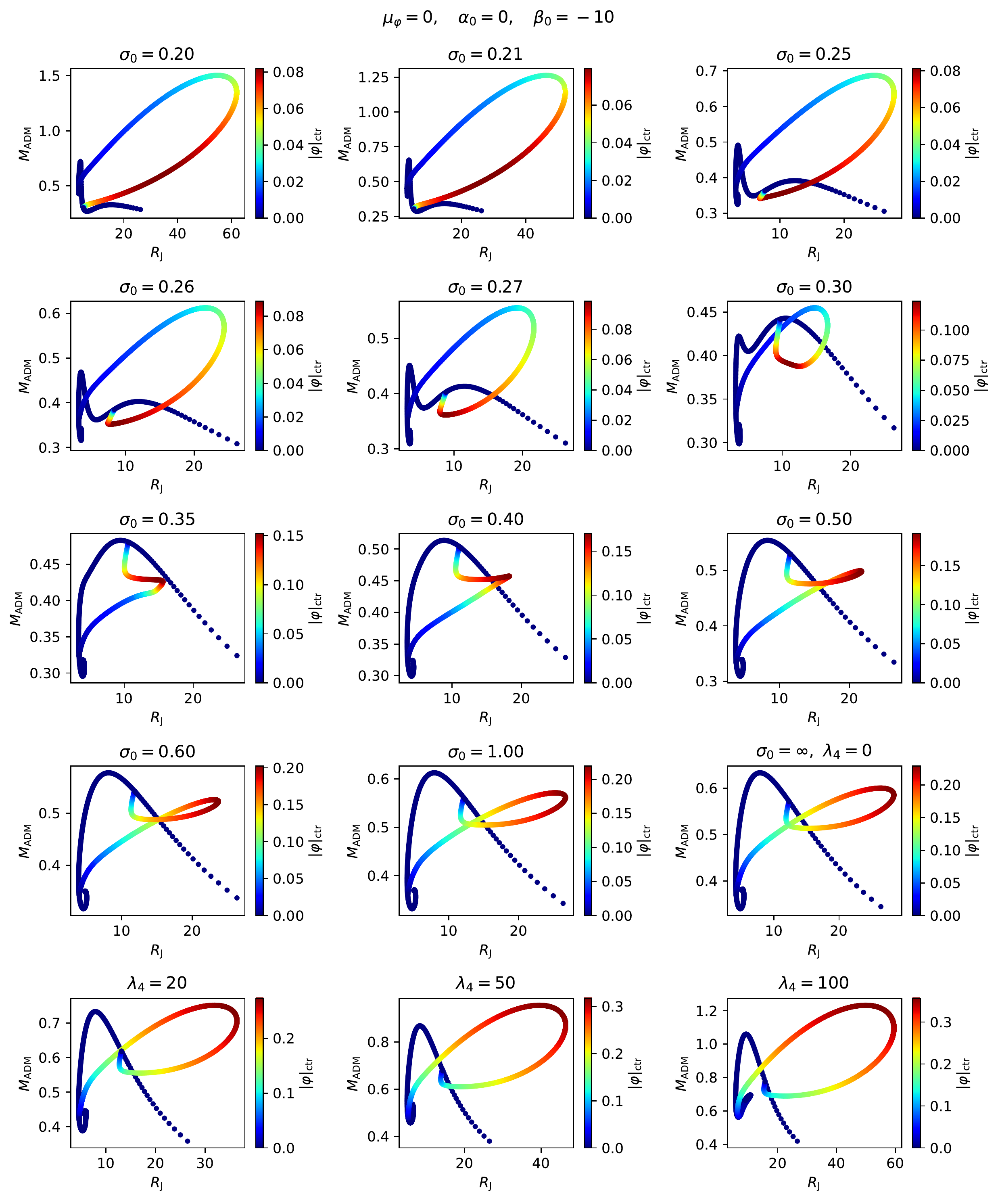}
  \caption{Same as Fig.~\ref{fig:BSfamilies} but with
  $|\varphi|_{\rm ctr}$ color coded.}
  \label{fig:families_vphctr}
\end{figure*}

\begin{figure*}
  \includegraphics[width=0.9\textwidth]{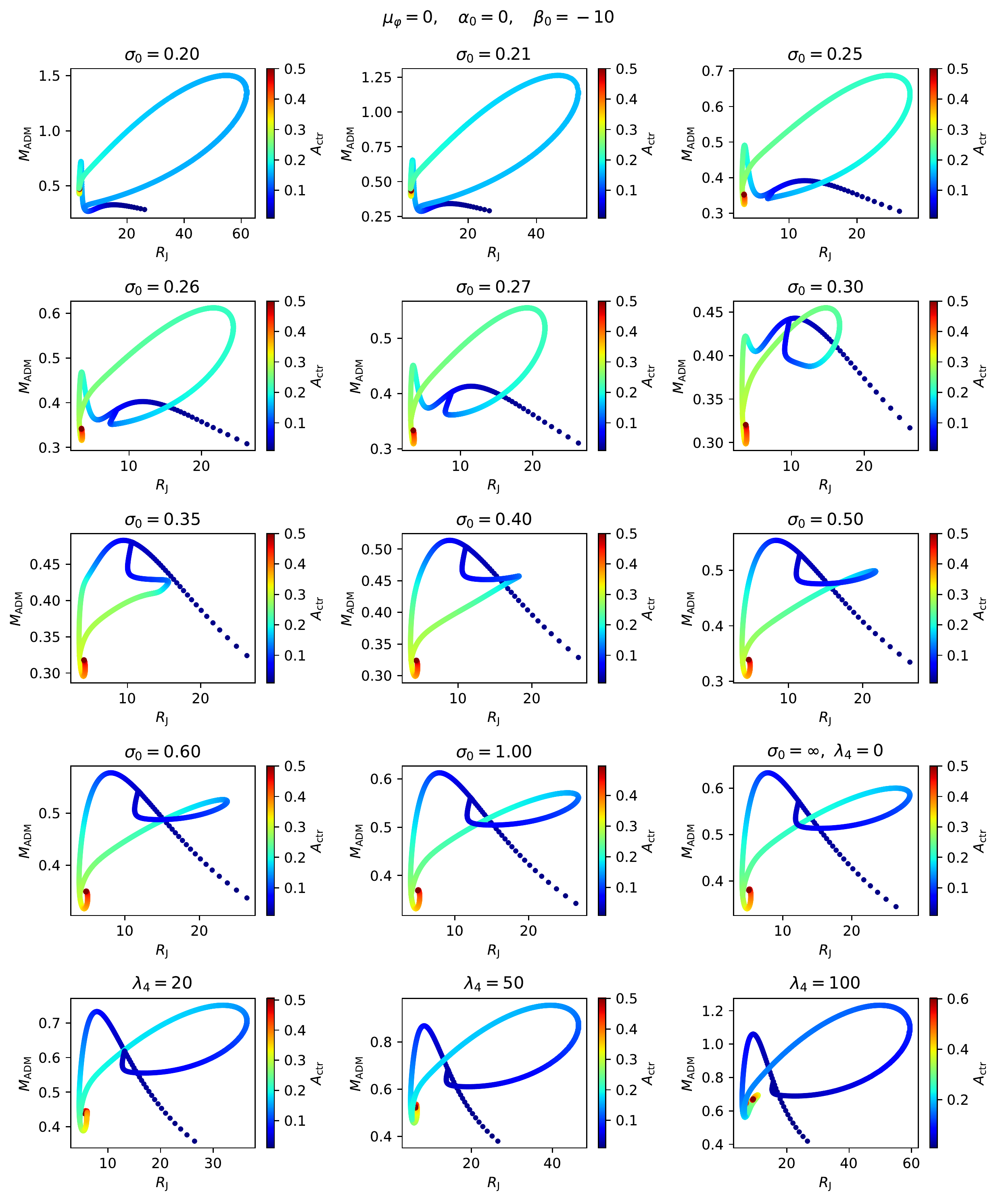}
  \caption{Same as Fig.~\ref{fig:BSfamilies} but with
  $A_{\rm ctr}$ color coded.}
  \label{fig:families_Actr}
\end{figure*}

\begin{figure*}
  \includegraphics[width=0.9\textwidth]{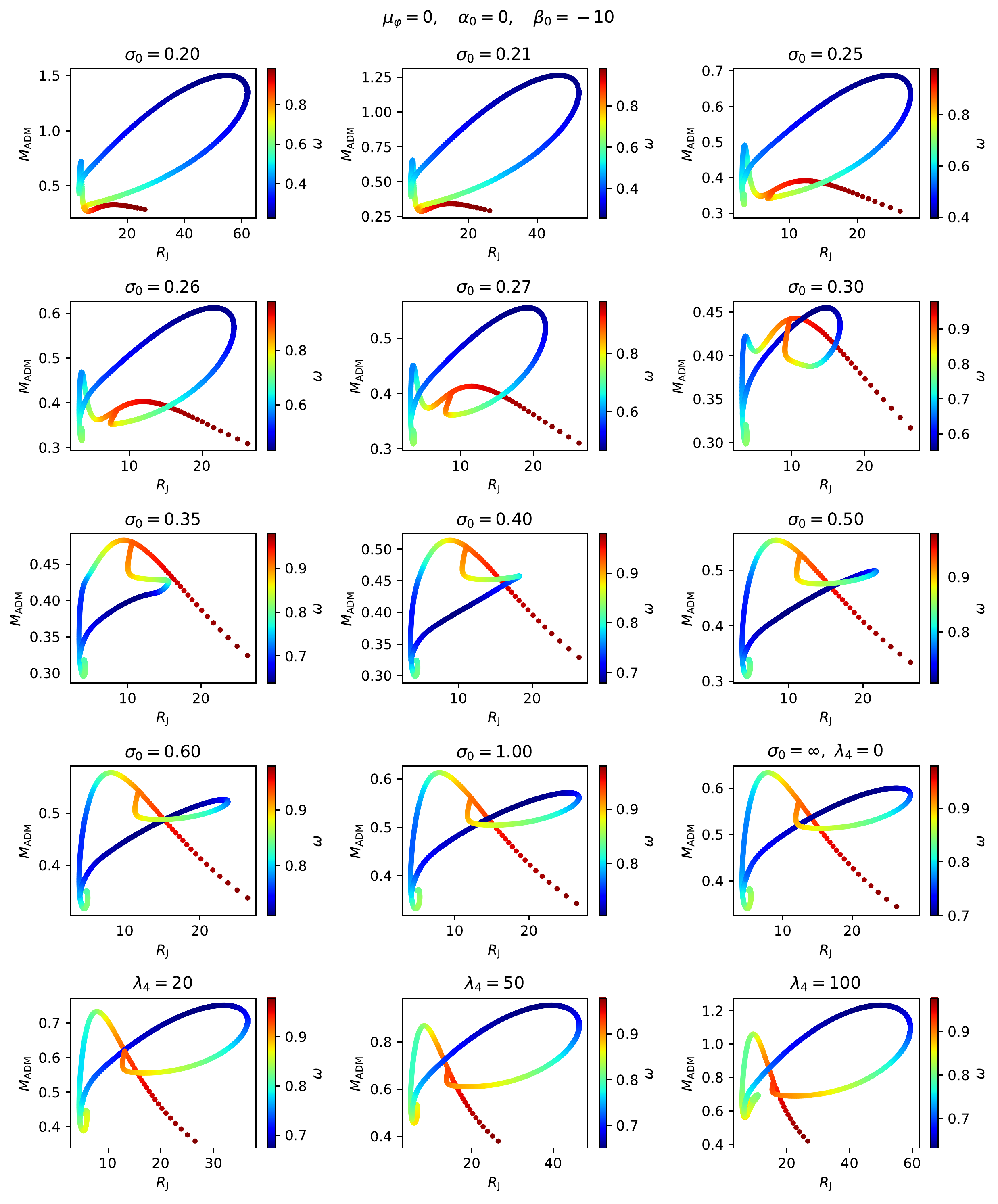}
  \caption{Same as Fig.~\ref{fig:BSfamilies} but with
  $\omega$ color coded.}
  \label{fig:families_omega}
\end{figure*}

In Figs.~\ref{fig:families_vphctr}-\ref{fig:families_omega}, we
display the same solutions but with $|\varphi|_{\rm ctr}$, $A_{\rm
ctr}$ and $\omega$ encoded in color. For solitonic potentials, the
maximal value $|\varphi|_{\rm max}$ generally exceeds the central
$|\varphi|_{\rm ctr}$, demonstrating that the gravitational scalar
tends to peak away from the origin. For the mini BS case and repulsive
potentials, on the other hand, $|\varphi|_{\rm max}$ and $|\varphi|_{\rm
ctr }$ are much closer which illustrates the trend towards $\varphi$
peaking near the origin.  For all potentials, Fig.~\ref{fig:families_omega}
exhibits lower frequency values along the scalarized branch as
compared to the GR models, confirming the observation made in the
bottom panel of Fig.~\ref{fig:thinshellMR} for thin-shell models.
As exemplified by the solitonic potentials for $\sigma_0\approx
0.35$, this is not simply a consequence of scalarized BSs being
larger, but a more complex phenomenon.

\begin{figure*}
  \includegraphics[width=0.9\textwidth]{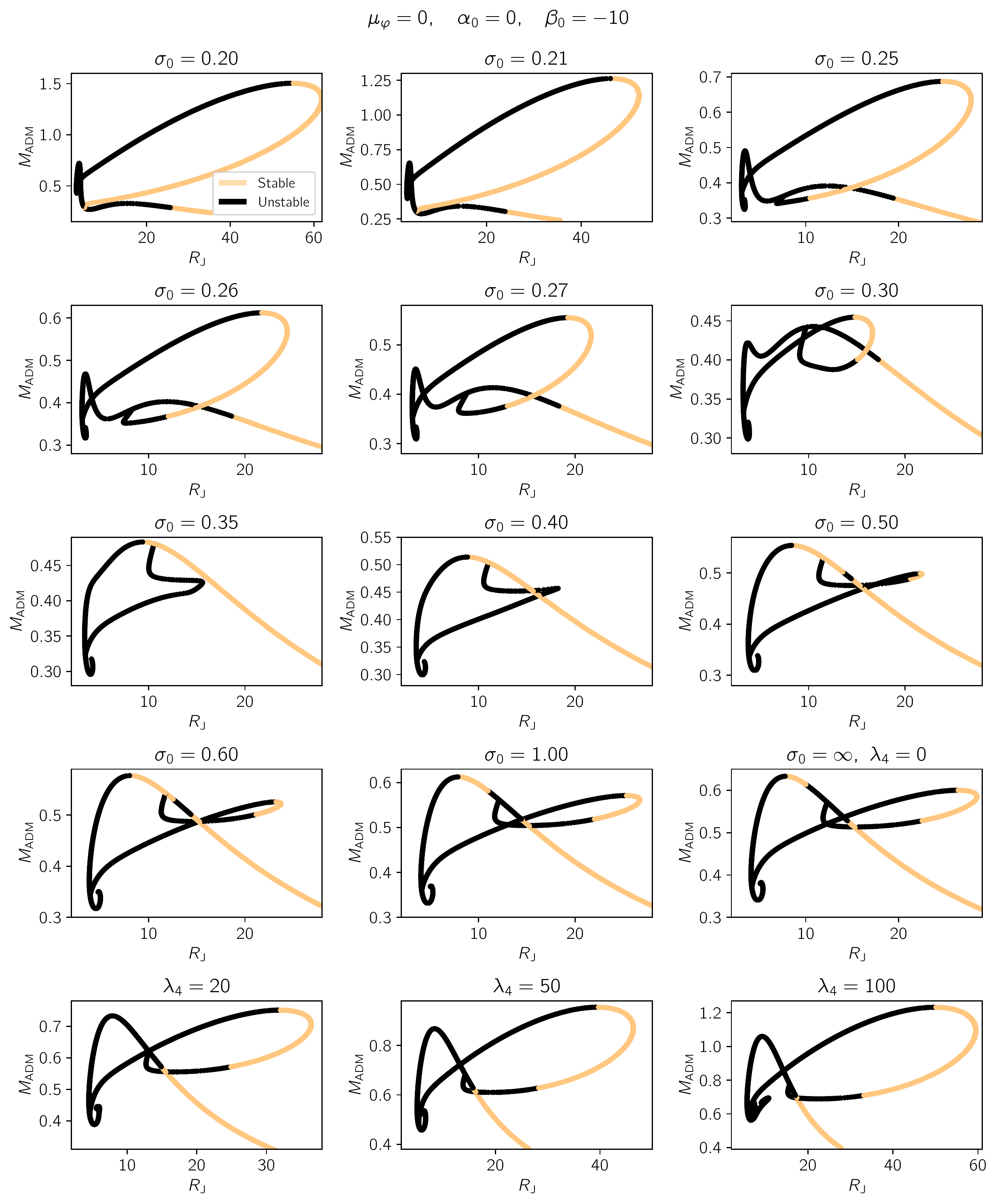}
  \caption{Same as Fig.~\ref{fig:BSfamilies} but with
  the stability coded in color, light (copper) for stable
  BSs and black for unstable models.
  }
  \label{fig:families_stability}
\end{figure*}

Finally, we mark in Fig.~\ref{fig:families_stability} the stable
and unstable models by light (copper) and dark (black) color. The
general trend is that scalarized models are energetically favored
as has also been observed for neutron stars. There are some exceptions
to this rule, however. (i) Scalarized stars need to be sufficiently
massive to become the stable member; cf.~the cases $\sigma_0=0.35$
and $0.4$ where scalarized BSs are mostly lighter than their GR
counterparts and are {\it not} stable. (ii) At the very low-mass
end, the GR models with very low compactness are favored over their
low-mass, high-compactness scalarized cousins. These features can
result in rather chequered stability curves as for example in the
case $\sigma_0=0.5$. Along both, GR and scalarized branches, the
transition from stable to unstable BS models often occurs at local
maxima in the mass viewed as a function of radius.
\end{document}